\documentclass[11pt,a4paper]{article}
\pdfoutput=1
\usepackage[utf8]{inputenc} 
\usepackage{jheppub}
\usepackage[T1]{fontenc}
\usepackage[english]{babel}
\usepackage{lmodern}
\usepackage{amsfonts}
\usepackage{url}
\usepackage{mmacells}
\usepackage{longtable}
\usepackage{mathtools}
\usepackage{xcolor}
\usepackage{slashed}
\usepackage{xspace}
\usepackage{braket}
\usepackage{placeins}
\allowdisplaybreaks

\newcommand*{\cf}{cf.\ }
\newcommand*{\ie}{i.e.\ }
\newcommand*{\eg}{e.g.\ }
\newcommand*{\Eq}{eq.\,}
\newcommand*{\Eqs}{eqs.\,}

\newcommand{\feyncalc}{\textsc{FeynCalc}\xspace}
\newcommand{\formcalc}{\textsc{FormCalc}\xspace}
\newcommand{\feynarts}{\textsc{FeynArts}\xspace}
\newcommand{\feynonium}{\textsc{FeynOnium}\xspace}
\newcommand{\feynrules}{\textsc{FeynRules}\xspace}
\newcommand{\pax}{\textsc{Package-X}\xspace}
\newcommand{\fire}{\textsc{FIRE}\xspace}

\newcommand{\feynhelpers}{\textsc{Feyn\-Help\-ers}\xspace}
\newcommand{\mma}{\textsc{Mathematica}\xspace}
\newcommand{\form}{\textsc{FORM}\xspace}

\newcommand{\reduce}{\textsc{Reduce}\xspace}

\newcommand{\formtracer}{\textsc{FormTracer}\xspace}
\newcommand{\maple}{\textsc{Maple}\xspace}
\newcommand{\qgraf}{\textsc{QGRAF}\xspace}
\newcommand{\hepmath}{\textsc{HepMath}\xspace}
\newcommand{\redberry}{\textsc{Redberry}\xspace}

\newcommand{\feyncalcformlink}{\textsc{FeynCalcFormLink}\xspace}

\newcommand{\bfeps}{\ensuremath{\boldsymbol{\varepsilon}}}
\newcommand{\bfSigma}{\ensuremath{\boldsymbol{\sigma}}}
\newcommand{\bfGamma}{\ensuremath{\boldsymbol{\gamma}}}

\newcommand{\bfDel}{\ensuremath{\boldsymbol{\nabla}}}
\newcommand{\bfPartial}{\ensuremath{\boldsymbol{\partial}}}
\newcommand{\bfE}{\ensuremath{\boldsymbol{E}}}
\newcommand{\bfA}{\ensuremath{\boldsymbol{A}}}
\newcommand{\bfB}{\ensuremath{\boldsymbol{B}}}
\newcommand{\bfD}{\ensuremath{\boldsymbol{D}}}
\newcommand{\bfk}{\ensuremath{\boldsymbol{k}}}
\newcommand{\bfl}{\ensuremath{\boldsymbol{l}}}
\newcommand{\bfp}{\ensuremath{\boldsymbol{p}}}
\newcommand{\bfn}{\ensuremath{\boldsymbol{n}}}
\newcommand{\bfx}{\ensuremath{\boldsymbol{x}}}
\newcommand{\bfq}{\ensuremath{\boldsymbol{q}}}
\newcommand{\bfr}{\ensuremath{\boldsymbol{r}}}
\newcommand{\bfR}{\ensuremath{\boldsymbol{R}}}
\newcommand{\bfO}{\ensuremath{\boldsymbol{O}}}
\newcommand{\bfP}{\ensuremath{\boldsymbol{P}}}
\newcommand{\bfK}{\ensuremath{\boldsymbol{K}}}

\DeclareMathOperator{\Tr}{Tr}

\mmaSet{morefv={gobble=2},morelst={breaklines=true}}
\mmaSet{morefv={gobble=2},morelst={breaklines=true}}

\title{FeynOnium: Using FeynCalc for automatic calculations in Nonrelativistic Effective Field Theories}

\author[a,b]{Nora Brambilla}
\author[a]{Hee Sok Chung}
\author[c,d,a]{Vladyslav Shtabovenko}
\author[a]{Antonio Vairo}

\emailAdd{nora.brambilla@ph.tum.de}
\emailAdd{heesok.chung@tum.de}
\emailAdd{v.shtabovenko@kit.edu}
\emailAdd{antonio.vairo@ph.tum.de}

\affiliation[a]{Technische Universität München, Physik-Department, James-Franck-Str. 1, 85748 Garching, Germany}
\affiliation[b]{Institute for Advanced Study, Technische Universität München, 
Lichtenbergstrasse 2a, 85748 Garching, Germany}
\affiliation[c]{Institut für Theoretische Teilchenphysik (TTP), Karlsruher Institut f\"ur Technologie (KIT), 76131 Karlsruhe, Germany}
\affiliation[d]{Zhejiang  Institute of Modern Physics, Department of Physics, Zhejiang University, Hangzhou 310027, China}

\abstract{
We present new results on \feynonium, an ongoing project to develop a general purpose software toolkit for semi-automatic symbolic calculations in nonrelativistic Effective Field Theories (EFTs). Building upon \feyncalc, an existing \mma package for symbolic evaluation of Feynman diagrams, we have created a powerful framework for automatizing calculations in nonrelativistic EFTs (NREFTs) at tree- and 1-loop level. This is achieved by exploiting the novel features of \feyncalc that support manipulations of Cartesian tensors, Pauli matrices and nonstandard loop integrals. Additional operations that are common in nonrelativistic EFT calculations are implemented in a dedicated add-on called \feynonium. While our current focus is on EFTs for strong interactions of heavy quarks, extensions to other systems that admit
a nonrelativistic EFT description are planned for the future. All our codes are open-source and publicly available. Furthermore, we provide several example calculations that demonstrate how \feynonium can be employed to reproduce known results from the literature.
}

\preprint{TUM-EFT 75/15, TTP19-021}

\begin{document}
\maketitle

\section{Introduction}

In the last decades we witnessed how Effective Field Theory (EFT) methods \cite{Wilson:1973jj,Weinberg:1978kz} were successfully applied to describe various phenomena governed by electromagnetic, weak, strong and gravitational interactions. A modern pedagogical introduction to the main ideas and techniques of EFTs can be found \eg in \cite{Petrov:2016azi,Pich:2018ltt,Manohar:2018aog}. Taking advantage of the hierarchy of widely separated dynamical scales found in many physical systems, we can construct suitable EFTs that precisely capture the behavior  of the given system at energies below a certain scale. The resulting EFT, which is based on the underlying symmetries, relevant degrees of freedom and power-counting rules, allows us to describe low energy phenomena in a simple but yet rigorous and systematic way.

From the technical point of view, calculations in EFTs are organized as expansions in small dimensionless parameters (\eg ratios of energy scales). The power-counting rules of the theory precisely tell us where the expansion should be truncated to achieve the precision we are aiming at. Even though the leading order predictions can be often obtained in a short pen and paper calculation, the usage of EFT methods does not imply that everything becomes trivial. On the contrary, the determination of higher order corrections routinely necessitates the usage of elaborate codes for automatic calculations. The time needed to develop such codes and subsequently run them on a sufficiently powerful computer often becomes a bottleneck in the task of obtaining higher order EFT predictions matching experimental accuracies.

Nonrelativistic Effective Field Theories (NREFTs) constitute 
a subbranch of EFTs for describing systems, where the relevant velocity scales are typically much smaller than the speed of light. Examples for such systems are nonrelativistic bound states (\eg positronium, muonium \cite{Caswell:1985ui}, heavy quarkonia \cite{Bodwin:1994jh,Brambilla:2004jw}) or systems made of nonrelativistic atoms \cite{Brambilla:2017ffe} and molecules \cite{Brambilla:2017uyf}. Even though NREFT methods are most commonly employed in nuclear and atomic physics, nowadays they are becoming increasingly popular for studying possible beyond the Standard Model scenarios such as nonrelativistic dark matter \cite{Hisano:2002fk,
Hisano:2003ec,Hisano:2004ds,Shepherd:2009sa,An:2015pva,Biondini:2018pwp,Biondini:2018xor,Beneke:2019vhz} or heavy neutrinos \cite{Biondini:2013xua}.

One can roughly distinguish between two ways to approach perturbative calculations in NREFTs. The first method attempts to ``hide'' the nonrelativistic nature of the theory by rewriting (whenever possible) operators and amplitudes in terms of Lorentz covariant quantities. In return, one hopes to benefit from existing codes for automatic calculations and to avoid dealing with nonrelativistic expressions as much as possible. The other approach is to embrace the loss of manifest Lorentz covariance and to perform the calculations directly, working with nonrelativistic integrals, Pauli matrices and Cartesian tensors.

In our view, the best strategy consists of finding the right balance between the two approaches without sacrificing any physical insight or 
computational convenience. In particular, we believe that depending on the size of the calculation and the questions one is trying to answer, it may be sometimes more advantageous to work with noncovariant quantities directly, rather than trying to eliminate them altogether.

However, since nowadays most calculations are carried out using computer codes, the choice between the two above mentioned approaches has also a technical dimension. Publicly available general purpose tools such as \feyncalc \cite{Mertig:1990an,Shtabovenko:2016sxi}, \formcalc \cite{Hahn:1998yk}, \pax \cite{Patel:2015tea,Patel:2016fam}, \hepmath \cite{Wiebusch:2014qba}, \formtracer \cite{Cyrol:2016zqb} and many others make it easy to automatize a manifestly Lorentz covariant tree- or 1-loop level calculation with comparably little effort and to obtain the desired analytic or numeric results. 

However, this is not the case once one becomes interested in performing the given calculation in a nonrelativistic fashion. While we do not see any intrinsic difficulties that make the automation of nonrelativistic calculations more challenging than the relativistic ones, one can hardly find any public codes applicable to this scenario. In principle, nothing prevents a programming-savvy user to implement the necessary operations in \form \cite{Vermaseren:2000nd,Kuipers:2012rf}, \mma, \maple or any other symbolic manipulation system. This is also what most practitioners usually do, when they face the necessity of carrying out a large-scale nonrelativistic calculation without being in the possession of suitable in-house codes. Unfortunately, the NREFT community suffers from a visible lack of interest in making such codes publicly available, which effectively means that a lot of people have no other choice than to write their codes from scratch.

In our view, this situation is very unfortunate and deserves to be improved. Our contribution to the solution of this problem and the novelty of this work is, therefore, to provide software packages that are publicly available, well documented, easy to use and most importantly suitable for automatizing nonrelativistic calculations. What is more, we will also explicitly show how these tools can be used to reproduce important results from the literature. 

The \feynonium project started in 2016 \cite{Shtabovenko:2017iqw} and its main focus is still directed towards tree- and 1-loop level calculations in Nonrelativistic QCD (NRQCD) \cite{Caswell:1985ui,Bodwin:1994jh} and potential Nonrelativistic QCD (pNRQCD) \cite{Pineda:1997bj,Brambilla:1999xf} as well as the electromagnetic counterparts Nonrelativistic QED (NRQED) and potential Nonrelativistic QED (pNRQED).
However, as it will become clear in the course of the paper, most of the provided routines are in no way limited to a particular theory and can be employed in very generic nonrelativistic calculations. Our key deliverables are a new version of the \feyncalc package \cite{Shtabovenko:2020gxv}, capable of dealing with nonrelativistic quantities out of the box and a special add-on (also called \feynonium) for NREFTs.
  
This paper is organized in the following way. In section \ref{sec:pnrqcd} we introduce NRQCD and pNRQCD, which will appear in many of our example calculations. To set the stage for our tools, we provide a brief overview of the existing codes for EFT calculations in section \ref{sec:tools}, while our technical implementation is described in section \ref{sec:feyncalc93}. The installation and usage of the packages are explained in section \ref{sec:usage}. In section \ref{sec:examples} we demonstrate how the presented codes can be employed to reproduce some well-known (NR)EFT results from the literature. Our conclusions and possible future directions of this work are summarized in section \ref{sec:summary}.
We provide useful formulas for algebraic manipulations of Pauli matrices 
in appendix\,\ref{sec:app-pauli}, 
and we list the Lorentz and Cartesian tensors in \feyncalc's internal (\texttt{FCI}-notation) and external (\texttt{FCE}-notation) notations in appendix\,\ref{sec:app-fcifce}. 
Finally, we present derivations of NRQCD and pNRQCD Feynman rules in appendix\,\ref{sec:app-frules}.

\section{Nonrelativistic QCD and potential nonrelativistic QCD} \label{sec:pnrqcd}

NRQCD is an effective field theory of QCD that is appropriate for describing
bound states of a heavy quark and a heavy antiquark, like heavy quarkonia. 
The heavy quark and the heavy antiquark with mass $m$ 
have a typical velocity $v$ inside the 
bound state, which is the small expansion parameter of this EFT. The degrees of
freedom of NRQCD are the two-component Pauli spinor fields $\psi$ and $\chi$ 
that describe the heavy quark and the heavy antiquark, respectively, 
which interact with the gluon field $A$ through the Lagrangian, given up to 
order $1/m$ by \cite{Caswell:1985ui,Bodwin:1994jh}
\begin{eqnarray}
\label{eq:NRQCD_Lagrangian}
\mathcal{L}_{\textrm{NRQCD}} &=&  
- \frac{1}{4} G^a_{\mu \nu} G^{\mu \nu a}
+ \sum_{i=1}^{n_f} \bar{q}_i i \slashed{D} q_i
\nonumber \\ && + 
\psi^\dagger \left(  i D^0 + c_k \frac{{\bf D}^2}{2m} 
+ \frac{c_F}{2 m} \bfSigma \cdot g {\bf B}^a T^a 
 \right ) \psi 
\nonumber \\ && + 
\chi^\dagger \left(  i D^0 - c_k \frac{{\bf D}^2}{2m}  
- \frac{c_F}{2 m} \bfSigma \cdot g {\bf B}^a T^a 
\right ) \chi 
+ \mathcal{O}\left( \frac{1}{m^2} \right),
\end{eqnarray}
where $i D^0 = i \partial^0 - g A^0$, 
$i {\bf D} = i {\bf \nabla} + g {\bf A}$, 
$G_{\mu \nu}^a = \partial_\mu A_\nu^a - \partial_\nu A_\mu^a + g f^{abc}
A_\mu^b A_\nu^c$, 
${\bf B}^{ai} = \frac{1}{2} \epsilon^{ijk} G^{akj}$, 
$q_i$ are massless quark fields with flavor $i$, 
and $c_n$ are the matching coefficients of NRQCD. 
The heavy-quark mass $m$ that appears in the NRQCD Lagrangian is the pole mass.
At order $1/m^2$ and beyond, operators of higher dimensions appear, which include heavy quark bilinears, four-quark operators, and gluonic operators. 

The velocity expansion in NRQCD is an expansion in powers of momentum divided by
the heavy quark mass $m$, where the momentum may scale like $mv$, the typical size of 3-momenta of the heavy quark and antiquark in 
the quarkonium, or $mv^2$, the typical size of the binding energy. 
This is similar to the heavy-quark effective theory (HQET) \cite{Isgur:1989vq,Isgur:1989ed,Eichten:1990vp,Georgi:1990um,Grinstein:1990mj},
where the expansion parameter is $\Lambda_{\textrm{QCD}}/m$. Both cases can be
regarded as a formal expansion in powers of $1/m$, and the two effective field
theories have the same Lagrangian in the two-fermion sector, 
although they have different power counting rules. 

The matching coefficients $c_n$ are determined by requiring the EFT to
reproduce QCD for processes involving nonrelativistic heavy quarks. That is, 
\begin{equation}
\label{eq:matching}
i \mathcal{M}_{\textrm{QCD}} (A \to B) = \sum_n c_n \braket{B| O_n |A}, 
\end{equation}
where $\mathcal{M}_{\textrm{QCD}}$ is a QCD amplitude, 
$O_n$ are NRQCD operators, with corresponding matching coefficients $c_n$. 
The NRQCD matrix elements $\braket{B|O_n|A}$ are computed with the same
initial and final states as the QCD amplitude. The NRQCD matrix elements 
scale in $v$, and hence, the sum over $n$ is organized in powers of $v$. 
In practice, in order to work with a finite number of NRQCD matrix elements, 
the sum is truncated at a given order in $v$. 

The matching coefficients $c_n$ can be determined perturbatively, by computing the QCD amplitude 
on the left-hand side of \Eq\eqref{eq:matching} in perturbative QCD (pQCD),
and the NRQCD matrix elements on the right-hand  side of \Eq\eqref{eq:matching} in perturbative NRQCD. The Feynman rules of perturbative NRQCD are listed in appendix \ref{sec:app-nrqcd}. 
Then, the $c_n$ are determined by requiring that the
right-hand side of \Eq\eqref{eq:matching} reproduces the perturbative 
QCD amplitude on the left-hand side to a desired accuracy in an expansion in
powers of the momenta of the heavy quarks, antiquarks, and soft gluons that
appear in the perturbative amplitude. 

Unlike what is usually done in perturbative QCD, 
perturbative NRQCD calculations are organized in terms of nonrelativistic
quantities like the 3-momenta of quarks and gluons. 
Two-component Pauli spinors and Pauli matrices handle the heavy-quark spin. 
On the other hand, amplitudes in perturbative QCD are usually given in terms
of relativistically covariant quantities, like 4-momenta, gamma matrices and
Dirac spinors. Hence, in order to compute $c_n$ from \Eq\eqref{eq:matching}, 
it is necessary to rewrite the pQCD amplitude in terms of
nonrelativistic quantities so that it can be compared with the NRQCD matrix
elements. That is, we need to rewrite 4-momenta in terms of 3-momenta,
gamma matrices in terms of Pauli matrices, and Dirac spinors in terms of
two-component Pauli spinors. 
This can involve a considerable amount of nonrelativistic algebra that is best
done on a computer. 

NRQCD involves two dynamical scales $mv$ and $mv^2$.
When $mv \gg  \Lambda_{\textrm{QCD}}$, the scale
$mv$ can be integrated out perturbatively to obtain a new effective
field theory called potential NRQCD (pNRQCD).
The degrees of freedom of pNRQCD are a singlet and an octet field,
low energy gluons and light quarks.
Since for the lowest quarkonium resonances the typical size of the
relative coordinate is smaller than the inverse on the confinement
scale $\Lambda_{\textrm{QCD}}$, one can employ the multipole expansion at the
Lagrangian level. The heavy quark sector of the pNRQCD Lagrangian in the weakly coupled case
($r \ll \Lambda_{\textrm{QCD}}^{-1}$) at
next-to-leading order in the multipole expansion
and at leading order in the $1/m$ expansion
is given by
\cite{Pineda:1997bj,Brambilla:1999xf}
\begin{align}
\label{eq:pNRQCD_Lagrangian}
\mathcal{L}_\textrm{pNRQCD} \biggl |_{\textrm{heavy quark}} &= \Tr \left \{ \textrm{S}^\dagger \left (i
\partial_0 - h_s (r) \right ) \textrm{S} + 
\textrm{O}^\dagger \left  (i D_0 -h_o(r) \right ) \textrm{O} \right \} 
\nonumber \\
& + g V_A (r) \Tr \{ \textrm{O}^\dagger \bfr \cdot \bfE \textrm{S}
+ \textrm{S}^\dagger \bfr \cdot \bfE  \textrm{O} \} 
+ g \frac{V_B (r)}{2} \Tr \left \{ \textrm{O}^\dagger 
\{ \bfr \cdot \bfE, \textrm{O} \} \right \}, 
\end{align}
where the $\textrm{S}$ and $\textrm{O}$ are the singlet and octet fields,
respectively, that depend on time, the relative coordinate $\bfr$ and the
center-of-mass coordinate $\bfR$. They have the following color indices:
\begin{equation}
\textrm{S}_{ij} (\bfr, \bfR,t) = \frac{\delta_{ij}}{\sqrt{N_c}} S(\bfr, \bfR,t),  \quad
\textrm{O}_{ij} (\bfr, \bfR,t) = \frac{T^a_{ij}}{\sqrt{T_F}} O^a(\bfr, \bfR,t),
\label{eq:sofields}
\end{equation}
where $N_c$ is the number of colors and $T_F = 1/2$. All gluon fields in \Eq\eqref{eq:pNRQCD_Lagrangian}, such as the
chromoelectric field $\bfE^i = G^{ai0} T^a$ and the covariant
derivative $i D_0 \textrm{O} = i \partial_0 \textrm{O} - g [A_0
(\bfR,t),\textrm{O}]$ are evaluated at the center-of-mass coordinate $\bfR$. 

The light part of the pNRQCD Lagrangian is the same as in
\Eq\eqref{eq:NRQCD_Lagrangian}.
The singlet and octet Hamiltonians $h_s$ and $h_o$ 
can be split into a kinetic term (for simplicity we just write the leading one) and a potential
\begin{eqnarray}
h_s (\bfr, \bfp, {\bf S}_1, {\bf S}_2) &=& 
\frac{\bfp^2}{m} + V_s (\bfr, \bfp, {\bf S}_1, {\bf S}_2), 
\\
h_o (\bfr, \bfp, {\bf S}_1, {\bf S}_2) &=& 
\frac{\bfp^2}{m} + V_o (\bfr, \bfp, {\bf S}_1, {\bf S}_2), 
\end{eqnarray}
where $V_s$ and $V_o$ are the color-singlet and color-octet potentials,
respectively. 
The functions $V_A(r)$ and $V_B(r)$, as well as the potentials 
$V_s(\bfr, \bfp, {\bf S}_1, {\bf S}_2)$ and 
$V_o(\bfr, \bfp, {\bf S}_1, {\bf S}_2)$ are the matching coefficients of
pNRQCD. 

The matching between NRQCD and pNRQCD in the weak coupling regime 
can be carried out by requiring the Green's functions in NRQCD and pNRQCD to be equal order by order in $1/m$, $\alpha_s$ and $r$.  In the case of perturbative matching, Green's functions in pNRQCD are computed using the pNRQCD Feynman rules, which are listed in appendix \ref{sec:app-pnrqcd}. As it was the case for the matching between QCD and NRQCD, also the matching between NRQCD and pNRQCD involves 3-dimensional vectors, which calls for a computer
environment that can deal with nonrelativistic algebra.

\section{Existing approaches to automatic EFT calculations} \label{sec:tools}

Many of the publicly available packages for EFT calculations aim at exploring the phenomenology of the Standard Model extended with nonrenormalizable operators, the so-called Standard Model Effective Field Theory (SMEFT) \cite{Buchmuller:1985jz,Grzadkowski:2010es}, at tree- or 1-loop level.
In practice one usually integrates out heavy fields from an assumed underlying theory of the physics beyond the Standard Model and constructs the corresponding EFT operators. Other common tasks include deriving operator bases from the given set of symmetries, switching between different  operator bases, computing the renormalization group evolution of Wilson coefficients or extracting Feynman rules from the effective Lagrangian.

Such calculations can be automatized using tools such as \textsc{Rosetta} \cite{Falkowski:2015wza}, \textsc{SMEFTsim} \cite{Brivio:2017btx}, \textsc{MatchingTools} \cite{Criado:2017khh}, \textsc{CoDEx} \cite{Bakshi:2018ics}, \textsc{Wilson} \cite{Aebischer:2018bkb}, \textsc{DEFT} \cite{Gripaios:2018zrz} \textsc{SmeftFR} \cite{Dedes:2019uzs}, \textsc{BasisGen} \cite{Criado:2019ugp}, \textsc{Sym2Int} \cite{Fonseca:2017lem}, \textsc{ECO} \cite{Marinissen:2020jmb}, \textsc{GrIP} \cite{Banerjee:2020bym} and many others. \textsc{WCxf} \cite{Aebischer:2017ugx} provides a special file format for exchanging Wilson coefficients of operators appearing in the SMEFT Lagrangian between different codes, while \textsc{FeynRules} \cite{Christensen:2008py,Alloul:2013bka} can be regarded as a multipurpose tool for the Feynman rule derivation. Automatic calculation of the UV-renormalization constants is made possible by the \textsc{NLOCT} \cite{Degrande:2014vpa} package.

An explicit evaluation of Feynman amplitudes for a given process usually lies beyond the scope of such packages. This part of the calculation can be accomplished \eg by exporting the Feynman rules for the relevant part of the effective Lagrangian to some common format
and then employing suitable codes for perturbative calculations. For example, a model in the \textsc{UFO} \cite{Degrande:2011ua} format can be imported into popular tools such as \textsc{MadGraph5\_aMC@NLO} \cite{Alwall:2014hca}, \textsc{GoSam} \cite{Cullen:2011ac,Cullen:2014yla}, \textsc{Herwig++} \cite{Bahr:2008pv}, \textsc{Sherpa} \cite{Gleisberg:2008ta}, \textsc{Whizard} \cite{Moretti:2001zz,Kilian:2007gr}, \textsc{CalcHep} \cite{Belyaev:2012qa}, \textsc{CompHep} \cite{Boos:2004kh} and many others. Diagrams from a \feynarts \cite{Hahn:2000kx} model can be directly computed with \eg \formcalc, \feyncalc or \hepmath.

The authors of the above-mentioned EFT codes often stress that their packages are not limited to SMEFT, but can be also employed for more generic theories. While this is certainly true, such theories are nonetheless expected to be manifestly Lorentz covariant, which is problematic for NREFT calculations. Additional limitations equally apply to EFTs that contain nonstandard propagators such as eikonal propagators in HQET, soft-collinear effective theory (SCET) \cite{Bauer:2000ew,Bauer:2000yr,Bauer:2001ct,Bauer:2001yt,Bauer:2002nz,Beneke:2002ph} or chiral perturbation theory (ChPT) \cite{Weinberg:1978kz,Gasser:1983yg,Gasser:1984gg}.

Some aspects of dark matter studies in an EFT framework can be automatized with 
\textsc{DirectDM} \cite{Bishara:2017nnn}, which can 
match the user-provided relativistic high-energy theory onto a low-energy EFT in which dark matter particles interact with nonrelativistic nucleons \cite{Bishara:2016hek,Bishara:2017pfq}. The matching is nonperturbative and is done at leading order (LO) in the chiral expansion (\ie one expands in the momentum transfer instead of the strong coupling constant). Furthermore, the determination of the Wilson coefficients is performed in a fully automatic fashion and does not require any explicit manipulations of nonrelativistic quantities. This is very different from the approach we follow in \feynonium, where the user is required to carry out the matching calculation explicitly but can do so in a much more flexible way.

As far as EFTs of strong interactions are concerned, the number of useful publicly available codes is rather low. For mesonic ChPT, the package \textsc{Ampcalculator} \cite{Unterdorfer:2005au} can be employed to automatically calculate selected processes up to 1-loop. \textsc{PHI}\footnote{\url{http://www.feyncalc.org/phi}}, a \feyncalc add-on developed by F.\,Orellana to generate and manipulate amplitudes in generic ChPT processes is, unfortunately, not compatible to the current version of \feyncalc. Integrals arising from propagator diagrams in HQET (up to 3-loops) can be automatically evaluated with \textsc{Grinder} \cite{Grozin:2000jv}, a special package available for \textsc{REDUCE} and \textsc{Axiom} computer algebra systems. In the case of SCET, numerical calculation of soft functions at NNLO is possible with \textsc{SoftSERVE} \cite{Bell:2018oqa}.
Regarding NRQCD, tree-level amplitudes for heavy quarkonium production and decay processes can be generated with \textsc{MadOnia} \cite{Artoisenet:2007qm} or \textsc{HELAC-ONIA} \cite{Shao:2012iz,Shao:2015vga}. A library of amplitudes for the heavy quarkonium hadroproduction at NLO
that were already evaluated with the private \textsc{FDC} \cite{Wang:2004du} code is available via the \textsc{FDCHQHP} package \cite{Wan:2014vka}.

When applying EFT methods to strong interactions, many practitioners prefer to rely on their in-house codes, which often combine multiple public and private tools in one framework. The first step usually involves the diagram generation, which can be accomplished with \feynarts or \qgraf \cite{Nogueira:1991ex}. After that, the output can be processed with suitable \form or \mma codes, although one might also want to use other computer algebra systems  such as \maple, \reduce or \redberry\cite{Bolotin:2013qgr}. The codes can be either completely self-written or based on publicly available tools like \feyncalc and \feyncalcformlink \cite{Feng:2012tk}. After having carried out all the necessary algebraic simplifications, one would like to evaluate the resulting loop integrals either symbolically or numerically. The loop integral calculus is an interesting topic on its own and we refer to \cite{Smirnov:2006ry} for a pedagogical introduction to the existing methods. Let us merely remark that the simpler (only few mass scales and legs) 1-loop EFT integrals can be very often calculated analytically via a direct application of the Feynman parametrization. Of course, more complicated cases might still require more elaborate techniques, such as Integration-By-Parts reduction (IBP) \cite{Chetyrkin:1981qh,Tkachov:1981wb}, differential equations \cite{Kotikov:1991pm,Kotikov:1990kg,Kotikov:1991hm,Bern:1993kr,Remiddi:1997ny,Gehrmann:1999as}, sector decomposition \cite{Binoth:2000ps,Binoth:2003ak,Binoth:2004jv} or Mellin-Barnes representation \cite{Smirnov:1999gc,Tausk:1999vh,Anastasiou:2005cb,Czakon:2005rk}. If the quantity one wants to calculate depends on the phase-space integration over a squared matrix element (possibly multiplied with other functions), it is common to do the evaluation using numerical methods. Unless the final state involves at most 2 or 3 legs, analytic results are very difficult to obtain, irrespective of whether one calculates the phase-space integrals directly or employs special methods such as reverse unitarity \cite{Anastasiou:2002yz,Anastasiou:2003yy}.

The obvious difference between the existing approaches that rely on private codes and \feynonium is not only that our codes are public but also that we are explicitly interested in providing the complete scripts required to reproduce a particular result. This should hopefully motivate other members of the EFT community to share their software tools and also make the EFT techniques more accessible to a broader audience, including students and researchers working in different areas of quantum field theory.

\section{\feyncalc 9.3} \label{sec:feyncalc93}

Turning \feyncalc into a tool that could support both relativistic and nonrelativistic calculations on the same footing was a challenging endeavor, both technically and conceptually. The reason is that \feyncalc was originally created to work with manifestly Lorentz covariant quantities, therefore it was not possible to design everything from scratch, but one had to ensure that the new features nicely fitted into the existing framework.

One of the main goals was to preserve backward compatibility and to allow the user to employ already familiar functions such as \texttt{Contract}, \texttt{Expand\-Scalar\-Product} or \texttt{Dirac\-Simplify} without worrying whether the input contained nonrelativistic expressions or not. New methods were added only for manipulations that were not available or not required in the previous \feyncalc version, \eg \texttt{Lorentz\-To\-Cartesian} for breaking manifest Lorentz covariance. This means that Cartesian tensors (just as Lorentz tensors) now belong to the most fundamental quantities that can be manipulated using \feyncalc.

While it is not our scope to give a full account of the implemented modifications, in the following we will describe the main design decisions that were taken to make \feyncalc useful for nonrelativistic calculations. This should hopefully help the reader to gain a better a feeling for the new abilities of the package.

\subsection{Lorentz and Cartesian indices and vectors}

The three fundamental \feyncalc objects used to manipulate 4-vectors, corresponding to scalar products, Levi-Civita tensors and Dirac matrices, are called \texttt{Pair}, \texttt{Eps} and \texttt{DiracGamma}, respectively. Essentially, \texttt{Pair} is a symmetric function with two slots. Each of the slots can accept two types of arguments, which are \texttt{LorentzIndex} (for Lorentz indices) and \texttt{Momentum} (for 4-momenta). Both of them also have two arguments. The first argument of \texttt{LorentzIndex} denotes the name of the corresponding index (\eg $\mu$, $\nu$, $\rho$ \ldots), while the first argument of \texttt{Momentum} specifies the name of the 4-momentum(\eg $p$, $q$, $l$ \ldots). The second argument of both functions fixes the spacetime dimension, which can be 4, $D$ or $D-4$.  Moreover, the second argument is optional and when it is missing the spacetime dimension defaults to 4. Depending on the combination of its arguments \texttt{Pair} may represent a Lorentz vector (\texttt{LorentzIndex} and \texttt{Momentum}), a metric tensor (twice \texttt{LorentzIndex}) or a scalar product (twice \texttt{Momentum}). As far as \texttt{DiracGamma} is concerned, its first argument (\texttt{LorentzIndex} or \texttt{Momentum}) specifies whether we have a Dirac matrix with a free Lorentz index $\gamma^\mu$ or a Feynman slash $\slashed{p}$, while the optional second slot is used for setting the spacetime dimension. The representation of Levi-Civita tensors follows the same pattern, with \texttt{Eps} being a function that has four slots for \texttt{LorentzIndex}- or \texttt{Momentum}-type arguments.

This symbolic representation of Quantum Field Theory (QFT) quantities within \feyncalc is called \emph{internal} or \texttt{FCI}-notation. In addition to that, \feyncalc is also equipped with an \emph{external} or \texttt{FCE}-notation, which consists of convenient shortcuts that are useful for the manual input or when exporting \feyncalc results to other programs. \feyncalc functions usually output the results in the internal notation but accept the input in both notations. The routines for switching between the two notations are \texttt{Feyn\-Calc\-Internal} (abbreviated with \texttt{FCI}) and \texttt{Feyn\-Calc\-External} (abbreviated with \texttt{FCE}). For example, to input a $D$-dimensional vector $p^\mu$ in the \texttt{FCI}-notation we need to write \texttt{Pair[Momentum[p,D],\-Lorentz\-Index[$\mu$,d]]}, while in the \texttt{FCE}-notation the same expression can be entered as \texttt{FVD[p,$\mu$]}. A summary of \feyncalc symbols that represent tensors and matrices in both notations can be found in appendix\,\ref{sec:app-fcifce}.

Automation of nonrelativistic calculations requires support for additional tensors that carry explicit temporal or spatial (Cartesian) indices. In particular, the code must be able to deal
not only with manifestly Lorentz covariant quantities (\eg $p^\mu$ or $l \cdot q$) but also with objects like $p^0$, $\bfp^i$, $l^0 q^0$ or $\bfl \cdot \bfq$. In \feyncalc 9.3 this has been achieved by extending the internal notation with the following symbols: \texttt{Cartesian\-Pair}, \texttt{Cartesian\-Momentum}, \texttt{Cartesian\-Index}, \texttt{Temporal\-Pair}, \texttt{Temporal\-Momentum} and \texttt{Pauli\-Sigma}. 

The first three are conceptually similar to the above-mentioned  \texttt{Pair}, \texttt{Momentum} and \texttt{Lorentz\-Index}. For example,  \texttt{Cartesian\-Pair} is a special pairing that accepts \texttt{Car\-tesian\-Momentum} or \texttt{Cartesian\-Index} as arguments for its two slots and can be used to represent 3-vectors, Cartesian scalar products or Kronecker deltas. It is important to stress that when \texttt{CartesianMomentum} and \texttt{CartesianIndex} have no second argument, their default dimension is 3. For calculations in dimensional regularization one should write them as $D-1$ dimensional quantities. This is because in \feyncalc every Cartesian tensor is always understood to be the spatial piece of the corresponding Lorentz tensor. Thus, if such a Lorentz tensor (\eg 4-vector) lives in $D$ dimensions, its spatial (a 3-vector) component must be a $D-1$-dimensional object.

As the name already suggests, the main purpose of introducing \texttt{Temporal\-Pair} and \texttt{Temporal\-Momentum} is to have a suitable representation for the temporal components of 4-vectors. Notice that we do not require a new dedicated symbol for the
0th index of a tensor, since it can be written using  the already existing \texttt{Explicit\-Lorentz\-Index[0]}.
Moreover, \texttt{Temporal\-Momentum} has only one argument that denotes the original 4-vector. This simply reflects the fact that in dimensional regularization the temporal component of a 4-vector still remains a 1-dimensional object, while its spatial components are analytically continued to $D-1$ dimensions.

\texttt{Cartesian\-Momentum} and \texttt{Cartesian\-Index} are equally valid arguments of \texttt{Dirac\-Gamma}. In this case they obviously represent a Dirac matrix contracted to a 3-vector (\eg $\bfGamma^i \bfp^i$) or a Dirac matrix with a Cartesian index (\eg $\bfGamma^i$). The same also applies for Levi-Civita tensors. For example, an \texttt{Eps} with three distinct \texttt{Cartesian\-Index} arguments stands for $\epsilon^{ijk}$. Symbolic Pauli matrices $\sigma^\mu$ and $\bfSigma^i$ are available via \texttt{PauliSigma}, which (similar to \texttt{DiracGamma}) can represent a Pauli matrix with a Lorentz or a Cartesian index as well as a Pauli matrix contracted to a Lorentz or a Cartesian vector. Here again we would like to refer to appendix\,\ref{sec:app-fcifce} for the list of available quantities and the commands to enter them in \feyncalc.

\subsection{Upper and lower indices} \label{sec:implementation-indices}

Many software frameworks for automatic QFT calculations do not explicitly distinguish between upper (contravariant) and lower (covariant) Lorentz indices. This simplification does not introduce any ambiguities, provided that all input expressions obey  Einstein's summing convention and are written in a Lorentz covariant fashion. Given that every pair of Lorentz indices appearing in a single term is understood to be contracted with each other, it is clear that one of the indices must appear upstairs, and the other one downstairs. For example, in
\begin{equation}
\mathtt{FV[p,}\mu\mathtt{]} \mathtt{FV[q,}\mu\mathtt{]} \equiv p^\mu q_\mu = p_\mu q^\mu
\end{equation}
it is irrelevant whether \texttt{FV[p,$\mu$]} stands for $p^\mu$ or $p_\mu$, as long as $\mu$ is understood to be a dummy index.

When dealing with free (\ie uncontracted) indices, it is enough to know that in a manifestly Lorentz covariant expression the position of the index on one side of the equation must match its position on the other side. Consider \eg the symbolic expression
\begin{mmaCell}[moredefined={SpinorUBar,SpinorU, GAD, GSD}]{Input}
  SpinorUBar[p].GAD[\(\mu\)].GSD[p].GAD[\(\nu\)].SpinorU[p]
\end{mmaCell}
being simplified to 
\begin{mmaCell}[moredefined={SpinorUBar,SpinorU, GAD, GSD,FV}]{Input}
  2\,SpinorUBar[p].GAD[\(\mu\)].SpinorU[p]*FV[p,\(\nu\)]
\end{mmaCell}
This can be interpreted as
\begin{equation}
\bar{u}(p) \gamma^\mu \slashed{p} \gamma^\nu u(p) = 2 p^\nu \bar{u}(p) \gamma^\mu u(p),
\end{equation}
but also 
\begin{equation}
\bar{u}(p) \gamma_\mu \slashed{p}  \gamma_\nu u(p) = 2 p_\nu \bar{u}(p) \gamma_\mu u(p),
\end{equation}
or
\begin{equation}
\bar{u}(p) \gamma^\mu \slashed{p}  \gamma_\nu u(p) = 2 p_\nu \bar{u}(p) \gamma^\mu u(p),
\end{equation}
as well as
\begin{equation}
\bar{u}(p) \gamma_\mu \slashed{p}  \gamma^\nu u(p) = 2 p^\nu \bar{u}(p) \gamma_\mu u(p).
\end{equation}
Notwithstanding the ambiguity of this symbolic notation, it does not lead to inconsistencies so that the calculations will always return sensible results. Moreover, once we mentally fix the positions of the free indices in the input expression, manifest Lorentz covariance guarantees that these positions will not change in the course of all intermediate symbolic manipulations and will be preserved in the final result. While such a ``mixing'' of covariant and contravariant indices may seem aesthetically unpleasant, this trick greatly helps to improve the performance of symbolic codes, which is especially important when working with very large expressions.

Things become more complicated once we want to handle expressions that contain both Lorentz and Cartesian tensors. Depending on the metric signature, moving a Cartesian or a temporal index into an opposite position may introduce a minus sign. For example, for
$g^{\mu \nu} = (1,-1,-1,-1)$ we have
\begin{equation}
p^0 = p_0, \quad \bfp^i = - \bfp_i.
\end{equation}
Furthermore, such indices are not constrained to appear in the same position on both sides of an equation, so that expressions like
\begin{equation}
\bfp^i A^{ij} = \bfp_i B_{ij},
\end{equation}
with $A$ and $B$ being some Cartesian tensors, are perfectly valid. Therefore, in order to make sense of  \feyncalc expressions such as \texttt{CV[p,i] CV[q,i]} or \texttt{CV[l,k] KD[j,k]} it is necessary to introduce additional rules that allow us to determine the positions of the Cartesian and temporal indices unambiguously. These rule read as follows

\begin{enumerate}
\item Every expression must satisfy Einsteins's summation convention, both for Lorentz and Cartesian indices. Single terms containing more than  two identical Lorentz or Cartesian indices are illegal and will lead to inconsistent results.
\item In a contraction of two Lorentz indices it is understood that one of them is upstairs and the other is downstairs.
\item In a contraction of two Cartesian indices, both indices are understood to be upper indices.
\item A free Lorentz or Cartesian index is always understood to be an upper index.
\end{enumerate}
While the first two rules merely formalize something that was always implicitly assumed in \feyncalc calculations, the last two rules are new. The third rule was never required before,
since earlier versions of \feyncalc could not deal with Cartesian tensors. The fourth rule helps to avoid ambiguities when interpreting expressions with free indices. Let us briefly illustrate how, by applying the above rules, we can interpret different \feyncalc expressions in a sensible way
\begin{subequations}
\begin{align}
\mathtt{CSP[p,q]} & \equiv \bfp \cdot \bfq, \\
\mathtt{CV[p,i] CV[q,i]} & \equiv \bfp^i \, \bfq^i, \\
\mathtt{CV[l,k] KD[j,k]} & \equiv \bfl^k \delta^{jk}.
\end{align}
\end{subequations}
Notice that our notation also applies to tensors that carry both Lorentz and Cartesian indices.
Such quantities often arise at different stages of nonrelativistic calculations and are, therefore, fully supported in \feyncalc 9.3. For example,
\begin{equation}
\mathtt{FV[p, mu] Pair[LorentzIndex[mu], CartesianIndex[i]] CV[q, i]} \equiv p_\mu g^{\mu i} \bfq^i,
\end{equation}
where we used the \texttt{FCI} notation to write down a metric tensor with mixed indices, since such an object has no corresponding \texttt{FCE} shortcut.

\subsection{Nonstandard integrals}
\label{sec:implementation-loops}

\feyncalc is very often employed as a convenient tool for symbolic manipulations of loop integrals, especially at 1-loop. Integrals with only one loop momentum and standard $1/(p^2-m^2)$-type propagators can be conveniently handled using the Passarino-Veltman reduction technique \cite{Passarino:1978jh}, which is available in \feyncalc since version 1.0. Indeed, tensor reduction and the subsequent analytic or numerical evaluation of the resulting Passarino-Veltman functions is sufficient for a large class of 1-loop calculations in the Standard Model and its extensions. 

Unfortunately, such methods often turn out to be inadequate when EFTs come into play. For example, eikonal propagators, as they appear in HQET or SCET, cannot be handled by the routines implemented in \feyncalc 9.2. The same is also true for Euclidean and Cartesian integrals as well as integrals involving temporal components of 4-vectors.

On the one hand, it is difficult to find a good strategy for treating such ``nonstandard'' integrals in \feyncalc in a generic fashion. As far as tensor reduction is concerned, even at 1-loop such integrals often cannot be directly rewritten in terms of scalar integrals with unit numerators. In the lack of a universal basis\footnote{A possible generalization of the Passarino-Veltman method to integrals without Lorentz invariance has been recently suggested in \cite{Chang:2020hii}.} similar to the Passarino-Veltman functions,\footnote{Even when certain types and families of nonstandard integrals can be reduced to a fixed set of master integrals, publicly available software libraries that already encode numerical or analytic results for those master integrals are very scarce.} the evaluation of the corresponding master integrals often proceeds on a case-by-case basis.

On the other hand, some algebraic manipulations that are needed in the course of the Passarino-Veltman reduction turn out to be applicable to almost all kinds of loop integrals. For example, partial fractioning and minimal tensor reduction to remove loop momenta with uncontracted indices can be straightforwardly applied to Cartesian and eikonal loop integrals such as
\begin{equation}
\int d^{D-1} \bfk \, \frac{4 (\bfk \cdot \bfp)}{\bfk^2 (\bfk+\bfp)^2 (\bfk-\bfp)^2} = \int  \frac{1}{\bfk^2 (\bfk-\bfp)^2} - \int  \frac{1}{\bfk^2 (\bfk+\bfp)^2}
\end{equation}
or
\begin{equation}
\int d^D k \frac{k^\mu k^\nu}{k^2 \, (k \cdot p - m^2)} =  \frac{m^4}{(D-1) p^4} (D \, p^\mu p^\nu  - p^2 g^{\mu \nu}) \int  \frac{d^D k}{k^2 \, (k \cdot p - m^2)}.
\end{equation}

The very first step in making \feyncalc useful for such calculations is to introduce new symbols to represent various nonstandard propagators. In the external notation this can be achieved by adding only 3 new shortcuts, which allow to cover a broad range of nonstandard loop integrals. These are \texttt{SFAD} (\texttt{Stan\-dard\-Feyn\-Amp\-Denominator}), \texttt{CFAD} (\texttt{Cartesian\-Feyn\-Amp\-Denominator}) and \texttt{GFAD} (\texttt{Generic\-Feyn\-Amp\-De\-nominator}), which stand for covariant, Cartesian or generic propagators respectively. Of course, for compatibility reasons the original symbol \texttt{FAD} (\texttt{Feyn\-Amp\-Denominator}) has been kept in \feyncalc as it has been employed there since the very beginning. For the sake of clarity, table \ref{tab:fads} summarizes all types of propagators that can be entered using the four above-mentioned shortcuts.
\clearpage
{
\small
\renewcommand*{\arraystretch}{1.5}
\setlength{\tabcolsep}{0.3cm}
\begin{longtable}[b]{|l|c|}
\hline
Shortcut in \feyncalc & Meaning \\
\hline
 $\mathtt{FAD[\{k - p_1 - \ldots,\, m,\, n\}]}$ & $\left [ \frac{1}{(k - p_1 - \ldots)^2 - m^2 + i \eta} \right ]^n$ \\
 $\mathtt{SFAD[\{\{k - p_1 - \ldots, \pm k.(q_1 + \ldots)\}, \{\pm m^2,\, \pm 1\}, \, n\}] }$ & 
 $\left [\frac{1}{(k - p_1 - \ldots)^2 \pm k.(q_1 + \ldots) \mp m^2 \pm i \eta  } \right ]^n$ \\
  $\mathtt{CFAD[\{\{k - p_1 - \ldots, \pm k.(q_1 + \ldots)\}, \{\pm m^2,\, \pm 1\}, \, n\}] }$ & 
 $\left [\frac{1}{(\bfk - \bfp_1 - \ldots)^2 \pm \bfk.(\bfq_1 + \ldots) \pm m^2 \pm i \eta  } \right ]^n$ \\
 
   $\mathtt{GFAD[\{\{x, \pm 1\}, n\}] }$ & 
 $\left [\frac{1}{x  \pm i \eta  } \right ]^n$ \\
 \hline
\caption{Implementation of the new propagator types using \texttt{Standard\-Feyn\-Amp\-Denominator}, 
\texttt{Cartesian\-Feyn\-Amp\-Denominator} and \texttt{Generic\-Feyn\-Amp\-Denominator}. Here $x$ can be an almost arbitrary function of loop-momentum dependent scalar products.} 
\label{tab:fads}
\end{longtable}
}
While the old \texttt{FAD} covers only a small fraction of propagators that are possible with the new \texttt{SFAD}, the former
is still somewhat better integrated into \feyncalc than the latter. This mainly concerns the use of the function \texttt{Feyn\-Amp\-Denominator\-Simplify} for detecting scaleless integrals and finding useful loop momentum shifts. These differences will be gradually eliminated in the future versions of the package, where the treatment of the new integral types will become more mature.

The syntax of \texttt{SFAD} may seem cumbersome at the first sight, but these inconveniences are more than compensated by the great flexibility encoded in this shortcut: Both quadratic and linear propagators are covered and the signs in front of the mass term $m^2$ and the causality parameter $i \eta$ can be chosen freely. Furthermore, some common propagator types can be entered faster as follows
\begin{subequations}
\begin{align}
\mathtt{SFAD[\{p, m^2\}]} & \equiv \frac{1}{p^2 - m^2 + i \eta}, \\
\mathtt{SFAD[\{p, \{-m^2, -1\}\}]} & \equiv \frac{1}{p^2 + m^2 - i \eta}, \\
\mathtt{SFAD[\{\{0, 2 \, p.q\}\}]} & \equiv \frac{1}{2 \, p \cdot q + i \eta}, \\
\mathtt{SFAD[\{\{p, -2 \, p.q\}, m^2\}]} & \equiv \frac{1}{p^2 - 2 \, p \cdot q - m^2 + i \eta}.
\end{align}
\end{subequations}
Notice that in the case of massless eikonal propagators \feyncalc takes special care to preserve the sign of  $i \eta$.
Rewriting of propagators as in
\begin{equation}
\frac{1}{- 2 \, p \cdot q + i \eta} = - \frac{1}{2 \, p \cdot q - i \eta},
\end{equation}
where the causality parameter switches its sign, is explicitly avoided.

\texttt{CFAD} can be regarded as the Cartesian counterpart of \texttt{SFAD} with the main difference being that the default signs of 
$m^2$ and $i \eta$ are opposite to that of \texttt{SFAD} \eg
\begin{align}
\mathtt{CFAD[\{p, m^2\}]} & \equiv \frac{1}{\bfp^2 + m^2 - i \eta}, \\
\mathtt{CFAD[\{\{0, 2 \, p.q\}\}]} & \equiv \frac{1}{2 \, \bfp \cdot \bfq - i \eta}.
\end{align}
Apart from this characteristic feature, \texttt{CFAD} has virtually the same syntax as \texttt{SFAD} and can be used to enter different types 
of Cartesian integrals.

Last but not least, one should also mention \texttt{GFAD} that acts as a generic placeholder for entering integrals that cannot be represented using \texttt{SFAD}s and \texttt{CFAD}s alone. For example, the singlet propagator in pNRQCD (\cf appendix\,\ref{sec:app-pnrqcd}) is a quantity that explicitly depends on the temporal component of a 4-momentum flowing through the corresponding line. Hence, we can write its denominator as
\begin{equation}
\mathtt{GFAD[TC[p] - En]} \equiv \frac{1}{p^0 - E_n + i \eta}.
\end{equation}
Since \texttt{GFAD} objects may represent almost arbitrary loop-momentum dependent denominators, \feyncalc will usually abstain from applying any kind of loop momentum shifts to integrals containing such propagators. This means that computations involving \texttt{GFAD}s will require significantly more user intervention at intermediate steps than those relying on the simpler but less versatile 
\texttt{SFAD}s and \texttt{CFAD}s. It is therefore advisable not to introduce \texttt{GFAD}s unless absolutely necessary. Having said that, we would also like to stress that partial fractioning and tensor reduction are nonetheless available also for integrals containing \texttt{GFAD} propagators.

An important limitation that the users should be aware of concerns types of integrals that can be manipulated using the built-in functions. Internally, \feyncalc always classifies input integrals into three possible categories:
\begin{enumerate}
\item Integrals in which loop momenta appear solely as 4-vectors, meaning that such expressions enjoy manifest Lorentz covariance.
\item Manifestly noncovariant integrals where each integration measure is split into temporal and spatial components \eg as in
\begin{equation}
\int d k^0 \, d^{D-1} \bfk \, f(k^0,\bfk),
\end{equation}
where the integrand $f(k^0,\bfk)$ explicitly depends on temporal and spatial components of the loop momentum $k$.
\item Integrals that are mixtures of covariant and noncovariant quantities \eg as in
\begin{equation}
\int d^D k \frac{1}{k^0 + x} \frac{1}{k^2} \frac{1}{(\bfp - \bfk)^2},
\label{eq:mixed-int}
\end{equation}
where $x$ is some c-number and $\bfp$ is an external 3-momentum.
\end{enumerate}
\feyncalc can readily handle integrals of type 1 or 2, but the ``mixed'' integrals of type 3
cannot be processed straightforwardly. This is because the underlying code heavily relies on working with linearly independent scalar products involving loop momenta. However, it is hardly possible to guarantee linear independence once 3-momenta and 4-momenta are allowed to appear in the same integral. For example, one might encounter zeros in the form
\begin{equation}
k \cdot q + \bfk \cdot \bfq, \quad \textrm{with } q = (0,\bfq)^T,
\end{equation}
which would remain undetected and hence lead to potentially disastrous consequences towards the end of the computation. Owing to the fact that integrals similar to those in \Eq\eqref{eq:mixed-int} do arise in many NREFT calculations, it is important to clarify how to handle
them properly. Here we propose three different strategies depending on the form of the involved integrals and the expected difficulties in calculating the resulting master integrals.

In some cases it might be possible to recast a mixed integral into a form that is manifestly Lorentz covariant \ie to convert a type 3 integral into a type 1 integral. As far as numerators are concerned, we can always introduce auxiliary vectors such as $n=(1,0,0,0)^T$ and $v = (p^0,0,0,0)^T$ to have
\begin{equation}
k^0 = k \cdot n, \quad \bfk \cdot \bfp = k \cdot (v - p)  \quad \textrm{with } p = (p^0,\bfp)^T,
\end{equation}
for a loop momentum $k$ and an arbitrary external 3-momentum $\bfp$. A 3-momentum vector with a free index can be written in a covariant fashion at the cost of introducing a metric tensor with Lorentz and Cartesian indices, \eg as in
\begin{equation}
\int d^D k \, \bfk^i f(k) = g^i_{\mu} \int d^D k \,  k^\mu  f(k).
\end{equation}
However, when applied to denominators, these methods may produce inconvenient propagators such as
\begin{equation}
\frac{1}{\bfk^2}  = \frac{1}{(k \cdot n)^2 -  k^2} 
\end{equation}
and alike. Furthermore, introducing too many auxiliary vectors will likely make the integral more complicated than it really is. This is why, in general, this approach is not always applicable or even sensible.

The other two strategies require us to convert a mixed integral into a type 2 integral first. This procedure is straightforward and unambiguous, since 
any scalar product of two 4-vectors can be always decomposed into its spatial and temporal components as in
\begin{equation}
k^2 = (k^0)^2 - \bfk^2, \quad k \cdot p = k^0 p^0 - \bfk \cdot \bfp.
\end{equation}
This means that each integration over a $D$-dimensional loop momentum $k$ splits into a 1-dimensional integration over the temporal component $k^0$ and a $D-1$-dimensional integration over the spatial comment $\bfk$. 

The second strategy would be to tensor reduce and partial fraction the $\bfk$-integrals, whereas $k^0$ will be regarded as an external parameter. The resulting integrals (that still depend both on $\bfk$ and $k^0$) are then declared to be master integrals.

When employing the third strategy we would, on the contrary, integrate over $k^0$ first, ending up with pure Cartesian $\bfk$-integrals. Those integrals 
may explicitly depend not only on the 3-vector $\bfk$ and its scalar products with external momenta, but also on the magnitudes of the 3-vectors, such as $|\bfk|$ and $|\bfk - \bfp|$. Furthermore, the $k^0$-integration requires some care in applying the residue theorem and picking up 
the correct poles. Nevertheless, this  procedure often leads to fewer and simpler master integrals as compared to the case where each master integral must be integrated in $\bfk$ and $k^0$.

In principle, \feyncalc can be useful in all 3 scenarios of dealing with mixed integrals, but the level of automation will
vary substantially. For example, integrations over the temporal components of loop momenta have to be performed by hand, since such
a procedure is too difficult to automatize in full generality. 

Another important restriction in the handling of tensor integrals is the requirement that those should not contain vanishing Gram determinants. Although this issue seems to be rarely discussed outside of the context of the Passarino-Veltman functions, the breakdown of naive tensor reduction for integrals with zero Gram determinants can, in principle, occur in all kinds of tensor integrals. For example, the result of the tensor reduction of the 3-point function
\begin{equation}
\int d^{D-1} \bfk \, \frac{\bfk^i}{\bfk^2 (\bfk-\bfp)^2 (\bfk-\bfq)^2}
\end{equation}
is proportional to the inverse of the Gram determinant $4((\bfp \cdot \bfq)^2 - \bfp^2 \, \bfq^2)$. Therefore, a naive attempt to tensor reduce this integral at the special kinematic point $\bfp \cdot \bfq = \bfp^2 = \bfq^2$ will inevitably fail. In general, it is well known (\cf \eg  \cite{Devaraj:1997es,Denner:2005nn}) that many of such cases can be worked around by considering a larger nonsingular system of linear equations and extracting necessary relations to reduce the original integral. However, as of now, such procedures are not yet implemented in \feyncalc.

More details on practical manipulations of nonstandard integrals in \feyncalc 9.3 can be found in section \ref{sec:usage-loops}.

\section{Installation and usage}\label{sec:usage}

\subsection{Installation}

The \feynonium project consists of two components: the recently released \feyncalc 9.3 \cite{Shtabovenko:2020gxv} that can be used for nonrelativistic calculations and a homonymous add-on that is  dedicated to NREFTs. Both \feyncalc and the add-on are
open source\footnote{Licensed under the General Public License (GPL) version 3.} with the source code hosted on \textsc{GitHub}.\footnote{The link to the repository is \url{https://github.com/FeynCalc}.} \feyncalc requires at least \mma 8 or later, while the add-on runs on top of \feyncalc 9.3 or later.
The most convenient way to setup the whole framework is to use the automatic online installer. The \feyncalc installer can be invoked by running the following code in a new \mma session
{\small
\begin{mmaCell}[index=1,moredefined={InstallFeynCalc}]{Input}
  Import@"https://raw.githubusercontent.com/FeynCalc/feyncalc/master/install.m"
  InstallFeynCalc[]
\end{mmaCell}
}
\noindent After that one can install the \feynonium add-on in a similar manner
{\small
\begin{mmaCell}[index=2,moredefined={InstallFeynOnium}]{Input}
  Import@"https://raw.githubusercontent.com/FeynCalc/feynonium/
  master/install.m"
  InstallFeynOnium[]
\end{mmaCell}
}
\noindent Although not strictly necessary, it is also recommended to install the \feynhelpers add-on \cite{Shtabovenko:2016whf}, which provides convenient and easy-to-use interfaces to other tools for evaluating Passarino-Veltman functions and performing IBP reductions of loop integrals. Last but not least, one should also consider downloading, \textsc{FeynRules}\footnote{The package can be obtained from \url{https://feynrules.irmp.ucl.ac.be}.} as it is used to create \feynarts model files that are employed in some of the \feyncalc and \feynonium examples.

Notice that an add-on can be activated only during the loading of \feyncalc.  This is why before loading the package the names of the add-ons (as strings) must be specified in a list assigned to the global variable \texttt{\$LoadAddOns}. For example, to use \feynonium and \feynhelpers one should run
{\small
\begin{mmaCell}[moredefined={LoadAddOns,FeynCalc}]{Input}
  \$LoadAddOns=\{"FeynOnium","FeynHelpers"\};
  <<FeynCalc`
\end{mmaCell}
}
at the very beginning of a \mma session.

\subsection{Basic nonrelativistic calculations}

Most standard \feyncalc routines for amplitude manipulations such as \texttt{Contract}, \texttt{Un\-contract}, \texttt{Expand\-Scalar\-Product}, \texttt{Momentum\-Combine}, \texttt{Complex\-Conjugate} etc. are directly applicable to expressions containing noncovariant quantities. Therefore, everyone who at least knows how to use \feyncalc for tree-level calculations should have no difficulties to master the new nonrelativistic capabilities of the package.

The basic Cartesian tensors required for nonrelativistic studies are 3-vectors (\eg $\bfp^i$ abbreviated as \texttt{CV[p,i]}), Kronecker deltas (\eg $\delta^{ij}$ abbreviated as \texttt{KD[i,j]}), scalar products of two 3-vectors (\eg $\bfp \cdot \bfq$ abbreviated as \texttt{CSP[p,q]}) and Cartesian Levi-Civita tensors  (\eg $\epsilon^{ijk}$ abbreviated as \texttt{CLC[i,j,k]}). Notice that all Cartesian vectors are typeset bold, with 3-dimensional vectors having a bar and $D-4$-dimensional vectors a hat. The vectors without a bar or a hat live in $D-1$ dimensions. This agrees with the existing \feyncalc
typesetting of 4-vectors that follows \cite{Buras:1989xd}. Explicitly, we have
\begin{equation}
\bfp^i = \bar{\bfp}^i + \hat{\bfp}^i,
\end{equation}
with
\begin{equation}
\textrm{dim} [\bfp^i] = D-1, \quad \textrm{dim} [\bar{\bfp}^i] = 3, \quad \textrm{dim} [\hat{\bfp}^i] = D-4.
\end{equation}
The same notation applies also to Dirac and Pauli matrices.

The shortcuts \texttt{CV}, \texttt{KD}, \texttt{CSP} and \texttt{CLC} correspond to 3-dimensional quantities. Their $D-1$-dimensional versions are obtained by attaching a \texttt{D} to the corresponding shortcut, \eg as in \texttt{CVD} or \texttt{KDD}. Attaching an \texttt{E} yields the respective $D-4$-dimensional symbol, \eg \texttt{CSE}.\footnote{As there is no Levi-Civita tensor in $D-4$ dimensions, \texttt{CLCE} is not defined.} In this respect the new Cartesian tensors behave in the same way as the existing Lorentz quantities.

The crucial task of manifestly breaking Lorentz covariance of tensors and matrices can be accomplished using \texttt{LorentzToCartesian}. This function rewrites the occurring tensors and matrices with Lorentz indices in terms of their temporal and spatial components as in
\begin{subequations}
\begin{align}
p^\mu &= g_{\nu}^{\mu} p^\nu  = g_0^\mu p^0  + g^{\mu}_i \bfp^i  =  g^{\mu 0} p^0 - g^{\mu i} \bfp^i, \\ \quad p \cdot q &= p^0 q_0 + \bfp^i \bfq_i = p^0 q^0 - \bfp \cdot \bfq,
\end{align}
\end{subequations}
which is important \eg when doing an amplitude-level matching between relativistic and nonrelativistic theories. For example, we can write
{\small
\begin{mmaCell}[moredefined={FV, LorentzToCartesian}]{Input}
  LorentzToCartesian[FV[p,\mmaUnd{\(\pmb{\mu}\)}]]
\end{mmaCell}

\begin{mmaCell}{Output}
  \mmaSup{p}{0} \mmaSup{\mmaOver{g}{_}}{0\(\mu\)}-\mmaSup{\mmaOver{{\bfp}}{_}}{$} \mmaSup{\mmaOver{g}{_}}{$\(\mu\)}
\end{mmaCell}
}
\noindent Here the dollar sign indicates an index contraction between the 3-vector $\bfp^i$ and the metric tensor with mixed indices $g^{i \mu}$. In the internal representation (currently there are no \texttt{FCE}-shortcuts for such objects) we have
\begin{equation}
\mathtt{Pair[CartesianMomentum[p], LorentzIndex[i]]} \equiv \bfp^i g^{i \mu} \equiv \bfp^{\$} g^{\$ \mu}.
\end{equation}
To avoid any misunderstandings, we kindly refer the reader to section \ref{sec:implementation-indices} that explains our treatment of covariant and contravariant indices in the program. We use the metric signature $(1,-1,-1,-1)$ and define the Cartesian scalar product as
$\bfp \cdot \bfq \equiv \bfp^i \bfq^i$.

At this stage one would often like to assign some specific values to the spatial and temporal components of 4-vectors and scalar products. This can be done via
direct assignments as in
{\small
\begin{mmaCell}[moredefined={TC, CSP}]{Input}
  TC[n] = 1;
  CSP[n,p] = 0;
\end{mmaCell}
}
\noindent where we set $n^0 = 1$ and $\bfn \cdot \bfp = 0$. As usual, these assignments can be removed via \texttt{FC\-Clear\-Scalar\-Products}. One can also exploit additional simplifications by extracting the magnitude (which could be \eg an expansion parameter in an EFT calculation) of a 3-vector or specifying relations between spatial components of some 4-vectors as in
\begin{equation}
\bfk^i = |\bfk| \hat{\bfk}^i.
\label{eq:momentum-assignments}
\end{equation}
To this end we can assign values directly to a particular \texttt{CartesianMomentum} and let
\feyncalc know that some symbols (\eg |\bfk|) are scalars and hence can be pulled out of 
expressions that denote vector contractions. The latter is done using the \texttt{Datatype} mechanism, where the corresponding symbols are defined to have datatype \texttt{FCVariable}.
For example, the relations given in \Eq\eqref{eq:momentum-assignments} can be implemented via
{\small
\begin{mmaCell}[moredefined={CSP, CV, KD, CartesianMomentum,DataType,FCVariable}]{Input}
  CartesianMomentum[k] = kv CartesianMomentum[khat];
  DataType[kv,FCVariable] = True;
  CSP[khat] = 1;
\end{mmaCell}
}
\noindent where we also account for the fact that the scalar product of a unit vector with itself is unity. In this case expressions such as $(\bfk + \bfp)^2$ or $\epsilon^{ijk} \bfk^i \bfp^j \bfl^k$ can be directly rewritten as
{\small
\begin{mmaCell}[moredefined={CSP, CLC, ExpandScalarProduct}]{Input}
  \{CSP[k + p], CLC[][k, p, l]\} // ExpandScalarProduct
\end{mmaCell}

\begin{mmaCell}{Output}
  \{2 kv (\mmaOver{\textbf{khat}}{_}\(\cdot\)\mmaOver{\textbf{p}}{_})+\mmaSup{\mmaOver{\textbf{p}}{_}}{2}+\mmaSup{kv}{2},kv \mmaSup{\mmaOver{\(\epsilon\)}{_}}{\mmaOver{ \textbf{khat}}{_} \mmaOver{\textbf{p}}{_} \mmaOver{\textbf{l}}{_}}\}
\end{mmaCell}
}
If one would like to differentiate with respect to a 3-vector, the corresponding routine is called \texttt{ThreeDivergence}. It works in exactly the same way as its 4-dimensional analogue \texttt{FourDivergence} \eg
{\small
\begin{mmaCell}[moredefined={ThreeDivergence, CSP, CV}]{Input}
  ThreeDivergence[1/(CSP[p, q] + a) (b + CSP[p]), CV[p, i]]
\end{mmaCell}

\begin{mmaCell}{Output}
  -\mmaFrac{b \mmaSup{\mmaOver{\textbf{q}}{_}}{\textbf{i}}}{\mmaSup{(\mmaOver{\textbf{p}}{_}\(\cdot\)\mmaOver{\textbf{q}}{_}+a)}{2}}+\mmaFrac{2 \mmaSup{\mmaOver{\textbf{p}}{_}}{\textbf{i}}}{\mmaOver{\textbf{p}}{_}\(\cdot\)\mmaOver{\textbf{q}}{_}+a}-\mmaFrac{\mmaSup{\mmaOver{\textbf{p}}{_}}{2} \mmaSup{\mmaOver{\textbf{q}}{_}}{\textbf{i}}}{\mmaSup{(\mmaOver{\textbf{p}}{_}\(\cdot\)\mmaOver{\textbf{q}}{_}+a)}{2}}
\end{mmaCell}
}
Let us now show how the introduced machinery can be employed in real-life calculations. To this end we can reproduce the value of the matching coefficient
\begin{equation}
\tilde{G}_1 (^3 P_0) = \frac{12}{64 \pi^2 s^2} \frac{1}{4} \sum_{\textrm{pols}} c_1^{J=0} (c_3^{\ast J=0}),
\label{eq:gcoeffnrqcd}
\end{equation}
which enters the $\mathcal{O}(\alpha_s^0 v^2)$ differential production cross-section for $e^+ (l_1) + e^- (l_2) \to \chi_{c_0} (P) + \gamma (k)$ in the NRQCD factorization formalism~\cite{Li:2013nna,Chao:2013cca,Brambilla:2017kgw}. Here $s$ stands for the square of the collision energy in the center of mass frame, while the summation sign implies that we must average over the polarizations of the leptons and sum over the polarizations of the photon. Explicit values of the short-distance coefficients $c_1^{J=0}$ and $c_3^{J=0}$ are given in \cite{Brambilla:2017kgw}: 
\begin{align}
c_1^{J=0} & =  \frac{i }{3} \frac{e^3 e_Q^2}{s}  \frac{1 - 3 r}{1-r} 
\bar{v}(l_2) \bfGamma^i u(l_1) \bfeps^{\ast i} (k), \\
c_3^{J=0} & = - \frac{i}{30} \frac{e^3 e_Q^2}{s}  \frac{9 - 24 r + 35 r^2}{(1-r)^2} \bar{v}(l_2) \bfGamma^i u(l_1) \bfeps^{\ast i} (k),
\end{align}
where $r = 4 m^2/s$, $\bfeps^{\ast i} (k)$ denotes the polarization 3-vector of the external photon and $\bar{v}(l_2) \gamma^i u(l_1)$ stands for the spatial piece of the leptonic current. The kinematics is chosen in such a way that
\begin{align}
l_1^2 &= l_2^2 = k^2 = 0, \quad l_1^0 = l_2^0= \frac{\sqrt{s}}{2}, \quad l_1 \cdot l_2 = \frac{s}{2}, \quad \nonumber \\
\bfl_1 &= - \bfl_2, \quad \bfk \cdot \bfl_1 = \frac{\sqrt{s} |\bfk|}{2} \cos \theta.
\end{align}
We would like to evaluate the photon polarization sum by introducing an auxiliary vector $n^\mu =(1,0,0,0)^T$ so that only physical degrees of freedom (transverse polarizations) are taken into account. Using \feyncalc, expressions similar to \Eq\eqref{eq:gcoeffnrqcd} can be computed as follows. First of all, we need to specify all kinematic constraints
{\small
\begin{mmaCell}[moredefined={FCClearScalarProducts, CP, SP, TC, CSP, CartesianMomentum}]{Input}
  FCClearScalarProducts[];
  SP[k] = 0;
  SP[l1, l1] = 0;
  SP[l2, l2] = 0;
  SP[l1, l2] = s/2;
  SP[k, n] = kv;
  TC[l1] = Sqrt[s]/2;
  TC[l2] = Sqrt[s]/2;
  CSP[k] = kv^2;
  CSP[k, l2] = -CSP[k, l1];
  CSP[k, l1] = kv cosTh*Sqrt[s]/2;
  SP[n] = 1;
  CartesianMomentum[n] = 0;
\end{mmaCell}
}
\noindent and define the already known short distance coefficients
{\small
\begin{mmaCell}[moredefined={c1J0,c3J0,FCClearScalarProducts, CP, SP, TC, CSP, SpinorU,SpinorVBar, CGA, CV, Polarization}]{Input}
  c1J0 = I/3 (1 - 3 r)/(1 - r)(el^2 eq^2/s) el *
  SpinorVBar[l2].CGA[i].SpinorU[l1]  CV[Polarization[k, -I], i];
  c3J0 = -I/30 (9 - 24 r + 35 r^2)/(1 - r)^2 (el^2 eq^2/s) el SpinorVBar[l2].CGA[i].SpinorU[l1] CV[Polarization[k, -I], i];
\end{mmaCell}
}
\noindent After that the calculation amounts to issuing a sequence of standard commands with self-explanatory names (\texttt{Complex\-Conjugate}, \texttt{Do\-Polarization\-Sums}, \texttt{Fermion\-Spin\-Sum}, \texttt{Dirac\-Simplify}) that should be familiar to \feyncalc practitioners from calculations in relativistic theories. In this respect there are no fundamental differences between manipulations of relativistic and nonrelativistic amplitudes in \feyncalc, at least at the tree-level. Evaluating
{\small
\begin{mmaCell}[moredefined={ComplexConjugate, CP, SP, TC, CSP, el, DiracSimplify,FermionSpinSum, DoPolarizationSums, Factor2, ExtraFactor,c1J0,c3J0}]{Input}
  12/(64 Pi^2 s^2) c1J0 ComplexConjugate[c3J0] // DoPolarizationSums[#, k, n] & // FermionSpinSum[#, ExtraFactor -> 1/2^2] & // DiracSimplify // Factor2 // ReplaceAll[#, el -> Sqrt[4 Pi al]] &
\end{mmaCell}
}
\noindent we readily obtain
{\small
\begin{mmaCell}{Output}
  -\mmaFrac{\(\pi\) \mmaSup{al}{3}(\mmaSup{cosTh}{2}+1)\mmaSup{eq}{4}(1-3 r)(35 \mmaSup{r}{2}-24 r+9)}{15\mmaSup{(1-r)}{3}\mmaSup{s}{3}}
\end{mmaCell}
}
\noindent which agrees with \Eq(58b) in \cite{Brambilla:2017kgw}.

\subsection{Dirac algebra}

Since \feyncalc 9.3 all routines related to Dirac algebra support manipulations of Dirac matrices with temporal or spatial indices. This allows the user to evaluate very generic noncovariant expressions involving Dirac matrices such as 
\begin{equation}
\bfGamma^i (\gamma^0 p^0 - \bfGamma^j \bfp^j -m) \bfGamma^i
\end{equation}
via
{\small
\begin{mmaCell}[moredefined={TGA, CGA, TC, CGS, DiracSimplify}]{Input}
  DiracSimplify[CGA[i].(TGA[] TC[p] - CGS[p] - m).CGA[i]]
\end{mmaCell}

\begin{mmaCell}{Output}
  -\mmaOver{\textbf{\(\gamma\)}}{_}\(\cdot\)\mmaOver{\textbf{p}}{_} + 3 \mmaSup{p}{0}\mmaSup{\mmaOver{\(\gamma\)}{_}}{0} + 3 m
\end{mmaCell}
}
or
\begin{equation}
\Tr (\gamma^\mu \bfGamma^j \gamma^0 \bfGamma^k \bfGamma^l \gamma^0 \gamma^5)
\end{equation}
using
{\small
\begin{mmaCell}[moredefined={TGA, DiracSimplify, DiracTrace, CGA, GA}]{Input}
  DiracTrace[GA[\(\mu\)].CGA[j].TGA[].CGA[k, l].TGA[].GA[5]]//DiracSimplify
\end{mmaCell}

\begin{mmaCell}{Output}
  4i\mmaSup{\mmaOver{\(\epsilon\)}{_}}{\textbf{j}\textbf{k}\textbf{l}\(\mu\)}
\end{mmaCell}
}
\noindent So far, Euclidean Dirac matrices are not yet supported, but given enough interest from the user side they might be added in the future.

\subsection{Pauli algebra}

Pauli matrices are a completely new class of algebraic objects introduced in \feyncalc 9.3 for the first time. For the sake of consistency and user convenience, their handling was modeled after the existing implementation of the Dirac algebra. Therefore, it should not come as a surprise that \feyncalc is equipped with routines called \texttt{PauliSimplify}, \texttt{PauliTrace} and \texttt{PauliOrder}.

If a chain of 3-dimensional Pauli matrices contains repeated Cartesian indices or contractions with identical 3-vectors as in
\begin{equation}
\bfSigma^i \bfSigma^j (\bfSigma \cdot \bfp) \bfSigma^i (\bfSigma \cdot \bfp),
\end{equation}
we can eliminate such pairs via
{\small
\begin{mmaCell}[moredefined={CSI,CSIS, PauliSimplify}]{Input}
  PauliSimplify[CSI[i, j].CSIS[p].CSI[i].CSIS[p]]
\end{mmaCell}

\begin{mmaCell}{Output}
  4\mmaSup{\mmaOver{\textbf{p}}{_}}{\textbf{j}}\mmaOver{\textbf{\(\sigma\)}}{_}\(\cdot\)\mmaOver{\textbf{p}}{_} - \mmaSup{\mmaOver{\textbf{p}}{_}}{2}\mmaSup{\mmaOver{\textbf{\(\sigma\)}}{_}}{\textbf{j}}
\end{mmaCell}
}
\noindent Trace calculations are possible using \texttt{PauliTrace} as in the following example for calculating
\begin{equation}
\Tr (\bfSigma^i \bfSigma^j \bfSigma^k \bfSigma^l)
\end{equation}
{\small
\begin{mmaCell}[moredefined={CSI,CSIS, PauliTrace,PauliSimplify}]{Input}
  PauliTrace[CSI[i, j, k, l]]//PauliSimplify
\end{mmaCell}

\begin{mmaCell}{Output}
  2\mmaSup{\mmaOver{\(\delta\)}{_}}{\textbf{i}\textbf{l}} \mmaSup{\mmaOver{\(\delta\)}{_}}{\textbf{j}\textbf{k}} - 2\mmaSup{\mmaOver{\(\delta\)}{_}}{\textbf{i}\textbf{k}} \mmaSup{\mmaOver{\(\delta\)}{_}}{\textbf{j}\textbf{l}} + 2\mmaSup{\mmaOver{\(\delta\)}{_}}{\textbf{i}\textbf{j}} \mmaSup{\mmaOver{\(\delta\)}{_}}{\textbf{k}\textbf{l}}
\end{mmaCell}
}
\noindent If it is necessary to reduce the number of matrices in a chain to at most one by repeatedly applying the relation
\begin{equation}
\bfSigma^i \bfSigma^j = \delta^{ij} + i \epsilon^{ijk} \bfSigma^k \label{eq:pmred},
\end{equation}
one should employ \texttt{PauliSimplify} with the option \texttt{PauliReduce} set to \texttt{True}. For the chain
\begin{equation}
\bfSigma^i \bfSigma^j \bfSigma^k
\end{equation}
we immediately find
{\small
\begin{mmaCell}[moredefined={CSI,CSIS, PauliTrace,PauliSimplify,PauliReduce}]{Input}
  PauliSimplify[CSI[i, j, k], PauliReduce -> True]
\end{mmaCell}

\begin{mmaCell}{Output}
  \mmaSup{\mmaOver{\textbf{\(\sigma\)}}{_}}{\textbf{i}}\mmaSup{\mmaOver{\(\delta\)}{_}}{\textbf{j}\textbf{k}} - \mmaSup{\mmaOver{\textbf{\(\sigma\)}}{_}}{\textbf{j}}\mmaSup{\mmaOver{\(\delta\)}{_}}{\textbf{i}\textbf{k}} + \mmaSup{\mmaOver{\textbf{\(\sigma\)}}{_}}{\textbf{k}}\mmaSup{\mmaOver{\(\delta\)}{_}}{\textbf{i}\textbf{j}} + i\mmaSup{\mmaOver{\(\epsilon\)}{_}}{\textbf{i}\textbf{j}\textbf{k}}
\end{mmaCell}
}
When doing loop calculations in dimensional regularization it becomes necessary to extend the definition of Pauli matrices to $D-1$ dimensions. It is well known that the anticommutator of two Pauli matrices can be consistently generalized to 
\begin{equation}
\{ \bfSigma^i, \bfSigma^j \} = 2 \delta^{ij},
\label{eq:pauli-anticomm}
\end{equation}
where $\delta^{ij}$ is a $D-1$-dimensional Kronecker delta with
\begin{equation}
(\delta^{ij})^2 = D-1.
\end{equation}
Then, using \Eq\eqref{eq:pauli-anticomm} one can derive relations for eliminating pairs of indices and vectors in a chain of Pauli matrices in $D-1$ dimensions. The same also applies for traces of an even number of matrices. A collection of such formulas can be found \eg in \cite{Hoang:2006ty}.

Yet the commutation relation of 3-dimensional Pauli matrices
\begin{equation}
[ \bfSigma^i, \bfSigma^j ] = 2 i \epsilon^{ijk} \bfSigma^k \label{eq:fo-man-comm-rel}
\end{equation}
becomes ambiguous in dimensional regularization, as the Levi-Civita tensor
$\epsilon^{ijk}$ is intrinsically a 3-dimensional quantity. Related issues are well known to the practitioners and a valuable discussion of this topic can be found in \cite{Hoang:2006ty}. In general, traces of odd numbers of Pauli matrices do not naively generalize to $D-1$ dimensions. Similar issues arise when trying to reduce products of Pauli matrices (\eg $\bfSigma^i \bfSigma^j \otimes \bfSigma^i \bfSigma^j$) to a finite 3-dimensional basis (\eg $1 \otimes 1$ and $\bfSigma^i \otimes \bfSigma^i$). Applying \Eq\eqref{eq:pmred} or any other projection method will also generate contributions that vanish in the limit $D \to 4$, the so-called evanescent operators. It is worth noting that evanescent operators multiplied by poles in $1/\varepsilon$ produce finite contributions to the final results and in general require dedicated treatments \cite{Dugan:1990df,Herrlich:1994kh}. Different prescriptions for dealing with Pauli matrices in $D-1$ dimensions can be found in the literature \cite{Braaten:1996rp,Pineda:1998kj,Hoang:2006ty,Gerlach:2019bso} and it is important to be aware of these issues to avoid inconsistencies.

As far as \feyncalc is concerned, the precise treatment of $D-1$-dimensional Pauli matrices can be specified via \texttt{FCSetPauliSigmaScheme[]}. The default value is \texttt{"None"}, meaning that only \Eq\eqref{eq:pauli-anticomm} is used to simplify chains of Pauli matrices, while $\bfSigma$-odd traces are left unevaluated. In this way the returned results are always unambiguous.

In order to obtain more compact (but also scheme-dependent) expressions, one may want to specify   a prescription for evaluating the remaining traces and reducing chains of Pauli matrices to a minimal basis. In the current version of \feyncalc the user can evaluate 
{\small
\begin{mmaCell}[moredefined={CSI,CSIS, FCSetPauliSigmaScheme}]{Input}
  FCSetPauliSigmaScheme["Naive"]
\end{mmaCell}
}
\noindent to allow the program to apply \Eq\eqref{eq:fo-man-comm-rel} in dimensional regularization
{\small
\begin{mmaCell}[moredefined={CSI,CSIS,CSID,PauliReduce,PauliSimplify}]{Input}
  PauliSimplify[CSID[i, j, k], PauliReduce -> True]
\end{mmaCell}

\begin{mmaCell}{Output}
  i\mmaSup{\mmaOver{\(\epsilon\)}{}}{\textbf{i}\textbf{j}\textbf{k}} + D\mmaSup{\textbf{\(\sigma\)}}{\textbf{i}} \mmaSup{\(\delta\)}{\textbf{j}\textbf{k}} - D\mmaSup{\textbf{\(\sigma\)}}{\textbf{j}} \mmaSup{\(\delta\)}{\textbf{i}\textbf{k}} - 3\mmaSup{\textbf{\(\sigma\)}}{\textbf{i}} \mmaSup{\(\delta\)}{\textbf{j}\textbf{k}} + 3\mmaSup{\textbf{\(\sigma\)}}{\textbf{j}} \mmaSup{\(\delta\)}{\textbf{i}\textbf{k}} + \mmaSup{\textbf{\(\sigma\)}}{\textbf{k}} \mmaSup{\(\delta\)}{\textbf{i}\textbf{j}}
\end{mmaCell}
}
\noindent The occurring products of $D-1$-dimensional Levi-Civita tensors are then calculated using 
\begin{equation}
\epsilon^{ijk} \epsilon^{lmn} = \begin{vmatrix}
 \delta^{il} &  \delta^{im} & \delta^{in} \\
  \delta^{jl} &  \delta^{jm} & \delta^{jn} \\
   \delta^{kl} &  \delta^{km} & \delta^{kn}
\end{vmatrix},
\end{equation}
where all Kronecker deltas are defined in $D-1$ dimensions. In particular, we have
\begin{equation}
\epsilon^{ijk} \epsilon^{ijm} = (D-3)(D-2) \delta^{km}.
\label{eq:clc-prod}
\end{equation}

We are looking forward to the feedback and suggestions from the NREFT community to implement more useful prescriptions in future iterations of the framework.

\subsection{Loop calculations} \label{sec:usage-loops}

The two main tools provided in \feyncalc for dealing with loop integrals are tensor reduction (via \texttt{TID} and \texttt{FC\-Multi\-Loop\-TID}) and partial fractioning (via \texttt{ApartFF}). Both operations employ \texttt{FeynAmp\-Denominator\-Simplify} (also abbreviated as \texttt{FDS}) for recognizing vanishing integrals and applying suitable loop momentum shifts. As has already been mentioned in section \ref{sec:implementation-loops}, \texttt{FDS} mainly relies on heuristics which may not work so well with nonstandard propagators, thus missing some obvious simplifications and not setting scaleless integrals to zero.

Let us discuss tensor reduction. If we are dealing with 1-loop tensor integrals that can be reduced to scalar integrals with unit numerators, it is advantageous to employ \texttt{TID}. Such a reduction is always possible for purely Lorentz or Cartesian integrals with quadratic propagators, but the support for Cartesian integrals is a new feature of \feyncalc 9.3. For simplicity, we can consider a massless Cartesian rank 2 tensor integral with one external momentum
\begin{equation}
\int d^{D-1} \bfk \, \frac{\bfk^i \bfk^j}{\bfk^2 (\bfk-\bfp)^2}
\end{equation}
that can be readily reduced to a massless 2-point function
{\small
\begin{mmaCell}[moredefined={TID,CVD,CFAD}]{Input}
  TID[CFAD[k, k - p] CVD[k, i] CVD[k, j], k]
\end{mmaCell}

\begin{mmaCell}{Output}
  \mmaFrac{\mmaSup{\textbf{p}}{2} \mmaSup{\(\delta\)}{\textbf{i}\textbf{j}}-(D-1) \mmaSup{\textbf{p}}{\textbf{i}} \mmaSup{\textbf{p}}{\textbf{j}}}{4 (2-D) (\mmaSup{\textbf{k}}{2}-i\,\(\eta\)).(\mmaSup{(\textbf{k}-\textbf{p})}{2}-i\,\(\eta\))}
\end{mmaCell}
}
Of course, more complicated integrals are also possible, as there are no formal limitations on the tensor rank and the number of external legs that can be processed by \texttt{TID}. The only practical limitation is the degrading performance when handling very complicated tensor integrals. 

When confronted with mixed integrals (\cf section \ref{sec:implementation-loops}), \texttt{TID} can often automatically perform tensor reduction with respect to the spatial part of the loop momentum. This is certainly true for integrals such as
\begin{equation}
\int d^{D} k \, \frac{\bfk^i}{k^2 (\bfk-\bfp)^2} \label{eq:mixedint1}
\end{equation}
{\small
\begin{mmaCell}[moredefined={TID,CVD,CFAD,SFAD,FVD}]{Input}
  TID[SFAD[k] CFAD[k - p] CVD[k, i], k]
\end{mmaCell}

\begin{mmaCell}{Output}
  -\mmaFrac{\mmaSup{(\mmaSup{k}{0})}{2}\mmaSup{\textbf{p}}{\textbf{i}}}{2\,\mmaSup{\textbf{p}}{2}(\mmaSup{\textbf{k}}{2}-i\,\(\eta\)).(\mmaSup{(\textbf{k}+\textbf{p})}{2}-\mmaSup{(\mmaSup{k}{0})}{2}-i\,\(\eta\))}+\mmaFrac{\mmaSup{\textbf{p}}{\textbf{i}}}{2\,\mmaSup{\textbf{p}}{2}(\mmaSup{\textbf{k}}{2}-\mmaSup{(\mmaSup{k}{0})}{2}-i\,\(\eta\))}-\mmaFrac{\mmaSup{\textbf{p}}{\textbf{i}}}{2(\mmaSup{\textbf{k}}{2}-i\,\(\eta\)).(\mmaSup{(\textbf{k}+\textbf{p})}{2}-\mmaSup{(\mmaSup{k}{0})}{2}-i\,\(\eta\))}
\end{mmaCell}
}
\noindent or 
\begin{equation}
\int d^{D} k \, \frac{k^\mu}{\bfk^2 (\bfk-\bfp)^2}
\end{equation}
{\small
\begin{mmaCell}[moredefined={TID,CVD,CFAD,SFAD,FVD}]{Input}
  TID[CFAD[k, k - p] FVD[k, i], k]
\end{mmaCell}

\begin{mmaCell}{Output}
  \mmaFrac{\mmaSup{k}{0} \mmaSup{\mmaOver{g}{_}}{0i}}{(\mmaSup{\textbf{k}}{2}-i\,\(\eta\)).(\mmaSup{(\textbf{k}-\textbf{p})}{2}-i\,\(\eta\))}-\mmaFrac{\mmaSup{\textbf{p}}{\textbf{$}} \mmaSup{g}{\textbf{$}i}}{2\,(\mmaSup{\textbf{k}}{2}-i\,\(\eta\)).(\mmaSup{(\textbf{k}-\textbf{p})}{2}-i\,\(\eta\))}
\end{mmaCell}
}
\noindent where \texttt{TID} essentially applies tricks described in section \ref{sec:implementation-loops}. Notice that the result for the integral in \Eq\eqref{eq:mixedint1} still contains a scaleless integral $\int d^D k/(\bfk^2 - (k^0)^2)$. This is an example of the difficulty of 
enhancing \texttt{FDS} with good heuristics for nonstandard integrals, especially when integrations in the temporal and spatial components of loop momenta must be treated separately.

Tensor reductions of integrals that are expected to contain irreducible denominators should be done using \texttt{FC\-Multi\-Loop\-TID}. 
Such denominators constitute a common feature of multiloop integrals, but they often arise already at 1-loop once propagators different from quadratic ones come into play. This is also the main reason why \texttt{TID} refuses to handle integrals with eikonal propagators: In such cases there is simply no guarantee that the reduction to integrals with unit numerators can succeed. \texttt{FC\-Multi\-Loop\-TID} is not affected by this problem, because it only considers loop momenta with free indices or those contracted to Dirac or Pauli matrices, Levi-Civita tensors and polarization vectors. For example, \texttt{FC\-Multi\-Loop\-TID}  does not regard
\begin{equation}
\int d^{D-1} \bfk \, \frac{\bfk \cdot \bfq}{\bfk^2 (\bfk-\bfp)^2}
\end{equation}
as a tensor integral and will therefore leave it unchanged. On the contrary, in the case of
\begin{equation}
\int d^{D-1} \bfk \, \frac{\bfSigma \cdot \bfk}{\bfk^2 (\bfk-\bfp)^2}
\end{equation}
the function will uncontract the scalar product of $\bfk$ and the Pauli matrix $\bfSigma$, producing a rank 1 tensor integral that will be subsequently reduced
{\small
\begin{mmaCell}[moredefined={CSISD,CVD,CFAD,SFAD,FCMultiLoopTID}]{Input}
  FCMultiLoopTID[CSISD[k] CFAD[k, k - p], {k}]
\end{mmaCell}

\begin{mmaCell}{Output}
  \mmaFrac{\textbf{\(\sigma\)}\(\cdot\)\textbf{p}}{2\,(\mmaSup{\textbf{k}}{2}-i\,\(\eta\)).(\mmaSup{(\textbf{k}-\textbf{p})}{2}-i\,\(\eta\))}
\end{mmaCell}
}
\noindent If we \emph{know} that uncontracting particular scalar products of loop momenta with other vectors may lead to a simpler 
result, we can use the option \texttt{Uncontract} to specify those vectors explicitly. One of such examples is
\begin{equation}
\int d^{D} k \, \frac{k \cdot q}{(k^2 - m^2) \, k \cdot p},
\end{equation}
where the default behavior of \texttt{FC\-Multi\-Loop\-TID} to leave this integral untouched is too restrictive. Using
{\small
\begin{mmaCell}[moredefined={FCMultiLoopTID,SPD,CFAD,SFAD,Uncontract}]{Input}
  FCMultiLoopTID[SFAD[\{k, m^2\}, \{\{0, k.p\}\}] SPD[k, q], \{k\}, Uncontract -> \{k\}]
\end{mmaCell}

\begin{mmaCell}{Output}
  \mmaFrac{p\(\cdot\)q}{\mmaSup{p}{2}(\mmaSup{k}{2}-m^2+i\,\(\eta\))}
\end{mmaCell}
}
\noindent we can nonetheless achieve the desired reduction.

The next thing we would like to discuss is partial fractioning. In \feyncalc 9.3 \texttt{ApartFF} has been extended to support the newly introduced nonstandard propagators, thus making it possible to handle many nontrivial cases such as
\begin{equation}
\int d^{D} k \, \frac{1}{k^2 (k^2 + k \cdot l) \, k \cdot (l - p) \, k \cdot (l + p)}
\end{equation}
{\small
\begin{mmaCell}[moredefined={FCMultiLoopTID,SPD,CFAD,SFAD,ApartFF}]{Input}
  ApartFF[SFAD[k, \{\{k, k.l\}\}, \{\{0, k.(l - p)\}\}, \{\{0, k.(l + p)\}\}], \{k\}]
\end{mmaCell}

\begin{mmaCell}{Output}
  -\mmaFrac{2}{\mmaSup{(k\(\cdot\)l-k\(\cdot\)p+i\,\(\eta\))}{2}.(k\(\cdot\)l+k\(\cdot\)p+i\,\(\eta\)).(\mmaSup{k}{2}+k\(\cdot\)l+i\,\(\eta\))}+\mmaFrac{2}{(\mmaSup{k}{2}+i\,\(\eta\)).\mmaSup{(k\(\cdot\)l-k\(\cdot\)p+i\,\(\eta\))}{2}.(k\(\cdot\)l+k\(\cdot\)p+i\,\(\eta\))}-\mmaFrac{1}{(\mmaSup{k}{2}+i\,\(\eta\)).\mmaSup{(k\(\cdot\)l-k\(\cdot\)p+i\,\(\eta\))}{2}.(\mmaSup{k}{2}+k\(\cdot\)l+i\,\(\eta\))}
\end{mmaCell}
}
\noindent However, we also observed that due to the specifics of \texttt{ApartFF}, some desirable decompositions cannot be obtained automatically. This mainly concerns integrals with propagators that do not form an overdetermined basis. Since \texttt{ApartFF} is applicable only to cases with overdetermined propagator bases (\cf section 3.3 in \cite{Shtabovenko:2016sxi} for more details), it would normally ignore such integrals altogether. The trick to overcome this behavior is to multiply the corresponding integral by unity \ie by a suitable propagator and its inverse. Given that the product of the original integral and the extra propagator contains an overdetermined basis of propagators, we may freely subject it to partial fractioning. At the end, multiplying back the so-obtained result with the inverse of the auxiliary propagator ensures that the final result is equivalent to the original expression.

For definiteness, let us consider the integral
\begin{equation}
\int d^{D-1} \bfk \, \frac{\bfk \cdot \bfp}{|\bfk| (\bfk-\bfp)^2},
\label{eq:apart-sqrt}
\end{equation}
where we would like to trade the numerator $\bfk \cdot \bfp$ for $\bfk^2$. We cannot achieve this neither with \texttt{FCMultiLoopTID} nor using the standard mode of \texttt{ApartFF}. This is why we extended the syntax of \texttt{ApartFF} to support the above-mentioned trick. When the second argument of the function is not a list, it is interpreted as the inverse of the auxiliary denominator that has already been added to the integral in the first argument. After having carried out such partial reduction, it is usually advisable to run \texttt{ApartFF} again (this time in the standard mode), to simplify the product of the intermediate result with the inverse denominator. In the case of the integral in \Eq\eqref{eq:apart-sqrt} we obviously need to introduce the unity as $\bfk^2 / \bfk^2 = 1$. Therefore, we multiply \Eq\eqref{eq:apart-sqrt} by $1/\bfk^2$ (written as \texttt{CFAD[k]}) and put $\bfk^2$ (as \texttt{CSPD[k]}) into the second slot of \texttt{ApartFF}. As far as the nonstandard propagator $1/|\bfk|$ is concerned, we can write it as a $\texttt{GFAD}$ with $\sqrt{\bfk^2}$. Owing to the abundance of such propagators in nonrelativistic calculations, we deliberately added support for square roots of Cartesian scalar products to \feyncalc. Putting everything together, we have
{\small
\begin{mmaCell}[moredefined={FCMultiLoopTID,CSPD,CFAD,SFAD,ApartFF,GFAD}]{Input}
  ApartFF[CFAD[\{\{k - p, 0\}, \{0, -1\}, 1\}] CSPD[k, p]*
  GFAD[\{\{Sqrt[CSPD[k, k]], 1\}, 1\}] CFAD[k], CSPD[k], \{k\}]//ApartFF[#, \{k\}] &
\end{mmaCell}

\begin{mmaCell}{Output}
  \mmaFrac{\mmaSqrt{\mmaSup{\textbf{k}}{2}}}{2\,(\mmaSup{(\textbf{k}-\textbf{p})}{2}-i\,\(\eta\))}+\mmaFrac{\mmaSqrt{\mmaSup{\textbf{k}}{2}} \mmaSup{\textbf{p}}{2}}{2\,(\mmaSup{\textbf{k}}{2}-i\,\(\eta\)).(\mmaSup{(\textbf{k}-\textbf{p})}{2}-i\,\(\eta\))}-\mmaFrac{\mmaSqrt{\mmaSup{\textbf{k}}{2}}}{2\,(\mmaSup{\textbf{k}}{2}-i\,\(\eta\))}
\end{mmaCell}
}
\noindent which indeed yields the desired form of the integral in \Eq\eqref{eq:apart-sqrt}.

Using the manipulations described in this section it should be possible to handle a wide range of (NR)EFT calculations, at least at 1-loop level. 

\subsection{FeynOnium}

The \feynonium extension builds upon the new symbols and routines introduced in the previous sections. It provides tools that help to streamline NREFT calculations by reducing the amount of code that needs to be written from scratch. Furthermore, \feynonium includes a number of worked out examples that explicitly reproduce selected NREFT results from the literature. This 
should not only help practitioners to quickly master the new framework but also lower the entry barrier for students and researchers from other branches of particle physics who would like to familiarize themselves with NREFT techniques.  

Most \feynonium functions tend to produce rather large output expressions, which are best viewed and processed within \mma.  Therefore, we prefer not to clutter this section by copying long code samples. Instead, we would like to explain the conceptual ideas behind those routines, making it clear where and why 
they should be used in practice. For explicit usage examples we refer to the \mma notebook accompanying this publication and scripts reproducing physical results that are provided together with the program.

In a matching calculation between a relativistic and a nonrelativistic theory with fermionic degrees of freedom it is often useful to rewrite Dirac spinor chains in terms of Pauli matrices and Pauli spinors. To this end \feynonium provides two special functions. \texttt{FM\-SpinorChain\-Explicit2} merely rewrites the chains using the Dirac representation of the Dirac matrices
\begin{equation}
\gamma^0 = \begin{pmatrix}
\mathbb{I} & 0 \\ 0 & -\mathbb{I} \end{pmatrix}, \qquad \bfGamma^i = \begin{pmatrix}
0 & \bfSigma^i \\ -\bfSigma^i & 0 \end{pmatrix}, \qquad \gamma^5 = \begin{pmatrix}
0 & \mathbb{I} \\ \mathbb{I} & 0 \end{pmatrix},
\end{equation}
without making any additional assumptions about the underlying process. In contrast, \texttt{FMSpinorChainExplicit} is specifically tailored for studying production or decay processes of heavy fermions in the rest or laboratory frame using Jacobi momenta. It implements the threshold expansion method from \cite{Braaten:1996jt} in 4 dimensions, where the small relative momentum between the two fermions in the rest frame can be used as an expansion parameter. The generalization to 3-body problems first derived in \cite{Brambilla:2017kgw} is also implemented. However, prior to applying \texttt{FMSpinorChainExplicit} it is necessary to perform an SPVAT (scalar, pseudoscalar, vector, axial-vector, tensor) decomposition of all Dirac chains using \texttt{DiracReduce}, convert the obtained spinor chains to a special notation via \texttt{FMToStandardSpinorChains} and employ \texttt{LorentzToCartesian} to break the manifest Lorentz covariance. 

In order to disentangle contributions from different angular momentum components $J$ in an amplitude one may want to explicitly project out the corresponding components of suitable tensors as in
\begin{equation}
\bfSigma^i \bfq^j \to
\begin{cases}
\displaystyle \frac{1}{3} \delta^{ij} (\bfSigma \cdot \bfq) \quad \textrm{for } J=0\\
\\
\displaystyle \frac{\bfSigma^i \bfq^j - \bfSigma^j \bfq^i}{2} \quad \textrm{for } J=1\\
\\
\displaystyle \frac{\bfSigma^i \bfq^j + \bfSigma^j \bfq^i}{2} - \frac{1}{3 } \delta^{ij} (\bfSigma \cdot \bfq) \quad \textrm{for } J=2
\end{cases}. \label{eq:cartProj}
\end{equation}
The routine \texttt{FM\-Cartesian\-Tensor\-Decomposition} encodes projections with $J=0, 1$ and $2$ for 3-dimensional tensors up to rank 5 and can be easily extended to contain more $J$-values  and higher rank tensors.

Another issue that regularly arises in complex nonrelativistic calculations are spurious terms that vanish by the virtue of the 3-dimensional Schouten identity
\begin{equation}
\epsilon^{i j k} \bfp^l -
\epsilon^{j k l} \bfp^i +
\epsilon^{k l i } \bfp^j -
\epsilon^{l i j } \bfp^k = 0,
\end{equation}
where $\bfp$ is an arbitrary Cartesian vector. In general, it is very difficult to apply this identity in  a systematic way, which is why \feynonium features a tool that facilitates this task.\footnote{Switching to a different basis spanned by 3 independent  vectors appearing in the calculation would be another possibility to deal with this problem.} \texttt{FMCartesianSchoutenBruteForce} tries out all possible combinations that can be formed out of the given list of Cartesian vectors and checks if this helps to reduce the number of terms in the expression. Although this approach may seem hopeless at first sight, in practice we observe that it works surprisingly well, eliminating most of the spurious terms after some number of iterations.

The use of covariant projectors for heavy nonrelativistic systems introduced in \cite{Bodwin:2002hg} can be automatized via \texttt{FMInsertCovariantProjector}. Production and decay projectors for spin singlet/triplet and color singlet/octet states can be thus applied straightforwardly.

Last but not least, we also implemented Feynman rules for pNRQCD vertices in the weak-coupling regime at order $r$ in the static limit as given in figure 5 of \cite{Brambilla:2004jw}. In the lack of a convenient way to generate pNRQCD Feynman diagrams automatically,\footnote{\texttt{FeynArts} does not support nonrelativistic theories, while \texttt{QGRAF} would require a separate interface to \feyncalc.} our implementation should significantly facilitate the tedious task of entering pNRQCD amplitudes by hand.

To sum up, let us once again clarify the distinction between the functionality available in \feyncalc 9.3 and \feynonium that is relevant for NREFTs. Here \feyncalc 9.3 provides the groundwork for nonrelativistic computations by introducing a new set of symbols that represent Cartesian tensors, Pauli matrices and nonrelativistic integrals. Those objects can be easily manipulated using existing \feyncalc routines. Such functions can be used in generic nonrelativistic QFT calculations, but they are not immediately useful for NREFTs. This shortcoming is explicitly addressed in \feynonium, where we provide specific tools for matching calculations in selected NREFTs: utilities for rewriting Dirac spinor chains into Pauli chains in specific kinematic frames, covariant projectors used in NRQCD, projections onto $J=0,1,2$ values of the angular momentum for Cartesian tensors or a routine to find and eliminate combinations of terms that vanish because of the Schouten identity in 3 dimensions. The practical usage of these tools can be inferred from the provided examples that are equally an integral part of \feynonium.

\section{Examples} \label{sec:examples}

In order to show how various functions of \feyncalc 9.3 and \feynonium can be put at work in real-life calculations, we include 8 example notebooks that reproduce various (NR)EFT calculations available in the literature. The notebooks are located in \texttt{FeynCalc\-/AddOns\-/FeynOnium\-/Examples}. This directory can be also accessed by clicking on the word \emph{examples} in the sentence ``Have a look at the supplied examples'' that appears when loading \feynonium
in \mma.

The 8 examples presented below are treated in the following scripts that can be found inside the \texttt{Examples} directory:
\begin{itemize}
\item Euler-Heisenberg Lagrangian
\begin{sloppypar}
 \quad \quad $\rightarrow$ \texttt{QED/OneLoop/GaGa-GaGa.m}
\end{sloppypar}
\item Heavy Baryon Effective Theory 
\begin{sloppypar}
 \quad \quad $\rightarrow$ \texttt{BChPT/OneLoop/N-N.m}
\end{sloppypar}
\item Dimension six 4-fermion operators in NRQCD (unequal mass case)
\begin{sloppypar}
 \quad \quad $\rightarrow$ \texttt{NRQCD\-/OneLoop\-/QiQjbar\--QiQjbar.m}
\end{sloppypar}
\item \texorpdfstring{$J/\psi \to 3 \gamma $}{JPsi to 3 gamma} decay in NRQCD 
\begin{sloppypar}
 \quad \quad $\rightarrow$ \texttt{NRQCD\-/Tree/QQbar-GaGaGa.m}
\end{sloppypar}
\item \texorpdfstring{$Q \bar{Q} \to \gamma \gamma$}{QQbat to 2 gamma} decays in NRQCD 
\begin{sloppypar}
 \quad \quad $\rightarrow$ \texttt{NRQCD\-/Tree/QQbar-GaGa.m}
\end{sloppypar}
\item Virtual corrections to inclusive hadronic decays of \texorpdfstring{$P$}{P}-wave quarkonia in NRQCD
\begin{sloppypar}
 \quad \quad $\rightarrow$ \texttt{NRQCD\-/OneLoop\-/QQbar-GlGl.m}
\end{sloppypar}
\item Relativistic corrections to quarkonium light-cone distribution amplitudes
\begin{sloppypar}
 \quad \quad $\rightarrow$ \texttt{NRQCD\-/Tree/\-H-QQbarGaGl-LCDA.m}
\end{sloppypar}
\item One-loop running of the chromoelectric dipole interaction in pNRQCD
\begin{sloppypar}
 \quad \quad $\rightarrow$ \texttt{pNRQCD\-/OneLoop\-/S-OG.m}
\end{sloppypar}
 \end{itemize}

\subsection{Euler-Heisenberg Lagrangian}

Although a systematic investigation of the EFT approach did not commence before the 70s of the last century, one can find many examples of much earlier applications of these techniques. One of them is the Euler-Heisenberg (EH) Lagrangian \cite{Heisenberg:1935qt}, an EFT of QED devised to describe photon-photon scattering at energies much below the electron mass $m_e$. The only degrees of freedom in this theory are low-energetic photons that interact with each other via 2$n$-photon vertices, with $n \geq 2$. These vertices arise from considering $2n$-photon scattering amplitudes in the full theory (QED) and integrating out the electrons. Since QED, unlike QCD, contains no tree-level gauge boson self-interactions, the matching starts at 1 loop. Effective vertices with an odd number of photons are forbidden by Furry's theorem \cite{Furry:1937zz}. 
At leading order in the $1/m_e$ expansion, the EH Lagrangian reads
\begin{equation}
\mathcal{L}_{\textrm{EH}} = - \frac{1}{4} F^{\mu \nu} F_{\mu \nu} + \frac{c_1}{m_e^4} \left ( F^{\mu \nu} F_{\mu \nu} \right )^2 + \frac{c_2}{m_e^4} F^{\mu \nu} F_{\nu \sigma} F^{\sigma \rho} F_{\rho \mu}.
\end{equation}
This theory is often presented in the introductory lectures to EFT (\cf \eg \cite{Manohar:2018aog}, \cite{Kaplan:2005es}), but the computation of the matching coefficients is either completely omitted or left as an exercise. More technical details can be found in \cite{Grozin:2009an}, yet the reader must still work out the missing steps on her or his own. 

Here we will follow the calculation of \cite{Grozin:2009an} and show how the matching coefficients (at leading order) can be determined in a semi-automatic fashion. On the QED side of the matching we need to consider the process
\begin{equation}
\gamma (p_1) + \gamma (p_2) \to \gamma (p_3) + \gamma (p_4),
\end{equation}
with $p_i^2 = 0$. It is sufficient to work with the forward scattering configuration $p_1 = p_3$, $p_2 = p_4$ which leaves us with only two kinematic invariants: $p_1 \cdot p_2$ and $m_e^2$. Then we can strip the QED amplitude of the polarization vectors and equate it to the corresponding amplitude in the EH EFT so that
\begin{equation}
T_{\textrm{QED}}^{\mu \nu \rho \sigma} = c_1 T_1^{\mu \nu \rho \sigma} + c_2 T_2^{\mu \nu \rho \sigma},
\end{equation}
where $c_1$ and $c_2$ are the unknown matching coefficients. This tensor equation can be converted into a system of two scalar linear equations by contracting it with $g^{\mu \nu} g^{\rho \sigma}$ and $g^{\mu \rho} g^{\nu \sigma}$. Then the task of determining $c_1$ and $c_2$ is reduced to the calculation of $T_{\textrm{QED}}^{\mu \nu \rho \sigma} g^{\mu \nu} g^{\rho \sigma}$ and $T_{\textrm{QED}}^{\mu \nu \rho \sigma} g^{\mu \rho} g^{\nu \sigma}$ expanded up to the third order in $p_1 \cdot p_2$ around 0.

To obtain the EFT amplitudes automatically, we need to create a \feynrules model of the EH Lagrangian at $\mathcal{O}(1/m_e^4)$ and export it to \feynarts. Since the Lagrangian is Lorentz covariant, this can be done in a straightforward fashion. The corresponding model file \texttt{EulerHeisenberg.fr} is already included in \feyncalc 9.3 and can be converted into a \feynarts model by evaluating the script \texttt{GenerateModelEulerHeisenberg.m}.\footnote{The script is located in \texttt{Examples/FeynRules/EulerHeisenberg} inside the \feyncalc directory.}

After having generated the QED and EH EFT amplitudes via \feynarts we need to convert them to the \feyncalc notation which is done with \texttt{FCFAConvert}. Then, \texttt{Contract} and \texttt{DiracSimplify} are employed to perform the contractions of the Lorentz indices and to simplify the Dirac algebra, including the evaluation of the Dirac traces. The quantities $T_{\textrm{QED}}^{\mu \nu \rho \sigma} g^{\mu \nu} g^{\rho \sigma}$ and $T_{\textrm{QED}}^{\mu \nu \rho \sigma} g^{\mu \rho} g^{\nu \sigma}$ are first reduced to scalar 1-loop integral by the means of \texttt{TID}. Then, \texttt{FIREBurn} calls \fire 5 \cite{Smirnov:2014hma} to eliminate propagators raised to integer powers using the IBP-reduction. The evaluation of the remaining scalar 1-loop integrals, including the expansion in $p_1 \cdot p_2$ is handled by \texttt{PaXEvaluate}, a frontend to \pax. As expected, the final result is free of UV and IR divergences, so that we can directly take the limit $d \to 4$. Upon substituting all the contributions back into the system of linear equations, we obtain the known result
\begin{equation}
c_1 = -\frac{\alpha^2}{36}, \quad c_2 = \frac{7 \alpha^2}{90},
\end{equation}
where $\alpha$ is the fine structure constant.

While one certainly could perform this calculation in a more efficient way using \form and the C++ version of \fire, the advantage of the presented approach is that it requires almost no familiarity with tools for automatic calculations and can be employed even by undergraduate students. On the other hand, a recent work \cite{Quevillon:2018mfl} that explored higher order operators in the EH Lagrangian and its QCD counterpart using \feyncalc and \feynhelpers clearly shows that these tools are very useful also in real research.

\subsection{Heavy Baryon Effective Theory}

\feyncalc's new ability to manipulate loop integrals with eikonal propagators can be handy even in 
comparably simple cases, such as the 1-loop correction to the heavy nucleon propagator in baryonic $\chi$PT \cite{Ecker:1995rk}. Following \cite{Scherer:2002tk}, we need to evaluate
\begin{equation}
 \int \frac{d^D k}{(2 \pi)^D} (- S_v \cdot k) \sigma^i \frac{1}{v \cdot (r-k) + i \eta} \frac{1}{k^2 - M^2 + i \eta} (S_v \cdot k) \sigma^i,
\label{eq:hbet}
\end{equation}
with 
\begin{equation}
S^\mu_v = - \frac{1}{2} \gamma_5 (\gamma^\mu \slashed{v} - v^\mu),
\end{equation}
which corresponds to a Feynman diagram with a nucleon emitting and absorbing a pion of mass $M$. Even though this calculation can be certainly done by pen and paper, with \feyncalc 9.3 it is effectively a one-liner that consists of applying \texttt{PauliSimplify} and 
\texttt{FCMultiLoopTID} to \Eq\eqref{eq:hbet} and readily yields the two master integrals $1/(k^2-M^2)$ and $1/[v \cdot (r-k) \, (k^2-M^2)]$ in agreement with the literature.

\subsection{Dimension six 4-fermion operators in NRQCD (unequal mass case)}

Matching coefficients that multiply NRQCD dimension six 4-fermion operators in the unequal mass case were originally obtained in \cite{Pineda:1998kj}. The corresponding operators are given by
\begin{align}
\delta \mathcal{L}_{\textrm{NRQCD}} &= \frac{d_{ss}}{m_1 m_2} \psi^\dagger \psi \chi^\dagger \chi + \frac{d_{sv}}{m_1 m_2} \psi^\dagger \bfSigma \psi \chi^\dagger \bfSigma \chi  \nonumber \\
& + \frac{d_{vs}}{m_1 m_2} \psi^\dagger T^a \psi \chi^\dagger T^a \chi
+ \frac{d_{vv}}{m_1 m_2} \psi^\dagger \bfSigma \psi \chi^\dagger \bfSigma \chi,
\end{align}
where $m_1$ ($m_2$) denotes the mass of a heavy quark (antiquark). The matching coefficients are determined by the hard momentum region (\ie loop momenta of order $m_1$, $m_2$) in the QCD box diagrams contributing to
\begin{equation}
Q (p_1) + \bar{Q}' (p_2) \to Q (p_3) + \bar{Q}' (p_4).
\end{equation}
It is convenient to rewrite the momenta $p_i$ as
\begin{subequations}
\begin{align}
p_1 &= \frac{1}{2} P + q, \quad p_2 = \frac{1}{2} P - q, \quad P \cdot q = 0, \\
p_3 &= \frac{1}{2} P' + q', \quad p_4 = \frac{1}{2} P' - q', \quad P' \cdot q' = 0, \\
\end{align}
\end{subequations}
with
\begin{equation}
p_1^2 = p_3^2 = m_1^2, \quad  p_2^2 = p_4^2 = m_2^2.
\end{equation}
To extract the values of $d_{ss}$, $d_{sv}$, $d_{vs}$ and $d_{vv}$, it is necessary to expand the QCD amplitudes at 0th order in the small relative momenta $q$ and $q'$. It is convenient to work in the center of mass frame, so that setting $q$ and $q'$ to zero is equivalent to setting
\begin{equation}
\bfp_1 = \bfp_2 = \bfp_3 = \bfp_4 = 0.
\end{equation}
In this case all scalar products between the 4-vectors $p_i$ can be expressed through polynomials in $m_1$ and $m_2$, \eg
\begin{equation}
(p_1 - p_2)^2 = (p_3 - p_4)^2 = (m_1 - m_2)^2.
\end{equation}

Having obtained the amplitude from \feynarts we carry out the tensor integral reduction as well as Dirac and color algebra simplifications. Employing the ``naive'' scheme for dealing with Pauli matrices in $D$-dimensions we can readily rewrite all Dirac structures in terms of $\xi^\dagger \bfSigma^i \xi$ and $\eta^\dagger \bfSigma^i \eta$. By using \texttt{Contract} with the option \texttt{EpsContract} set to \texttt{False} we actively prevent contractions of products of Levi-Civita tensors and can therefore implement 
the prescription of \cite{Pineda:1998kj} via a replacement rule. Explicitly, this amounts to using
\begin{equation}
\epsilon^{ijk} \epsilon^{ijk'} = (D-2) \delta^{k k'} \label{eq:fo-applications-ps-eps}.
\end{equation}
The loop integral structure of the amplitude has already been rewritten in terms of Passarino-Veltman functions which can be directly evaluated via \texttt{PaXEvaluateUVIRSplit} for $D=4-2 \varepsilon$ and expanded around $\varepsilon = 0$. Using following Fierz identities for color matrices \cite{Bodwin:1994jh}
\begin{subequations}
\begin{align}
T^a T^b \otimes T^b T^a &=
\frac{C_F}{2 N_c} 1 \otimes 1 + \frac{N_c^2-2}{2 N_c} T^a \otimes T^a, \\
T^a T^b \otimes T^a T^b &=
\frac{C_F}{2 N_c} 1 \otimes 1 - \frac{1}{ N_c} T^a \otimes T^a,
\end{align}
\end{subequations}
where $C_F = (N_c^2-1)/(2 N_c)$, we can easily separate color singlet and color octet contributions from each other. The separation into spin singlet and spin triplet pieces is even simpler, the former being proportional to $\xi^\dagger \xi$ or $\eta^\dagger \eta$ and the latter to $\xi^\dagger \bfSigma^i \xi$ or $\eta^\dagger \bfSigma^i \eta$ respectively.

Thus we finally recover the known results from the literature given by \cite{Pineda:1998kj} (confirmed also in \cite{Brambilla:2005yk})
\begin{subequations}
\begin{align}
d_{ss} &= - C_F  \left (\frac{N_c}{2} - C_F \right ) \frac{\alpha_s^2}{m_1^2-m_2^2} \left [ m_1^2 \left ( \log \frac{m_2^2}{\mu^2} + \frac{1}{3} \right ) - m_2^2 \left ( \log \frac{m_1^2}{\mu^2} + \frac{1}{3} \right )   \right ], \\
d_{sv} &=  C_F  \left (\frac{N_c}{2} - C_F \right ) \frac{\alpha_s^2}{m_1^2-m_2^2} m_1 m_2 \log \frac{m_1^2}{m_2^2}, \\
d_{vs} &= - \frac{2 C_F \alpha_s^2}{m_1^2-m_2^2} \left [ m_1^2 \left ( \log \frac{m_2^2}{\mu^2} + \frac{1}{3} \right ) - m_2^2 \left ( \log \frac{m_1^2}{\mu^2} + \frac{1}{3} \right )   \right ] \nonumber \\
& + \frac{N_c \alpha_s^2}{4(m_1^2-m_2^2)} \left [
3 \left (m_1^2 \left ( \log \frac{m_2^2}{\mu^2} + \frac{1}{3} \right ) - m_2^2 \left ( \log \frac{m_1^2}{\mu^2} + \frac{1}{3} \right ) \right) \right. \nonumber \\
& \left. + \frac{1}{m_1 m_2} \left (m_1^4 \left ( \log \frac{m_2^2}{\mu^2} + \frac{10}{3} \right ) - m_2^4 \left ( \log \frac{m_1^2}{\mu^2} + \frac{10}{3} \right )  \right ) \right ], \\
d_{vv} &=  \frac{2 C_F \alpha_s^2}{m_1^2 - m_2^2} m_1^2 m_2^2 \log \frac{m_1^2}{m_2^2} \nonumber \\
& + \frac{N_c \alpha_s^2}{4(m_1^2-m_2^2)} \left [
\left ( m_1^2 \left ( \log \frac{m_2^2}{\mu^2} + 3 \right ) - m_2^2 \left ( \log \frac{m_1^2}{\mu^2} + 3 \right ) \right ) - 3 m_1 m_2  \log \frac{m_1^2}{m_2^2} \right ],
\end{align}
\end{subequations}
where $\alpha_s$ is the strong coupling constant.

\subsection{\texorpdfstring{$J/\psi \to 3 \gamma $}{JPsi to 3 gamma} decay in NRQCD}

The LO (both in velocity and $\alpha_s$) NRQCD prediction for the decay $J/\psi \to 3 \gamma$ (or $\Upsilon(1S) \to 3 \gamma$) can be extracted by adapting
the corresponding calculation for orthopositronium \cite{Ore:1949te}. Nonetheless, it is also instructive to explicitly repeat this calculation by matching the QCD tree-level amplitude
\begin{equation}
Q (p_1) + \bar{Q} (p_2) \to \gamma (k_1) + \gamma (k_2) + \gamma (k_3)
\end{equation}
to NRQCD. The kinematics is
\begin{equation}
p_1 = \frac{1}{2} P + q, \quad p_2 = \frac{1}{2} P - q, \quad P \cdot q = 0
\end{equation}
and
\begin{equation}
p_{1,2}^2  = m_Q^2, \quad k_{1,2,3}^2 = 0,
\end{equation}
which implies
\begin{equation}
p_{1,2}^0 =  E_{\bfq}, \quad k_{1}^0 = |\bfk_1|, \quad k_{2}^0 = |\bfk_2|, \quad k_{3}^0 = 2 E_{\bfq} - |\bfk_1| - |\bfk_2|,
\end{equation}
with $E_{\bfq} = \sqrt{\bfq^2 + m_Q^2}$. It is also convenient to parametrize $|\bfk_1|$ and $|\bfk_2|$ as
\begin{equation}
|\bfk_1| = E_{\bfq} x_1, \quad |\bfk_2| = E_{\bfq} x_2
\end{equation}
where $x_2$ ranges from $0$ to $1- x_1$, while $x_1$ will be eventually integrated from 0 to 1.

It is sufficient to expand the 6 QCD diagrams to 0th order in $|\bfq|$ and, after switching to Pauli matrices and spinors via \texttt{Lorentz\-To\-Cartesian} and \texttt{FM\-Spinor\-Chain\-Explicit2}, we can square the QCD amplitude and sum over the polarizations of the photons. The angular integration of the 3-body phase space can be replaced by the $J=0$ projection with respect to the unit vectors $\hat{\bfk}_1$ and $\hat{\bfk}_2$  via \texttt{FM\-Cartesian\-Tensor\-Decomposition}. Upon integrating over $x_1$ and $x_2$ (here it can be done analytically using \mma's \texttt{Integrate}) and multiplying with the corresponding prefactor we 
obtain the total decay rate in QCD at LO given by
\begin{equation}
\Gamma_{\textrm{QCD}} ( Q \bar{Q} \to 3 \gamma) = \frac{8 (\pi^2 - 9) \alpha^3 e_Q^6}{9 m_Q^2} \eta^\dagger \bfSigma^i \xi \, \xi^\dagger \bfSigma^i \eta,
\end{equation}
where $e_Q$ is the fractional electric charge of the heavy quark $Q$. Comparing it to the corresponding perturbative NRQCD expression
\begin{equation}
\Gamma_{\textrm{pert. NRQCD}}  = \frac{2 \textrm{Im} f_{\textrm{em}} (^3 S_1 )}{m_Q^2} \eta^\dagger \bfSigma^i \xi \, \xi^\dagger \bfSigma^i \eta,
\end{equation}
we correctly conclude that \cite{Bodwin:1994jh}
\begin{equation}
 \textrm{Im} f_{\textrm{em}} (^3 S_1 ) = \frac{4}{9} (\pi^2 - 9) \alpha^3 e_Q^6.
\end{equation}

\subsection{\texorpdfstring{$Q \bar{Q} \to \gamma \gamma$}{QQbat to 2 gamma} decays in NRQCD}

Let us consider at tree-level the QCD process 
\begin{equation}
Q (p_1) + \bar{Q} (p_2) \to \gamma (k_1) + \gamma (k_2),
\end{equation}
with
\begin{equation}
p_1 = \frac{1}{2} P + q, \quad p_2 = \frac{1}{2} P - q, \quad P \cdot q = 0,
\end{equation}
and
\begin{equation}
p_{1,2}^2  = m_Q^2, \quad k_{1,2}^2 = 0,
\end{equation}
where we want to expand the amplitude in the relative momentum of the heavy quark pair, $q$, up to 4th order. To this end it is necessary to spell out all kinematic dependence on $|\bfq|$, \eg to specify
\begin{equation}
q^0 = 0, \quad p_{1,2}^0 = k_{1,2}^0 = E_{\bfq},
\end{equation}
with $E_{\bfq} = \sqrt{\bfq^2 + m_Q^2}$.
This nonrelativistic expansion requires us not only to distinguish between spin singlet and spin triplet contributions but also to 
explicitly project components of the amplitude corresponding to the total angular momentum values $J=0,1,2$. Of course, the $J=1$ contribution must vanish due to the Landau-Yang theorem \cite{Landau:1948kw,Yang:1950rg}.

These steps can be directly performed with the aid of the \feynonium functions \texttt{FM\-Spinor\-ChainExplicit2}, \texttt{Pauli\-Simplify}
and \texttt{FM\-Cartesian\-Tensor\-Decomposition}. Furthermore, by applying \texttt{FM\-Cartesian\-Schouten\-BruteForce} to the $J=2$ contribution we can readily remove terms that vanish by Schouten's identity.

The so-obtained QCD amplitudes can be then matched to NRQCD, as it was done in \cite{Brambilla:2006ph}, to obtain the matching coefficients relevant for the
quarkonium decay processes $\eta_Q \to \gamma \gamma$, $\chi_{Q0} \to \gamma \gamma$ and $\chi_{Q2} \to \gamma \gamma$, where
the heavy quark flavor $Q$ can be $c$ (for charmonia) or $b$ (for bottomonia). Since we cannot generate NRQCD amplitudes automatically, they must be entered by hand. We write the NRQCD amplitudes up to order $|\bfq|^4$. In the next step, we square the NRQCD amplitudes and sum over the polarizations of the photons to arrive to the final heavy quarkonia decay rates in perturbative NRQCD. From there we can read off the matching coefficients of NRQCD decay operators that contribute through the leading heavy quarkonium Fock state $\ket{Q \bar{Q}}$. These are at LO (in agreement with \cite{Brambilla:2006ph})
\begin{subequations}
\begin{align}
\textrm{Im} f_{\textrm{em}} (^1 S_0) &= \alpha^2 e_Q^4 \pi, \\
\textrm{Im} g_{\textrm{em}} (^1 S_0) &= - \frac{4}{3} \alpha^2 e_Q^4 \pi, \\
\textrm{Im} f_{\textrm{em}} (^3 P_0) &= 3 \alpha^2 e_Q^4 \pi, \\
\textrm{Im} f_{\textrm{em}} (^3 P_2) &= \frac{4}{5} \alpha^2 e_Q^4 \pi, \\
\textrm{Im} h_{\textrm{em}} (^1 D_2) &= \frac{2}{15} \alpha^2 e_Q^4 \pi, \\
\textrm{Im} h'_{\textrm{em}} (^1 S_0) + \textrm{Im} h''_{\textrm{em}} (^1 S_0) &= \frac{68}{45} \alpha^2 e_Q^4 \pi, \\
\textrm{Im} g_{\textrm{em}} (^3 P_0) &= - 7 \alpha^2 e_Q^4 \pi, \\
\textrm{Im} g_{\textrm{em}} (^3 P_2) &= - \frac{8}{5} \alpha^2 e_Q^4 \pi, \\
\textrm{Im} g_{\textrm{em}} (^3 P_2, ^3 F_2) &= - \frac{20}{21} \alpha^2 e_Q^4 \pi.
\end{align}
\end{subequations}
The definitions of operators multiplying these coefficients can be found in the appendix of \cite{Brambilla:2006ph}.

\subsection{Inclusive hadronic decays of \texorpdfstring{$P$}{P}-wave quarkonia in NRQCD}

The inclusive decay of $\chi_{QJ}$ ($Q=c$ or $b$) into light hadrons (LH) at LO in the velocity in the framework of NRQCD can be written as \cite{Bodwin:1994jh}
\begin{equation}
\Gamma (\chi_{QJ} \to \textrm{LH}) = \frac{2 \textrm{Im} f_1 (^3 P_J)}{m_Q^4} \braket{\chi_{QJ}|\mathcal{O}_1 (^3 P_J) |\chi_{QJ}} +  \frac{2 \textrm{Im} f_8 (^3 S_1)}{m_Q^2} \braket{\chi_{QJ}|\mathcal{O}_8 (^3 S_1) |\chi_{QJ}},
\end{equation}
where $\braket{\chi_{QJ}|\mathcal{O}_1 (^3 P_J) |\chi_{QJ}}$ and $\braket{\chi_{QJ}|\mathcal{O}_8 (^3 S_1) |\chi_{QJ}}$ denote NRQCD matrix elements.
Here we would like to consider virtual next-to-leading order (NLO) loop corrections in the 2 gluon channel to $\textrm{Im} f_1 (^3 P_{0,2})$, which were first calculated in \cite{Barbieri:1980yp} (with IR divergences regularized with a gluon mass) and later in \cite{Petrelli:1997ge} using the NRQCD formalism and
explicitly distinguishing between UV and IR poles in dimensional regularization.

To extract the Born level contribution we need to consider three\footnote{The diagram involving the three-gluon vertex does not contribute to the color singlet
state but is added for the sake of completeness.} tree-level diagrams describing the process 
\begin{equation}
Q (p_1) + \bar{Q} (p_2) \to g (k_1) + g (k_2) \label{eq:qqbartogg}
\end{equation}
in QCD, where all external particles are put on-shell. The momenta of the heavy quarks can be rewritten as 
\begin{equation}
p_1 = \frac{1}{2} P + q, \quad p_2 = \frac{1}{2} P - q, \quad P \cdot q = 0,
\end{equation}
where $P$ is the total heavy quarkonium momentum and $q$ is the relative momentum. Since we are not interested in relativistic corrections
we can simplify the kinematics by setting
\begin{equation}
P^2 \approx 4 m_Q^2, \quad P \cdot k_{1,2} \approx 2 m_Q^2, \quad q \cdot k_2 = - q \cdot k_1, \quad k_1 \cdot k_2 \approx 2 m_Q^2.
\end{equation}
Following  \cite{Petrelli:1997ge} we project on spin-triplet $P$-wave states via a suitable spin-triplet color singlet covariant projector
\begin{equation}
\mathcal{A}_{S=1,L=1} = \mathcal{E}_{\alpha \beta} \frac{d}{d q^\beta} \Tr \left [ \frac{\mathbb{I}_c}{\sqrt{N_c}} \Pi_1^\alpha \mathcal{A} \right ] \biggl |_{q=0},
\end{equation}
with 
\begin{equation}
\Pi_1^\alpha = \frac{1}{\sqrt{8 m_Q^3}} \left ( \frac{\slashed{P}}{2} - \slashed{q}- m_Q \right ) \gamma^\alpha \left ( \frac{\slashed{P}}{2} + \slashed{q} + m_Q  \right ),
\end{equation}
where $\mathcal{A}$ is the original amplitude, $\mathcal{E}_{\alpha \beta}$ stands for the polarization of the quarkonium, and $\mathbb{I}_c$ is a unit matrix in color space. The trace is understood to be taken over spinor and color indices.

The trace of $\mathcal{A}$ can be implemented with \texttt{FM\-Insert\-Covariant\-Projector}, while the derivative with respect to $q^\beta$ can be obtained using \texttt{Four\-Divergence}. When squaring the amplitude $\mathcal{A}_{S=1,L=1}$ using \texttt{Complex\-Conjugate}, we need to sum over the polarizations of the quarkonia with different $J$-values using
\begin{subequations}
\begin{align}
\sum_{J_z} \mathcal{E}^{(J=0)}_{\alpha \beta}  \mathcal{E}^{(J=0) \ast}_{\alpha' \beta'} &= \frac{1}{D-1} \Pi_{\alpha \beta} \Pi_{\alpha' \beta'}, \\
\sum_{J_z} \mathcal{E}^{(J=1)}_{\alpha \beta}  \mathcal{E}^{(J=1) \ast}_{\alpha' \beta'} &= \frac{1}{2} \left ( \Pi_{\alpha \alpha'} \Pi_{\beta \beta'} -  \Pi_{\alpha \beta'} \Pi_{\alpha' \beta} \right ), \\
\sum_{J_z} \mathcal{E}^{(J=2)}_{\alpha \beta}  \mathcal{E}^{(J=2) \ast}_{\alpha' \beta'} &= \frac{1}{2} \left ( \Pi_{\alpha \alpha'} \Pi_{\beta \beta'} + \Pi_{\alpha \beta'} \Pi_{\alpha' \beta} \right ) - \frac{1}{D-1} \Pi_{\alpha \beta} \Pi_{\alpha' \beta'},
\end{align}
\end{subequations}
where
\begin{equation}
\Pi_{\alpha \beta} = - g_{\alpha \beta} + \frac{P_\alpha P_\beta}{4m_Q^2}.
\end{equation}
Doing so we recover ($\varepsilon$ is the dimensional regularization parameter from $D = 4 - 2 \varepsilon$)
\begin{subequations}
\begin{align}
\Gamma_{\textrm{Born}} ( ^3 P_0^{[1]} \to gg) &= C_F \frac{144 \alpha_s^2 \mu^{4 \varepsilon} \pi^2}{m_Q^4} \Phi_{(2)} \frac{1-\varepsilon}{3-2\varepsilon} \braket{H|\mathcal{O}_1 (^3 P_0) |H}, \\
\Gamma_{\textrm{Born}} ( ^3 P_1^{[1]} \to gg) &= 0, \\
\Gamma_{\textrm{Born}} ( ^3 P_2^{[1]} \to gg) &= C_F \frac{32 \alpha_s^2 \mu^{4 \varepsilon} \pi^2}{m_Q^4} \Phi_{(2)} \frac{6-13\varepsilon + 4 \varepsilon^2}{(3-2\varepsilon)(5-2\varepsilon)} \braket{H|\mathcal{O}_1 (^3 P_0) |H},
\end{align}
\end{subequations}
from appendix\,B.1 of \cite{Petrelli:1997ge}, where
\begin{equation}
 \Phi_{(2)} = \frac{1}{8 \pi} \left ( \frac{\pi}{m_Q^2}\right )^\varepsilon  \frac{\Gamma(1-\varepsilon)}{\Gamma(2-2\varepsilon)}.
\end{equation}

The calculation of the virtual corrections proceeds along the same lines as above, but is technically more challenging. We need to evaluate QCD 1-loop corrections to
the process given in \Eq\,\eqref{eq:qqbartogg}. The results for the $J=0$ and $J=2$ contributions for each Feynman diagram are available in tables 2 and 3 of \cite{Petrelli:1997ge}. Contrary to the approach chosen in the original publication, we choose to evaluate loop integrals \emph{after} and not before applying covariant projectors and expanding in $q$. As it has been observed in \cite{Butenschon:2009zza}, in this case we do not encounter any Coulomb singularities. Furthermore, to speed up the calculation, we choose to reduce the number of integrals that need to be evaluated by using IBP reduction. This obscures however the distinction between UV and IR divergences in dimensional regularization, so that we denote both kind of poles with $\varepsilon$. Despite using solely \mma we can obtain the final result within half an hour on a modern laptop. Having assembled the virtual color singlet contributions to the decay of $^3 P_J$ quarkonia into two gluons from the evaluated 1-loop diagrams and the tree-level amplitude we find
\begin{equation}
\Gamma (^3 P_J^{[1]} \to gg)  = \Gamma_{\textrm{Born}} (^3 P_J^{[1]} \to gg) \frac{\alpha_s}{\pi} f_\varepsilon \left ( - \frac{N_c}{\varepsilon^2} + B^{[1]}_{^3 P_J}  \right ),
\end{equation}
with 
\begin{subequations}
\begin{align}
B^{[1]}_{^3 P_0} &= C_F \left ( - \frac{7}{3} + \frac{\pi^2}{4} \right ) + N_c \left ( \frac{1}{3} + \frac{5}{12} \pi^2 \right ), \\
B^{[1]}_{^3 P_1} &= 0, \\
B^{[1]}_{^3 P_2} &= - 4 C_F + N_c \left ( \frac{1}{3} + \frac{5}{3} \ln 2 + \frac{\pi^2}{6} \right ),
\end{align}
\end{subequations}
where
\begin{equation}
f_\varepsilon = \left ( \frac{\mu}{m_Q} \right )^{2 \varepsilon} \Gamma(1+\varepsilon).
\end{equation}
The result from \cite{Petrelli:1997ge} agrees with our expression upon setting $\varepsilon_{\textrm{UV}} = \varepsilon_{\textrm{IR}}$ and dropping the Coulomb singularity.

\subsection{Relativistic corrections to 
quarkonium light-cone distribution amplitudes}

\feynonium's capability of handling nonrelativistic objects can be combined with
the built-in 4-dimensional algebra of \feyncalc to perform nonrelativistic
expansion of QCD hadronic matrix elements that involve lightlike collinear
momenta. 
In this section, we show how \feynonium can be used to compute light-cone
distribution amplitudes (LCDAs) of $J^{PC} = 1^{--}$ heavy quarkonia in NRQCD 
to relative order $v^4$ accuracy at leading order in $\alpha_s$, 
which was first done in \cite{Brambilla:2019fmu}. 
Following \cite{Brambilla:2019fmu}, we compute the LCDA for
the $J/\psi$ state with momentum $P$, which is defined by the matrix element
\begin{equation}
\label{eq:LCDAdef}
\braket{ J/\psi | \epsilon \cdot \mathcal{Q} (x) | 0 } =
\braket{ J/\psi | 
\int \frac{d \omega}{2 \pi} 
e^{-i (x-1/2) \omega \bar{n} \cdot P} 
(\bar{Q} W_c) (\omega \bar{n}/2) \slashed{\bar{n}} \varepsilon \cdot \gamma
(W_c^\dag Q) (-\omega \bar{n}/2)
| 0},
\end{equation}
where $\varepsilon$ is the polarization 4-vector of the $J/\psi$, and $Q(x)$ is the QCD heavy-quark field, 
which is a four-component Dirac spinor field. The Wilson line
\begin{equation}
W_c(x)=\mathcal{P} \exp \left[-i g \int_{-\infty}^0 ds \, \bar{n} \cdot A(x+s \bar{n})
\right],
\end{equation}
where $\mathcal{P}$ is the path ordering operator, ensures the
gauge invariance of the nonlocal operator $\mathcal{Q}^\alpha (x)$.
The light-cone vectors $n$ and $\bar{n}$ are lightlike vectors
that satisfy $n \cdot \bar{n} = 2$.

To relative order $v^4$ accuracy, the $J/\psi$ LCDA is 
given in the NRQCD factorization formalism by 
\begin{equation}
\label{eq:LCDANRQCDfac}
- \braket{ J/\psi | \epsilon \cdot \mathcal{Q} (x) | 0 }
= \sum_n \frac{\tilde{c}_n (x)}{m^{d_n-3}} \bfeps \cdot 
\braket{J/\psi | \mathcal{\bfO}_n | 0},
\end{equation}
where $d_n$ is the dimension of the NRQCD operator $\mathcal{\bfO}_n$, and 
$\tilde{c}_n (x)$ are perturbative short-distance coefficients. 
To relative order $v^4$ accuracy, 
the sum over $n$ involves the NRQCD operators with $J^{PC} = 1^{--}$ 
of dimensions up to 7 that are listed in \Eqs(4) and (6) of 
\cite{Brambilla:2019fmu}. 
Additionally, NRQCD operators with $J^{PC} = 1^{+-}$ of dimensions up to 7 
can be found in appendix\,C of \cite{Brambilla:2019fmu}.

Our goal is to compute the $\tilde{c}_n(x)$ at leading order in 
$\alpha_s$. The coefficients $\tilde{c}_n(x)$ can be determined from the 
matching conditions 
\begin{subequations}
\label{eq:matching-lcda}
\begin{eqnarray} 
- \braket{ Q \bar{Q} (J^{PC}=1^{--}) | \varepsilon \cdot \mathcal{Q} (x) | 0}
&=& \sum_n \frac{\tilde{c}_n (x)}{m^{d_n-3}} \bfeps \cdot 
\braket{ Q \bar{Q} | \mathcal{\bfO}_n | 0 },
\\
- \braket{ Q \bar{Q} g (J^{PC}=1^{--})| \varepsilon \cdot \mathcal{Q} (x) | 0}
&=& \sum_n \frac{\tilde{c}_n (x)}{m^{d_n-3}} \bfeps \cdot 
\braket{ Q \bar{Q} g | \mathcal{\bfO}_n | 0 }.
\end{eqnarray}
\end{subequations}
On the left-hand sides of \Eqs\eqref{eq:matching-lcda}, we project onto the 
$J^{PC} = 1^{--}$ state, because unlike the NRQCD operators on the right-hand 
sides of \Eqs\eqref{eq:matching-lcda}, the operator $\mathcal{Q}^\alpha (x)$
does not have a definite $J^{PC}$. Here, $\varepsilon$ is the polarization vector of the perturbative $Q \bar{Q}$ or $Q \bar{Q} g$ state; a state $\ket{Q \bar{Q} g}$ is  a state made of a heavy quark ($Q$), a heavy antiquark ($\bar{Q}$) and a gluon ($g$).

While the NRQCD matrix elements on the right-hand sides of \Eqs\eqref{eq:matching-lcda} can be computed straightforwardly at orders $g^0$ and $g^1$, respectively~\cite{Brambilla:2017kgw}, the calculation of the QCD matrix elements on the left-hand sides of \Eqs\eqref{eq:matching-lcda} is much more involved.
\feynonium can provide significant simplifications of the calculation: 
the complicated algebraic manipulations that are needed to compute 
the QCD matrix elements on the left-hand sides of \Eqs\eqref{eq:matching-lcda} 
can be done by using \feynonium's built-in functions. 

The initial step of the calculation involves the computation of the tree-level
diagrams that contribute to 
$\braket{ Q \bar{Q} | \mathcal{Q}^\alpha (x) | 0 }$ 
and $\braket{ Q \bar{Q} g | \mathcal{Q}^\alpha (x) | 0 }$, and expanding in
powers of the small relative 3-momenta of the $Q$, $\bar{Q}$ and $g$. 
The temporal and spatial components of the 4-momenta that appear in the
matrix elements are made explicit in the expressions using the \texttt{LorentzToCartesian} command. In this way complicated 4-dimensional 
tensors can be rewritten in terms of Cartesian tensors. The gamma matrices are now expressed in terms of Pauli matrices, and the spins of the heavy quark and the heavy antiquark combine into spin singlets
and spin triplets. Then, the expansion in powers of the small momenta 
can be done by standard Mathematica commands. 

The calculation of the left-hand sides of \Eqs\eqref{eq:matching-lcda} is completed
by projecting onto $J=1$, $C=-1$ and $P=-1$. 
The projection onto $C=-1$ is straightforward: since the charge conjugation of
the operator $\mathcal{Q}^\alpha (x)$ is $-\mathcal{Q}^\alpha (1-x)$, 
the $C=-1$ contributions of 
$\braket{ Q \bar{Q} | \mathcal{Q}^\alpha (x) | 0 }$
and $\braket{ Q \bar{Q} g | \mathcal{Q}^\alpha (x) | 0 }$
are the contributions symmetric in $x \leftrightarrow 1-x$. 

The projection onto $P=-1$ is also straightforward. The parity
transform involves reversing the signs of momenta and gluon polarization
3-vectors, as well as the spins of the heavy quark and the heavy antiquark in
singlet and triplet combinations. This can be done simply by using \mma's
\texttt{Replace} command. Alternatively, instead of projecting onto $P=-1$, 
we can also include NRQCD operators with $P = +1$ on the right-hand sides of 
\Eqs\eqref{eq:matching-lcda}. 

The projection onto $J=1$ can be complicated, as it requires the
application of the Cartesian tensor reduction algorithm developed in 
\cite{Coope1970} to reduce the tensors of rank up to 5 built from 
3-vectors to tensors of rank 1. 
This reduction can be done automatically with \feynonium's built-in 
command \texttt{FMCartesianTensorDecomposition}.

Once $\braket{ Q \bar{Q} (J^{PC} = 1^{--}) | \mathcal{Q}^\alpha (x) | 0 }$ 
and $\braket{ Q \bar{Q} g (J^{PC} = 1^{--})| \mathcal{Q}^\alpha (x) | 0 }$ 
have been expressed in terms of 3-vectors, the short-distance coefficients $\tilde{c}_n(x)$ can be obtained from \Eqs\eqref{eq:matching-lcda}. 
These results for the $\tilde{c}_n(x)$ can then be plugged into 
\Eq\eqref{eq:LCDANRQCDfac} to obtain the quarkonium LCDAs. 
In \cite{Brambilla:2019fmu}, the $J/\psi$ and $\Upsilon(nS)$ LCDAs 
were computed to relative order $v^4$ accuracy using \feynonium, and the 
LCDAs were then used to compute Higgs boson decays to a heavy quarkonium plus 
a photon. 

\subsection{One-loop running of the chromoelectric dipole interaction in pNRQCD}

To illustrate the practical usefulness of the nonrelativistic integral manipulations presented in section \ref{sec:usage-loops}, we can reproduce the 1-loop renormalization group equations (RGEs) for the running of the matching coefficients $V_A(r)$ and $V_B(r)$ in weakly coupled pNRQCD (\cf \Eq\eqref{eq:pNRQCD_Lagrangian}) \cite{Brambilla:1999xf,Pineda:2000gza}.

The calculation proceeds by first entering 1-loop amplitudes for the pNRQCD processes $S \to O g$ (for $V_A$) and $O \to O g$ (for $V_B$). The corresponding two diagrams are shown in figures 7 and 8 of ref.\,\cite{Brambilla:1999xf}, respectively. The \feynonium shortcuts for the pNRQCD Feynman rules render this procedure less error-prone and more convenient than copying expressions from a pen and paper calculation. We choose to handle mixed integrals such as $\int d^D k \,/{[k^2 \bfk^2 (\bfp - \bfk)^2]}$ by explicitly integrating over $k^0$ and closing the contour below, so that we enclose the pole at $k^0 = |\bfk| - i \eta$.

To apply this operation in \feyncalc we employ the function \texttt{FCLoopExtract}, which gives us a list of all loop integrals present in the expression. Then we implement the residue integration in form of a custom function that automatically generates a suitable replacement rule for each of the loop integrals. Upon substituting these results back into the original amplitude we end up
with purely Cartesian integrals.

Our next step consists of employing \texttt{FCMultiLoopTID} for the tensor reduction and using \texttt{ApartFF} to enforce
some custom partial fractionings. To this end we multiply the amplitude with $\bfk^2/\bfk^2$ and $|\bfk|/|\bfk|$ as explained in section \ref{sec:usage-loops}. After this, we end up with 14 resulting 1-loop integrals that need to be calculated by hand. Some of them are obviously scaleless and can be put to zero immediately. Notice that to obtain
the running we care only about UV singularities, so that all UV-finite parts can be discarded. This significantly simplifies the evaluation of the master integrals.

We find that both for $V_A$ and  $V_B$ the pole of the first diagram contributing to the running is canceled by the pole of the second diagram.\footnote{
The result in \cite{Pineda:2000gza} corrects a sign error in the earlier computation in \cite{Brambilla:1999xf}.} This implies that at $\mathcal{O}(\alpha_s)$ it holds that
\begin{subequations}
\begin{align}
\mu \frac{d V_A}{d \mu} &= 0, \\
\mu \frac{d V_B}{d \mu} &= 0,
\end{align}
\end{subequations}
which reproduces the results of \cite{Pineda:2000gza}.

\section{Summary} \label{sec:summary}

In this work we have presented a software named \feynonium that works on top of \textsc{Wolfram} \mma
and \textsc{FeynCalc}. The main purpose of \feynonium is to facilitate the application of EFTs to various particle physics phenomena at tree- and 1-loop level by providing a large set of useful routines for tasks that typically arise in such calculations.

One of the highlights of the package is the ability to perform calculations in nonrelativistic QFTs, a feature not readily available in other public codes. To achieve this, it was necessary to perform extensive modifications of the \textsc{FeynCalc} package, which is now capable of directly manipulating Cartesian tensors and integrals. 

\feynonium is open-source, publicly available, flexible and easy to use. The large number of included examples illustrates how this software can be used to quickly reproduce many well known EFT results from the literature, in particular in the frameworks of NRQCD, pNRQCD, HQET and ChPT.

Conceptually, the usage of \feynonium is very similar to a pen and paper calculation, in the sense that everything can be organized as a sequence of simple operations (contractions, expansions, algebraic simplifications etc.) and all intermediate expressions are easily accessible for plausibility checks or comparisons to existing results.

\feynonium should not be confused with other packages that target EFT practitioners but attempt to hide the technical side of the calculation by presenting to the user only the final results such as matching coefficients or cross sections. Our design philosophy was not to create another ``black box'', but a ``toolbox'' that naturally assumes the familiarity of the user with the EFT methods and provides her or him with means to investigate the relevant questions in a very flexible and convenient way.

We readily admit that performance-wise \feynonium can hardly compete with codes that were specifically tailored and optimized for a particular calculation, especially if they are based on \form. However, while such high performance private codes are usually accessible only to a very small subset of EFT practitioners, our package is available for everyone. This promotes good scientific practice by sharing tools that are beneficial for the whole EFT community and in particular by encouraging researchers from other branches to embrace the EFT techniques.

\acknowledgments

The research of N.\,B. is supported by the Deutsche Forschungsgemeinschaft (DFG, German Research Foundation) Grant No. BR 4058/2-2.
The work of V.\,S. and A.\,V. has been supported by the DFG and the NSFC through funds provided to the Sino-German CRC 110 ``Symmetries and the Emergence of Structure in QCD'' (NSFC Grant No. 11261130311). V.\,S. also acknowledges the support from the DFG under grant 396021762 - TRR 257 ``Particle Physics Phenomenology after the Higgs Discovery'', National Science Foundation of China (11135006, 11275168, 11422544, 11375151, 11535002) and the Zhejiang University Fundamental Research Funds for the Central Universities (2017QNA3007). N.\,B., H.\,S.\,C., V.\,S and A.\,V. acknowledge support from the DFG cluster of excellence ``ORIGINS'' under Germany's Excellence Strategy - EXC-2094 - 390783311. H.\,S.\,C. also acknowledges support from the Alexander von Humboldt Foundation.

All \mma expressions in this work were converted to \LaTeX\,  using the \textsc{CellsToTeX} (\url{https://github.com/jkuczm/MathematicaCellsToTeX}) package. V.\,S would like to thank the author of the package Jakub Kuczmarski for his help with the automatic conversion of some typographically involved expressions (resulting from typesetting rules implemented in \feyncalc).
\appendix

\section{Useful formulas for Pauli algebra}
\label{sec:app-pauli}

In this appendix, we collect some formulas for Pauli algebra calculations
in $4$- and $D$-dimensions that we find particularly useful and that 
can be readily implemented in a computer algebra system.

Chains of Pauli matrices in $4$- or $D$-dimensions can be simplified as follows
\begin{subequations}
\begin{align}
\bfSigma^i \bfSigma^{j_1} \ldots \bfSigma^{j_{n}} \bfSigma^i &= (-1)^n (D-3) \bfSigma^{j_1} \ldots \bfSigma^{j_n} \nonumber \\
&  + 2 \sum_{i=1}^{n-1} (-1)^{i+1} \bfSigma^{j_1} \ldots \bfSigma^{j_{i-1}} \bfSigma^{j_{i+1}} \ldots \bfSigma^{j_{n}} \bfSigma^{j_{i}}, \\
(\bfSigma \cdot \bfp) \bfSigma^{j_1} \ldots \bfSigma^{j_{n}} (\bfSigma \cdot \bfp) &= (-1)^n \bfp^2 \bfSigma^{j_1} \ldots \bfSigma^{j_n} \nonumber \\
&  + 2 \sum_{i=1}^{n} (-1)^{i+1} \bfp^{j_i}\bfSigma^{j_1} \ldots \bfSigma^{j_{i-1}} \bfSigma^{j_{i+1}} \ldots \bfSigma^{j_{n}} (\bfSigma \cdot \bfp).
\end{align}
\end{subequations}
A trace of an even number of Pauli matrices in $4$- or $D$-dimensions can be evaluated via the following recursive relation
\begin{equation}
\Tr(\bfSigma^{i_1} \ldots \bfSigma^{i_{2n}}) = \sum_{j=2}^{2n} \delta^{i_1 i_j} (-1)^j \Tr(\bfSigma^{i_2} \ldots \bfSigma^{i_{j-1}} \bfSigma^{i_{j+1}} \ldots \bfSigma^{i_{2n}}).
\end{equation}
A trace of an odd number of Pauli matrices is not well defined in $D$-dimensions, but in 4-dimensions one can use
\begin{equation}
\Tr(\bfSigma^{i_1} \dots \bfSigma^{i_{2n}} \bfSigma^{i_{2n+1}}) = \delta^{i_1 i_2} \Tr(\bfSigma^{i_3} \ldots \bfSigma^{i_{2n}} \bfSigma^{i_{2n+1}}) + i \epsilon^{i_1 i_2 k}  \Tr(\bfSigma^{k} \bfSigma^{i_3} \ldots \bfSigma^{i_{2n}} \bfSigma^{i_{2n+1}}).
\end{equation}

\section{Tensors and matrices in \feyncalc}
\label{sec:app-fcifce}

For the sake of completeness, we list here the implementation of various 
tensors with Lorentz and Cartesian indices in the internal and external representations of \feyncalc. These are
useful for interpreting \feyncalc results and writing codes that rely on the package.

\FloatBarrier
\clearpage
\subsection{Lorentz and Cartesian tensors in the internal (\texttt{FCI}) notation}

{
\small
\renewcommand{\arraystretch}{1.3}
\begin{longtable}[b]{|l|c|}
\hline
Command in \feyncalc & Meaning \\
\hline
 \texttt{Pair[LorentzIndex[$\mu$],LorentzIndex[$\nu$]]} & $\bar{g}^{\mu \nu}$ \\
  \texttt{Pair[LorentzIndex[$\mu$,$D$],LorentzIndex[$\nu$],$D$]} & $g^{\mu \nu}$ \\
    \texttt{Pair[LorentzIndex[$\mu$,$D-4$],LorentzIndex[$\nu$],$D-4$]} & $\hat{g}^{\mu \nu}$ \\
 \texttt{Pair[Momentum[$p$],LorentzIndex[$\mu$]]} & $\bar{p}^\mu$ \\
 \texttt{Pair[Momentum[$p$,$D$],LorentzIndex[$\mu$,$D$]]} & $p^\mu$ \\
  \texttt{Pair[Momentum[$p$,$D-4$],LorentzIndex[$\mu$,$D-4$]]} & $\hat{p}^\mu$ \\
\texttt{Pair[Momentum[$p$],Momentum[$q$]]} & $\bar{p} \cdot \bar{q}$ \\
\texttt{Pair[Momentum[$p$,$D$],Momentum[$q$,$D$]]} & $p \cdot q$ \\  \texttt{Pair[Momentum[$p$,$D-4$],Momentum[$q$,$D-4$]]} & $\hat{p} \cdot \hat{q}$ \\
\hline
\caption{Lorentz structures in $4$, $D$ and $D-4$ dimensions, which can be represented using \texttt{Pair}.}
\label{tab:fo-overview-lorentz}
\end{longtable}
}

{
\small
\renewcommand{\arraystretch}{1.3}
\setlength{\tabcolsep}{0.4cm}
\begin{longtable}[b]{|p{8.3cm} p{1.4cm}|}
\hline
Command in \feyncalc & Meaning \\
\hline
 \texttt{Eps[}\texttt{LorentzIndex[$\mu$]}, \texttt{LorentzIndex[$\nu$]},  \hfill\phantom{,}
 \texttt{LorentzIndex[$\rho$]} \texttt{LorentzIndex[$\sigma$]}\texttt{]}  & $\bar{\epsilon}^{\mu \nu \rho \sigma}$ \\
\texttt{Eps[}\texttt{LorentzIndex[$\mu$,$D$]}, \texttt{LorentzIndex[$\nu$,$D$]}, \hfill\phantom{,}
 \texttt{LorentzIndex[$\rho$,$D$]} \texttt{LorentzIndex[$\sigma$,$D$]}\texttt{]}  & $\epsilon^{\mu \nu \rho \sigma}$ \\
 \texttt{Eps[}\texttt{LorentzIndex[$\mu$]}, \texttt{LorentzIndex[$\nu$]}, \hfill\phantom{,}
 \texttt{Momentum[$p$]} \texttt{Momentum[$q$]}\texttt{]}  & $\bar{\epsilon}^{\mu \nu \rho \sigma} \bar{p}_\rho \bar{q}_\sigma$ \\
 \texttt{Eps[}\texttt{LorentzIndex[$\mu$,$D$]}, \texttt{LorentzIndex[$\nu$,$D$]},  \hfill\phantom{,}
 \texttt{Momentum[$p$,$D$]} \texttt{Momentum[$q$,$D$]}\texttt{]}  & $\epsilon^{\mu \nu \rho \sigma} p_\rho q_\sigma$ \\
\hline
\caption{$4$- and $D$-dimensional Levi-Civita symbols in \feyncalc. For brevity, we do not list all possible combinations of
\texttt{LorentzIndex} and \texttt{Momentum} in the arguments of \texttt{Eps}.}
\label{tab:fo-overview-eps}
\end{longtable}
}

\clearpage
{
\small
\renewcommand{\arraystretch}{1.3}
\setlength{\tabcolsep}{0.5cm}
\begin{longtable}[b]{|p{11.5cm} p{1.2cm}|}
\hline
Command in \feyncalc & Meaning \\
\hline
 \texttt{Pair[}\texttt{LorentzIndex[$\mu$]}, \texttt{LorentzIndex[0]}\texttt{]} & $\bar{g}^{\mu 0}$ \\
 \texttt{Pair[}\texttt{LorentzIndex[$\mu$]},\texttt{CartesianIndex[$i$]}\texttt{]} & $\bar{g}^{\mu i}$ \\
  \texttt{Pair[}\texttt{LorentzIndex[$\mu$,D]},\texttt{CartesianIndex[$i$,D-1]}\texttt{]} & $g^{\mu i}$ \\
  \texttt{Pair[}\texttt{LorentzIndex[$\mu$,D-4]},\texttt{CartesianIndex[$i$,D-4]]} & $\hat{g}^{\mu i}$ \\

  \texttt{Pair[}\texttt{LorentzIndex[0]},\texttt{LorentzIndex[0]}\texttt{]} & $\bar{g}^{00}$ \\
  \texttt{Pair[}\texttt{CartesianIndex[$i$]},\texttt{CartesianIndex[$j$]}\texttt{]} & $\bar{g}^{i j}$ \\
  \texttt{Pair[}\texttt{CartesianIndex[$i$,D-1]},\texttt{CartesianIndex[$j$,D-1]}\texttt{]} & $g^{i j}$ \\
  \texttt{Pair[}\texttt{CartesianIndex[$i$,D-4]},\texttt{CartesianIndex[$j$,D-4]}\texttt{]} & $\hat{g}^{i j}$ \\

    \texttt{CartesianPair[}\texttt{CartesianIndex[$i$]},\texttt{CartesianIndex[$j$]}\texttt{]} & $\bar{\delta}^{i j}$ \\
  \texttt{CartesianPair[}\texttt{CartesianIndex[$i$,D-1]}, \texttt{CartesianIndex[$j$,D-1]}\texttt{]} & $\delta^{i j}$ \\
  \texttt{CartesianPair[}\texttt{CartesianIndex[$i$,D-4]}, \texttt{CartesianIndex[$j$,D-4]}\texttt{]} & $\hat{\delta}^{i j}$ \\

 \texttt{TemporalPair[}\texttt{TemporalMomentum[$p$]},\texttt{LorentzIndex[0]}\texttt{]} & $\bar{p}^0$ \\
 \texttt{Pair[}\texttt{CartesianMomentum[$p$]},\texttt{LorentzIndex[$\mu$]}\texttt{]} & $\bar{\bfp}^i \bar{g}^{i \mu}$ \\
  \texttt{Pair[}\texttt{CartesianMomentum[$p$,D-1]},\texttt{LorentzIndex[$\mu$,D]}\texttt{]} & $\bfp^i g^{i \mu}$ \\
    \texttt{Pair[}\texttt{CartesianMomentum[$p$,D-4]},\texttt{LorentzIndex[$\mu$,D-4]}\texttt{]} & $\hat{\bfp}^i \hat{g}^{i \mu}$ \\
     \texttt{CartesianPair[}\texttt{CartesianMomentum[$p$]}, \texttt{CartesianIndex[$i$]}\texttt{]} & $\bar{\bfp}^i$ \\
      \texttt{CartesianPair[}\texttt{CartesianMomentum[$p$,D-1]}, \texttt{CartesianIndex[$i$,D-1]}\texttt{]} & $\bfp^i$ \\
            \texttt{CartesianPair[}\texttt{CartesianMomentum[$p$,D-4]},\hfill\phantom{,}\texttt{CartesianIndex[$i$,D-4]}\texttt{]} & $\hat{\bfp}^i$ \\
   \texttt{CartesianPair[}\texttt{CartesianMomentum[$p$]}, \texttt{CartesianMomentum[$q$]}\texttt{]} & $\bar{\bfp} \cdot \bar{\bfq}$ \\
      \texttt{CartesianPair[}\texttt{CartesianMomentum[$p$,D-1]},\texttt{CartesianMomentum[$q$,D-1]}\texttt{]} & $\bfp \cdot \bfq$ \\
    \texttt{CartesianPair[}\texttt{CartesianMomentum[$p$,D-4]}, \texttt{CartesianMomentum[$q$,D-4]}\texttt{]} & $\hat{\bfp}\cdot \hat{\bfq}$\\
\hline
\caption{New tensors with Cartesian and mixed (Lorentz and Cartesian) indices using \texttt{Pair}, \texttt{CartesianPair} and \texttt{TemporalPair}.}
\label{tab:fo-overview-cartesian}
\end{longtable}
}
\clearpage

{
\small
\renewcommand{\arraystretch}{1.3}
\setlength{\tabcolsep}{0.3cm}
\begin{longtable}[b]{|p{10.0cm} p{1.2cm}|}
\hline
Command in \feyncalc & Meaning \\
\hline 
\parbox[t]{6.7cm}{
 \texttt{Eps[}\texttt{LorentzIndex[0]}, \texttt{LorentzIndex[$\nu$]}, \texttt{LorentzIndex[$\rho$]} \texttt{LorentzIndex[$\sigma$]}\texttt{]}} & $\bar{\epsilon}^{0 \nu \rho \sigma}$ \\

 \parbox[t]{6.2cm}{\texttt{Eps[}\texttt{LorentzIndex[0]}, \texttt{CartesianIndex[$i$]}, \texttt{LorentzIndex[$\mu$]}, \texttt{LorentzIndex[$\nu$]}\texttt{]} }  & $\bar{\epsilon}^{0 i \mu \nu}$ \\

\parbox[t]{6.2cm}{
 \texttt{Eps[}\texttt{CartesianIndex[$i$]}, \texttt{CartesianIndex[$j$]}, \texttt{LorentzIndex[$\mu$]}, \texttt{LorentzIndex[$\nu$]}\texttt{]}}  & $\bar{\epsilon}^{i j \mu \nu}$ \\

\parbox[t]{9.1cm}{
 \texttt{Eps[}\texttt{CartesianIndex[$i$,$D-1$]}, \texttt{CartesianIndex[$j$,$D-1$]}, \texttt{LorentzIndex[$\mu$,$D$]} \texttt{LorentzIndex[$\nu$,$D$]}\texttt{]}}  & ${\epsilon}^{i j \mu \nu}$ \\

\parbox[t]{6.5cm}{
 \texttt{Eps[}\texttt{LorentzIndex[$\mu$]}, \texttt{LorentzIndex[$\nu$]}, \texttt{LorentzIndex[$\rho$]}\texttt{]}}  & $\bar{\epsilon}^{\mu \nu \rho}$ \\

\parbox[t]{6.9cm}{
 \texttt{Eps[}\texttt{LorentzIndex[$\mu$,$D$]} \texttt{LorentzIndex[$\nu$,$D$]}, \texttt{LorentzIndex[$\rho$,$D$]}\texttt{]}}  & ${\epsilon}^{\mu \nu \rho}$ \\
 
\parbox[t]{6.7cm}{
 \texttt{Eps[}\texttt{CartesianIndex[$i$]}, \texttt{LorentzIndex[$\mu$]},
 \texttt{LorentzIndex[$\nu$]}\texttt{]}}  & $\bar{\epsilon}^{i \mu \nu}$ \\

\parbox[t]{8.7cm}{
 \texttt{Eps[}\texttt{CartesianIndex[$i$,$D-1$]}, \texttt{LorentzIndex[$\mu$,$D$]},
 \texttt{LorentzIndex[$\nu$,$D$]}\texttt{]}}  & ${\epsilon}^{i \mu \nu}$ \\

\parbox[t]{7.2cm}{
 \texttt{Eps[}\texttt{CartesianIndex[$i$]}, \texttt{CartesianIndex[$j$]},
 \texttt{CartesianIndex[$k$]}\texttt{]}}  & $\bar{\epsilon}^{i j k}$ \\

\parbox[t]{9.4cm}{
 \texttt{Eps[}\texttt{CartesianIndex[$i$,$D-1$]},  \texttt{CartesianIndex[$j$,$D-1$]},
 \texttt{CartesianIndex[$k$,$D-1$]}\texttt{]}}  & ${\epsilon}^{i j k}$ \\

\parbox[t]{8.1cm}{
 \texttt{Eps[}\texttt{CartesianIndex[$i$]},  \texttt{CartesianMomentum[$p$]},
 \texttt{CartesianMomentum[$q$]}\texttt{]}}  & $\bar{\epsilon}^{i j k} \bar{\bfp}^j \bar{\bfq}^k$  \\

\parbox[t]{9.8cm}{
 \texttt{Eps[}\texttt{CartesianIndex[$i$,$D-1$]}, \texttt{CartesianMomentum[$p$,$D-1$]},
 \texttt{CartesianMomentum[$q$,$D-1$]}\texttt{]}} & ${\epsilon}^{i j k} \bfp^j \bfq^k$ \\

\hline
\caption{Representation of Levi-Civita symbols with spatial and temporal indices in \feyncalc. For brevity, we do not list all possible combinations of \texttt{LorentzIndex}, \texttt{Momentum}, \texttt{CartesianIndex}, \texttt{CartesianMomentum} and \texttt{TemporalIndex} in the arguments of \texttt{Eps}.}
\label{tab:fo-overview-cartesian-eps}
\end{longtable}
}

{
\small
\renewcommand{\arraystretch}{1.3}
\begin{longtable}[b]{|l|c|}
\hline
Command in \feyncalc & Meaning \\
\hline
 \texttt{DiracGamma[LorentzIndex[$\mu$]]} & $\bar{\gamma}^\mu$ \\
 \texttt{DiracGamma[LorentzIndex[$\mu$,$D$],$D$]} & $\gamma^\mu$ \\
  \texttt{DiracGamma[LorentzIndex[$\mu$,$D-4$],$D-4$]} & $\hat{\gamma}^\mu$ \\
   \texttt{DiracGamma[Momentum[$p$]]} & $\bar{\gamma} \cdot \bar{p}$ \\
 \texttt{DiracGamma[Momentum[$p$,$D$],$D$]} & $\gamma \cdot p$ \\
  \texttt{DiracGamma[Momentum[$p$,$D-4$],$D-4$]} & $\hat{\gamma} \cdot \hat{p}$ \\
\hline
\caption{Representation of Dirac matrices in $4$-, $D$- and $D-4$ dimensions using \texttt{DiracGamma}.}
\label{tab:fo-overview-dirac}
\end{longtable}
}

{
\small
\renewcommand{\arraystretch}{1.3}
\begin{longtable}[b]{|l|c|}
\hline
Command in \feyncalc & Meaning \\
\hline
 \texttt{DiracGamma[LorentzIndex[0]]} & $\bar{\gamma}^0$ \\
 \texttt{DiracGamma[CartesianIndex[$i$]]} & $\bar{\bfGamma}^i$ \\
 \texttt{DiracGamma[CartesianIndex[$i$,$D-1$],$D$]} & $\bfGamma^i$ \\
  \texttt{DiracGamma[CartesianIndex[$i$,$D-4$],$D-4$]} & $\hat{\bfGamma}^i$ \\
   \texttt{DiracGamma[CartesianMomentum[$p$]]} & $\bar{\bfGamma} \cdot \bar{\bfp}$ \\
 \texttt{DiracGamma[CartesianMomentum[$p$,$D-1$],$D$]} & $\bfGamma \cdot \bfp$ \\
  \texttt{DiracGamma[CartesianMomentum[$p$,$D-4$],$D-4$]} & $\hat{\bfGamma} \cdot \hat{\bfp}$ \\
\hline
\caption{Representation of Dirac matrices with temporal or Cartesian indices in $4$-, $D$- and $D-4$ dimensions using \texttt{DiracGamma}.}
\label{tab:fo-overview-cartesian-dirac}
\end{longtable}
}

{
\small
\renewcommand{\arraystretch}{1.3}
\begin{longtable}[b]{|l|c|}
\hline
Command in \feyncalc & Meaning \\
\hline
 \texttt{PauliSigma[LorentzIndex[$\mu$]]} & $\bar{\sigma}^\mu$ \\
 \texttt{PauliSigma[LorentzIndex[$\mu$,$D$],$D-1$]} & $\sigma^\mu$ \\
  \texttt{PauliSigma[LorentzIndex[$\mu$,$D-4$],$D-4$]} & $\hat{\sigma}^\mu$ \\
   \texttt{PauliSigma[Momentum[$p$]]} & $\bar{\sigma} \cdot \bar{p}$ \\
 \texttt{PauliSigma[Momentum[$p$,$D$],$D-1$]} & $\sigma \cdot p$ \\
  \texttt{PauliSigma[Momentum[$\mu$,$D-4$],$D-4$]} & $\hat{\sigma} \cdot \hat{p}$ \\

 \texttt{PauliSigma[CartesianIndex[$i$]]} & $\bar{\bfSigma}^i$ \\
 \texttt{PauliSigma[CartesianIndex[$i$,$D-1$],$D-1$]} & $\bfSigma^i$ \\
  \texttt{PauliSigma[CartesianIndex[$i$,$D-4$],$D-4$]} & $\hat{\bfSigma}^i$ \\
   \texttt{PauliSigma[CartesianMomentum[$p$]]} & $\bar{\bfSigma} \cdot \bar{\bfp}$ \\
 \texttt{PauliSigma[CartesianMomentum[$p$,$D$],$D$]} & $\bfSigma \cdot \bfp$ \\
  \texttt{PauliSigma[CartesianMomentum[$\mu$,$D-4$],$D-4$]} & $\hat{\bfSigma} \cdot \hat{\bfp}$ \\

\hline
\caption{Representation of Pauli matrices in $4$-, $D$- and $D-4$ dimensions using \texttt{PauliSigma}.}
\label{tab:fo-overview-pauli}
\end{longtable}
}

\clearpage

\subsection{Lorentz and Cartesian tensors in the external (\texttt{FCE}) notation}

{
\small
\renewcommand{\arraystretch}{1.3}
\begin{longtable}[b]{|l|c|}
\hline
Shortcut in \feyncalc & Meaning \\
\hline
 \texttt{MT[$\mu$,$\nu$]},  \texttt{MTD[$\mu$,$\nu$]}  \texttt{MTE[$\mu$,$\nu$]} & $\bar{g}^{\mu \nu}$, $g^{\mu \nu}$, $\hat{g}^{\mu \nu}$ \\
 \texttt{FV[$p$,$\mu$]}, \texttt{FVD[$p$,$\mu$]}, \texttt{FVE[$p$,$\mu$]} & $\bar{p}^\mu$, $p^\mu$, $\hat{p}^\mu$ \\
\texttt{SP[$p$,$q$]}, \texttt{SPD[$p$,$q$]}, \texttt{SPE[$p$,$q$]} & $\bar{p} \cdot \bar{q}$, $p \cdot q$, $\hat{p} \cdot \hat{q}$ \\
\texttt{GA[$\mu$]}, \texttt{GAD[$\mu$]}, \texttt{GAE[$\mu$]} & $\bar{\gamma}^\mu$, ${\gamma}^\mu$, $\hat{\gamma}^\mu$ \\
\texttt{GS[$p$]}, \texttt{GSD[$p$]}, \texttt{GSE[$p$]} & $\bar{\gamma} \cdot \bar{p}$, $\gamma \cdot p$, $\hat{\gamma} \cdot \hat{p}$ \\
\texttt{LC[$\mu$,$\nu$,$\rho$,$\sigma$]}, \texttt{LC[$\mu$,$\nu$][$p$,$q$]} & $\bar{\epsilon}^{\mu \nu \rho \sigma}$, $\bar{\epsilon}^{\mu \nu \rho \sigma} p_\rho  q_\sigma$ \\
\texttt{LCD[$\mu$,$\nu$,$\rho$,$\sigma$]}, \texttt{LCD[$\mu$,$\nu$][$p$,$q$]}  & ${\epsilon}^{\mu \nu \rho \sigma}$, ${\epsilon}^{\mu \nu \rho \sigma} \hat{p}_\rho  \hat{q}_\sigma$ \\
\hline
\caption{Some of the existing \feyncalc shortcuts.}
\label{tab:fo-overview-shortcuts-old}
\end{longtable}
}

{
\small
\renewcommand{\arraystretch}{1.3}
\begin{longtable}[b]{|l|c|}
\hline
Shortcut in \feyncalc & Meaning \\
\hline
 \texttt{KD[$i$,$j$]},  \texttt{KDD[$i$,$j$]},  \texttt{KDE[$i$,$j$]} & $\bar{\delta}^{ij}$, $\delta^{ij}$, $\hat{\delta}^{ij}$  \\
 \texttt{CV[$p$,$i$]},  \texttt{CVD[$p$,$i$]},  \texttt{CVE[$p$,$i$]} & $\bar{\bfp}^i$, $\bfp^i$, $\hat{\bfp}^i$ \\
\texttt{CSP[$p$,$q$]}, \texttt{CSPD[$p$,$q$]}, \texttt{CSPE[$p$,$q$]} & $\bar{\bfp} \cdot \bar{\bfq}$, $\bfp \cdot \bfq$, $\hat{\bfp} \cdot \hat{\bfq}$  \\
\texttt{TGA[]} & $\bar{\gamma}^0$ \\
\texttt{CGA[$i$]}, \texttt{CGAD[$i$]}, \texttt{CGAE[$i$]} & $\bar{\bfGamma}^i$, ${\bfGamma}^i$, $\hat{\bfGamma}^i$  \\
\texttt{CGS[$p$]}, \texttt{CGSD[$p$]}, \texttt{CGSE[$p$]} & $\bar{\bfGamma} \cdot \bar{p}$, $\bfGamma \cdot \bfp$, $\hat{\bfGamma} \cdot \hat{\bfp}$ \\
\texttt{CLC[$i$,$j$,$k$]}, \texttt{CLC[$i$,$j$][$p$]} & $\bar{\epsilon}^{ijk}$, $\bar{\epsilon}^{ijk} \bar{\bfp}^k$ \\
\texttt{CLCD[$i$,$j$,$k$]}, \texttt{CLCD[$i$,$j$][$p$]} & ${\epsilon}^{ijk}$, ${\epsilon}^{ijk} {\bfp}^k$ \\
\texttt{SI[$\mu$]}, \texttt{SID[$\mu$]}, \texttt{SIE[$\mu$]} & $\bar{\sigma}^\mu$,  ${\sigma}^\mu$, $\hat{\sigma}^\mu$ \\
\texttt{SIS[$p$]}, \texttt{SISD[$p$]}, \texttt{SISE[$p$]} & $\bar{\sigma} \cdot \bar{p}$, $\sigma \cdot p$, $\hat{\sigma} \cdot \hat{p}$  \\
\texttt{CSI[$i$]}, \texttt{CSID[$i$]}, \texttt{CSIE[$i$]} & $\bar{\bfSigma}^i$, ${\bfSigma}^i$, $\hat{\bfSigma}^i$ \\
\texttt{CSIS[$p$]}, \texttt{CSISD[$p$]}, \texttt{CSISE[$p$]}  & $\bar{\bfSigma} \cdot \bar{\bfp}$, $\bfSigma \cdot \bfp$, $\hat{\bfSigma} \cdot \hat{\bfp}$ \\
\hline
\caption{New \feyncalc shortcuts for nonrelativistic calculations.}
\label{tab:fo-overview-shortcuts-new}
\end{longtable}
}

\section{Derivation of Feynman rules in NREFTs}
\label{sec:app-frules}

The extraction of Feynman rules from a given Lagrangian is an important ingredient
of every particle physics calculation. Since every EFT Lagrangian formally contains an infinite number of operators, it is clearly not possible to derive those rules once and for all, as it can be done \eg for QED or QCD. Instead, every EFT calculation that aims for higher precision in the expansion parameter(s) must take into account new operators that show up at that accuracy.

On the one hand, the procedure of deriving Feynman rules for new operators can be automatized using dedicated software packages such as \textsc{LanHEP} \cite{Semenov:1996es,Semenov:1997qm}, \textsc{SARAH} \cite{Staub:2009bi,Staub:2010jh,Staub:2012pb,Staub:2013tta} or \textsc{FeynRules}. On the other hand, those tools mainly focus on Lorentz covariant theories and are therefore less useful for nonrelativistic calculations. For example, operators containing Pauli spinors, Cartesian tensors, spatial and temporal derivatives or (chromo)electric and (chromo)magnetic fields are not supported out-of-the-box. For this reason Feynman rules for NREFTs are still often derived by hand. Although the technicalities behind this procedure are certainly very familiar to the practitioners, they are rarely discussed at length in the literature. In the following we would like to treat this subject in a more detailed and pedagogical way, including explicit examples and useful recipes for practical calculations. 

Path integral formalism and canonical field quantization are the two most popular methods for deriving Feynman rules. We choose to employ the latter procedure, owing to its conceptual simplicity and the fact that it is straightforward to automatize in almost any symbolic manipulation framework. A very concise description of the method can be found in the
\textsc{FeynRules} manual \cite{Christensen:2008py}, which we will follow here.

However, when dealing with NREFTs we also need to account for fields that directly annihilate the vacuum \ie satisfy $\phi \ket{0} = 0$. This happens if a field that contains both particle and antiparticle components is transformed in such a way, that both components decouple from each other and are then treated as separate fields that create/annihilate a single particle/antiparticle. Therefore, we will slightly modify the rules from \cite{Christensen:2008py}, making them applicable to relativistic and nonrelativistic theories alike. The main recipe for deriving the Feynman rule associated to a given operator $O$ can be then summarized in the following 3 steps:

\begin{enumerate}
\item For fields that do not annihilate the vacuum \ie $\phi \ket{0} \neq 0$ and $\bra{0} \phi  \neq 0$, multiply $O$ by creation operators for the fields from the right. For fields that annihilate the vacuum, multiply $O$ by creation operators for $\phi$ from the right and by annihilation operators for $\phi^\dagger$ from the left. This ensures that the matrix element under consideration does not vanish.  The fermion creators and annihilators should be ordered in a reversed way as compared to the ordering of the corresponding fermion fields in the operator. 
\item Put the resulting expression between the vacuum states $\bra{0}$ and $\ket{0}$. 
Move the creation (annihilation) operators to the left (right) where they annihilate $\bra{0}$ ($\ket{0}$).
\item Replace $\braket{0|0}$ by unity, remove the overall exponential and the external states (\eg spinors and polarization vectors) and multiply the rest by $i$. Finally, reverse the sign of each momentum that stems from an annihilation operator that was multiplying $O$ from the left, so that all momenta are incoming.
\end{enumerate}
Conceptually, this technique is very similar to the calculation of matrix elements in the so-called ``old fashioned perturbation theory'' that was the main way of doing QFT calculations even before the invention of Feynman diagrams. Regarding the treatment of operators with field derivatives, the \textsc{FeynRules} approach is to pull the derivatives outside of the matrix element and apply them to the exponentials after the second step. Alternatively, one can also work out the (anti)commutation relations for fields with derivatives in advance and use them already \emph{during} the second step. From our experience, the latter is often more convenient in pen and paper calculations, while what is done in \textsc{FeynRules} is naturally more useful for automatic codes.

The main ingredient of the provided recipe is the process of moving creation and annihilation operators past field operators, where we need to apply suitable (anti)commutation relations. Once we have quantized the free part of our EFT in the operator formalism, these relations are obtained straightforwardly.

\subsection{Feynman rules for NRQCD}
\label{sec:app-nrqcd}

For definiteness, let us start with NRQCD. The free part of the NRQCD Lagrangian at $\mathcal{O}(1/m)$ reads

\begin{equation}
\mathcal{L}_{\textrm{free}} =  \psi^\dagger \left(  i \partial^0 + \frac{\nabla^2}{2m}  \right ) \psi +
\chi^\dagger \left(  i \partial^0 - \frac{\nabla^2}{2m}  \right ) \chi - \frac{1}{4} \widehat{G}^a_{\mu \nu} \widehat{G}^{\mu \nu a},
\end{equation}
where $\psi$ ($\chi$) is a Pauli field that annihilates (creates) a heavy quark (antiquark), while $\widehat{G}^a_{\mu \nu} 
= \partial_\mu A_\nu^a - \partial_\nu A_\mu^a$ is the noninteracting part of the field strength tensor $G^a_{\mu \nu} $, with $A^a_\mu$ being the gluon field. Notice that $\psi \ket{0} = 0$ and $\bra{0} \chi = 0$. The Fourier decompositions of the free NRQCD fields are
\begin{subequations} \label{eq:fourier-decs}
\begin{align}
\psi^n_i (x) &= \int \frac{d^3 p}{(2\pi)^3} \sum_{s=1}^{2} \sum_{c=1}^3 a(\bfp,s,c) \, \xi^n_i (s,c) \, e^{- i p \cdot x}, \label{eq:psifd} \\
\chi^n_i (x) &= \int \frac{d^3 p}{(2\pi)^3} \sum_{s=1}^{2} \sum_{c=1}^3  b^{\dagger} (\bfp,s,c) \, \eta^n_i (s,c) \,  e^{ i p \cdot x}, \label{eq:chifd} \\
A^{a \mu} (x) & = \int \frac{d^3 p}{(2\pi)^3} \sum_{\lambda=1}^{2} \sum_{d=1}^8 \left ( g(\bfp,\lambda,d) \epsilon^{a \mu}(p,\lambda,d)  e^{-ip \cdot x}  + g^{\dagger}(\bfp,\lambda,d) \epsilon^{\ast a \mu}(p,\lambda,d)  e^{ip \cdot x}  \right ) \label{eq:gfd},
\end{align}
\end{subequations}
where $p_0 = \bfp^2/(2 m)$ in \Eqs\eqref{eq:psifd} and \eqref{eq:psifd}, and $p^0 = |\bfp|$ in \Eq\eqref{eq:gfd}. Moreover, $s$, $\lambda$, $c$ and $d$ denote the spin, polarization and color quantum numbers respectively. 
A quark field carries one spinor index $i$ and one fundamental color index $n$, whereas a gluon field has an adjoint color index $a$ and a Lorentz index $\mu $ attached to it. The quantities $\xi^n_i (s,c)$ and $\eta^n_i (s,c)$ should be understood as a product of a 2-spinor and a 3-color vector, \ie
\begin{align}
\xi^n_i (s,c) \equiv  \xi_i (s) v^n_3(c), \quad \eta^n_i (s,c) \equiv  \eta_i (s) v_3^n(c),
\end{align}
with a possible choice for the spinors ($\xi,\eta$) and 3-color vectors $v_3$ being
\begin{equation}
\xi (1) = \eta (1) = (1,0)^T, \quad \xi (2) = \eta (2) = (0,1)^T,
\end{equation}
and
\begin{equation}
v_3 (1) = (1,0, 0)^T, \quad v_3 (2) = (0,1, 0)^T, \quad v_3 (3) = (0,0,1)^T.
\end{equation}
Likewise, for the gluon field we have 
\begin{align}
\varepsilon^{a \mu}(p,\lambda,d) \equiv \varepsilon^{\mu}(p,\lambda) v^a_8(d),
\end{align}
where $\varepsilon^{\mu}(p,\lambda)$ is an ordinary polarization vector, while $v^a_8(d)$ describes an 8-color vector with \eg $v_8(1) = (1,0,0,0,0,0,0,0)^T$. In order not to clutter the notation further, here we chose to suppress the flavor indices of the quark fields. Those can be made explicit exactly in the same manner as the color indices, \ie by introducing corresponding 6-dimensional vectors.

Furthermore, in order to avoid dealing with unphysical degrees of freedom of massless vector bosons, we let the gluon field possess only transverse polarizations (radiation gauge). This is not an issue here, since we employ the operator formalism only as a convenient shortcut to derive Feynman rules. Repeating the same exercise using the BRST construction \cite{Becchi:1974md,Tyutin:1975qk} would only complicate the derivation but yield exactly the same Feynman rules, so that we do not consider it useful here. We also omit the treatment of the gluon-ghost interactions, since the corresponding Feynman rules are exactly the same as in ordinary QCD.

The creation and annihilation operators of the NRQCD fields satisfy following nonvanishing (anti)commutation relations
\begin{subequations}
\begin{align}
\{ a(\bfp,s,c), a^\dagger (\bfp',s',c') \} &= (2\pi)^3 \delta^{(3)} (\bfp - \bfp') \delta_{ss'} \delta_{cc'}, \\
\{ b(\bfp,s,c), b^\dagger (\bfp',s',c') \} &= (2\pi)^3 \delta^{(3)} (\bfp - \bfp') \delta_{ss'} \delta_{cc'}, \\
[ g(\bfp,\lambda,d), g^\dagger (\bfp',\lambda',d') ] &=  (2\pi)^3 \delta^{(3)} (\bfp - \bfp') \delta_{\lambda \lambda'} \delta_{dd'}.
\end{align}
\end{subequations}
As it is customary in NRQCD, we define the 1-particle Fock states to have nonrelativistic normalization, so that
\begin{subequations}
\begin{align}
\ket{Q(p,s,c)} &= a^\dagger (\bfp,s,c)  \ket{0}, \\
\ket{\bar{Q}(p,s,c)} &=  b^{\dagger} (\bfp,s,c) \ket{0}, \\
\ket{g(p,\lambda,d)} &= g^\dagger (\bfp,\lambda,d) \ket{0}.
\end{align}
\end{subequations}
The nonvanishing (anti)commutators between fields and creation operators are given by
\begin{subequations}
\begin{align}
\{ \psi^n_i(x), a^\dagger (\bfp,s,c) \} &= \xi^n_i (s,c) e^{- i p \cdot x}, \\
\{ \chi^{\dagger n}_i(x), b^\dagger (\bfp,s,c) \} &= \eta^{\dagger n}_i (s,c) e^{- i p \cdot x}, \\
[ A^{a \mu}(x), g^\dagger (\bfp,\lambda,d) ] &= \varepsilon^{a \mu}(p,\lambda,d)  e^{- i p \cdot x}.
\end{align}
\label{eq:nrqcd-field-op-ac}
\end{subequations}
By differentiating the exponentials in \Eqs\eqref{eq:nrqcd-field-op-ac}  one can also obtain the
corresponding relations with temporal or spatial derivatives applied to the field operators \eg
\begin{subequations}
\begin{align}
\{ \bfDel^i \psi(x), a^\dagger (\bfp,s,c) \} &= i \bfp^i \xi (s,c)  e^{- i p \cdot x}, \\
\{ \bfDel^i \bfDel^j \psi(x), a^\dagger (\bfp,s,c) \} &= -  \bfp^i \bfp^j \xi (s,c) e^{- i p \cdot x}, \\
[ \partial_0 \bfA^{a i}(x), g^\dagger (\bfp,\lambda,d) ] &= -i p^0 \bfeps^{a i}(p,\lambda,d)  e^{- i p \cdot x},
\end{align}
with $\bfDel^i  = \boldsymbol{\partial}_i =  \partial/ \partial \bfx^i$.
\end{subequations}
It is also easy to derive auxiliary formulas containing products of fields, which may be convenient for pen and paper calculations \eg
\begin{equation}
[ A^{a \mu}(x) A^{b \nu}(x) , g^\dagger (\bfp,\lambda,d) ] = \left ( A^{a \mu}(x) \varepsilon^{b \nu}(p,\lambda,d) + A^{b \nu}(x) \varepsilon^{a \mu}(p,\lambda,d) \right )  e^{- i p \cdot x}.
\end{equation}
Introducing chromoelectric and chromomagnetic fields as
\begin{subequations}
\begin{align}
\bfE^{ai}  &= G^{ai0} = - \bfDel^i A^{a0} - \partial^0 \bfA^{ai} + g f^{abc} \bfA^{bi} A^{c0}, \\
\bfB^{ai} & = \frac{1}{2} \epsilon^{ijk} G^{akj} = \epsilon^{ijk} \bfPartial_j \bfA^{ak} = (\bfDel \times \bfA^a)^i,
\end{align}
\end{subequations}
we find
\begin{subequations}
\begin{align}
[ \bfE^{a i}(x), g^\dagger (\bfp,\lambda,d) ] &= i( -  \bfp^i \varepsilon^{a 0}(p,\lambda,d)
+  p^0 \bfeps^{a i}(p,\lambda,d)) e^{-i p \cdot x} \nonumber \\
&+ g f^{abc} \left ( \bfA^{b i}(x) \varepsilon^{c 0}(p,\lambda,d) + A^{c 0}(x) \bfeps^{b i}(p,\lambda,d) \right )  e^{- i p \cdot x},  \\
[ \bfB^{a i}(x), g^\dagger (\bfp,\lambda,d) ] &= i \epsilon^{ijk} \bfp^j \bfeps^{a k}(p,\lambda,d)  e^{- i p \cdot x}.
\end{align}
\end{subequations}
Having written down all required relations between operators and states, we are now in the position to work out some explicit examples. Our intention is not to provide a complete list of NRQCD Feynman rules up to some order in the $1/m$ expansion (\cf \cite{Pineda:2011dg} for a comprehensive summary), but rather to demonstrate how to obtain such rules in an algorithmic-like fashion for arbitrary operators. 

Let us concentrate on the 2-fermion sector of the theory and consider the operator $\psi^\dagger \bfD^2/(2 m) \psi$. Observe that
\begin{equation}
\bfD^2 \psi = \nabla^2 \psi - i g (\nabla \cdot \bfA ) \psi - 2  i g \bfA \cdot (\nabla \psi) - g^2 \bfA^2 \psi,
\end{equation}
where the second and third terms on the r.h.s give rise to an interaction of two heavy quarks and one gluon, while the last term generates a seagull vertex with two quarks and two gluons. Applying our prescriptions we obtain
\begin{subequations}
\begin{align}
& - \frac{i g}{2m} T^a_{n_1 n_2} \delta_{s_1 s_2} \braket{0| a(\bfp',s',c') \psi^{\dagger n_1}_{s_1} (\nabla \cdot \bfA^a) \psi^{n_2}_{s_2} \, a^\dagger(\bfp,s,c)   g^\dagger (\bfk,\lambda,d) |0} \nonumber \\
& = - \frac{i g}{2m} T^a_{n_1 n_2} \delta_{s_1 s_2}  \{a(\bfp',s',c'), \psi^{\dagger n_1}_{s_1} \}  [ (\nabla \cdot \bfA^a),   g^\dagger (\bfk,\lambda,d) ] \{ \psi^{n_2}_{s_2} , a^\dagger(\bfp,s,c) \}  \nonumber \\
& = - \frac{i g}{2m} T^a_{n_1 n_2} \delta_{s_1 s_2} \xi^{\dagger n_1}_{s_1}(s',c') \xi^{n_2}_{s_2}(s,c) i \bfk \cdot \bfeps^a(k,\lambda,d) e^{-i (p+k-p') \cdot x} \nonumber \\
& \to  \frac{i g}{2m} T^a_{n_1 n_2} \delta_{s_1 s_2} \bfk^i,
\end{align}

\begin{align}
& - \frac{i g}{m} T^a_{n_1 n_2} \delta_{s_1 s_2} \braket{0| a(\bfp',s',c') \psi^{\dagger n_1}_{s_1} \bfA^{ai} (\nabla^i \psi^{n_2}_{s_2}) \, a^\dagger(\bfp,s,c)   g^\dagger (\bfk,\lambda,d) |0} \nonumber \\
& = - \frac{i g}{m} T^a_{n_1 n_2} \delta_{s_1 s_2}  \{a(\bfp',s',c'), \psi^{\dagger n_1}_{s_1} \}  [ \bfA^{ai},   g^\dagger (\bfk,\lambda,d) ] \{ (\nabla^i \psi^{n_2}_{s_2}) , a^\dagger(\bfp,s,c) \}  \nonumber \\
& = - \frac{i g}{m} T^a_{n_1 n_2} \delta_{s_1 s_2} \xi^{\dagger n_1}_{s_1}(s',c') \xi^{n_2}_{s_2}(s,c) i \bfp \cdot \bfeps^a(k,\lambda,d) e^{-i (p+k-p') \cdot x} \nonumber \\
& \to  \frac{i g}{m} T^a_{n_1 n_2} \delta_{s_1 s_2} \bfp^i,
\end{align}

\begin{align}
& - \frac{ g^2}{2m} T^a_{n_1 n_2} T^b_{n_2 n_3} \delta_{s_1 s_2} \braket{0| a(\bfp',s',c') \psi^{\dagger n_1}_{s_1} \bfA^{ai} \bfA^{bi} \psi^{n_3}_{s_2} \, a^\dagger(\bfp,s,c)   g^\dagger (\bfk_1,\lambda_1,d_1) g^\dagger (\bfk_2,\lambda_2,d_2) |0} \nonumber \\
& = - \frac{ g^2}{2m} T^a_{n_1 n_2} T^b_{n_2 n_3} \delta_{s_1 s_2}  \{a(\bfp',s',c'), \psi^{\dagger n_1}_{s_1} \} \bigl ( [\bfA^{ai}, g^\dagger (\bfk_1,\lambda_1,d_1)] [ \bfA^{bi}, g^\dagger (\bfk_2,\lambda_2,d_2)] \nonumber \\
& + [\bfA^{ai}, g^\dagger (\bfk_2,\lambda_2,d_2)] [ \bfA^{bi}, g^\dagger (\bfk_1,\lambda_1,d_1)]  \bigr ) \{  \psi^{n_3}_{s_2} , a^\dagger(\bfp,s,c) \}  \nonumber \\
& = - \frac{ g^2}{2m} T^a_{n_1 n_2} T^b_{n_2 n_3} \delta_{s_1 s_2} \xi^{\dagger n_1}_{s_1}(s',c') \xi^{n_3}_{s_2}(s,c) \nonumber \\
& \times  \left ( \bfeps^{ai}(k_1,\lambda_1,d_1) \bfeps^{bi}(k_2,\lambda_2,d_2) + \bfeps^{bi}(k_1,\lambda_1,d_1) \bfeps^{ai}(k_2,\lambda_2,d_2) \right ) e^{-i (p+k_1+k_2) \cdot x} \nonumber \\
& \to - \frac{i g^2}{2m} \left ( T^a_{n_1 n_2} T^b_{n_2 n_3} + T^b_{n_1 n_2} T^a_{n_2 n_3}   \right ) \delta_{s_1 s_2} \delta^{ij}. 
\end{align}
\end{subequations}
In the case of the operator $-\chi^\dagger \bfD^2/(2 m) \chi$ we find
\begin{align} 
&  \frac{i g}{2m} T^a_{n_1 n_2} \delta_{s_1 s_2} \braket{0| b(\bfp',s',c') \chi^{\dagger n_1}_{s_1} (\nabla \cdot \bfA^a) \chi^{n_2}_{s_2} \, b^\dagger(\bfp,s,c)   g^\dagger (\bfk,\lambda,d) |0} \nonumber \\
& =  \frac{i g}{2m} T^a_{n_1 n_2} \delta_{s_1 s_2} [(\nabla \cdot \bfA^a),g^\dagger (\bfk,\lambda,d)]   \braket{0| \chi^{\dagger n_1}_{s_1} b(\bfp',s',c') b^\dagger(\bfp,s,c) \chi^{n_2}_{s_2} |0} \nonumber \\
& = - \frac{i g}{2m} T^a_{n_1 n_2} \delta_{s_1 s_2} [(\nabla \cdot \bfA^a),g^\dagger (\bfk,\lambda,d)]   \braket{0| \{\chi^{\dagger n_1}_{s_1}, b^\dagger(\bfp,s,c)\} \{ b(\bfp',s',c'), \chi^{n_2}_{s_2}\} |0} \nonumber \\
& = - \frac{i g}{2m} T^a_{n_1 n_2} \delta_{s_1 s_2} \eta^{\dagger n_1}_{s_1}(s,c) \eta^{n_2}_{s_2}(s',c') i \bfk \cdot \bfeps^a(k,\lambda,d) e^{-i (p+k-p') \cdot x} \nonumber \\
& \to  \frac{i g}{2m} T^a_{n_1 n_2} \delta_{s_1 s_2} \bfk^i.
\end{align}
Another example in the 2-fermion sector is the operator $\psi^\dagger( g(\bfSigma \cdot \bfB)/ 2m) \psi$ with
\begin{align}
& \frac{ g}{2m} T^a_{n_1 n_2} \bfSigma^i_{s_1 s_2} \braket{0| a(\bfp',s',c') \psi^{\dagger n_1}_{s_1} \bfB^{ai} \psi^{n_2}_{s_2} \, a^\dagger(\bfp,s,c)   g^\dagger (\bfk,\lambda,d) |0} \nonumber \\
& = \frac{ g}{2m} T^a_{n_1 n_2} \bfSigma^i_{s_1 s_2}  \{a(\bfp',s',c'), \psi^{\dagger n_1}_{s_1} \} [\bfB^{ai},g^\dagger (\bfk,\lambda,d)] \{\psi^{n_2}_{s_2} , a^\dagger(\bfp,s,c) \}  \nonumber \\
& = \frac{ g}{2m} T^a_{n_1 n_2} \bfSigma^i_{s_1 s_2} \xi^{\dagger n_1}_{s_1}(s',c') \xi^{n_2}_{s_2}(s,c) \, i \epsilon^{ijk} \bfk^j \epsilon^{a k}(k,\lambda,d) e^{- i (p+k-p')\cdot x} \nonumber \\
& \to -\frac{ g}{2m} T^a_{n_1 n_2} \epsilon^{ijk} \bfSigma^i_{s_1 s_2}  \bfk^j.
\end{align}
Feynman rules in the 4-fermion sector can be derived in exactly the same fashion. One can also automatize the derivation using one's favorite symbolic manipulation system. This is particularly useful when dealing with higher dimensional operators that contain products of $\bfD$, $\bfE$, $\bfB$ and $\bfSigma$ contracted with each other in different ways.

\subsection{Feynman rules for pNRQCD}
\label{sec:app-pnrqcd}

Finally, we analyze pNRQCD in the same manner as we did it for NRQCD. The main reason for doing so is to highlight some interesting aspects of the theory that make pNRQCD conceptually similar to ordinary nonrelativistic quantum mechanics. 

In weakly coupled pNRQCD our degrees of freedom are bilocal color singlet and color octet fields as well as multipole expanded gluons. As it has already been explained in section \ref{sec:pnrqcd}, the bilocal fields depend both on $\bfr$ and $\bfR$, while gluons are sensitive only to $\bfR$. The free part of the pNRQCD Lagrangian at $\mathcal{O}(1/m)$ and $\mathcal{O}(\bfr^0)$ in the multipole expansion is given by
\begin{equation}
\mathcal{L}_\textrm{pNRQCD} = \Tr \left \{ \textrm{S}^\dagger \left (i \partial_0 - \frac{\bfDel^2}{m} \right ) \textrm{S} \right \} + 
\Tr \left \{ \textrm{O}^\dagger \left  (i \partial_0 - \frac{\bfDel^2}{m} \right ) \textrm{O} \right \} - \frac{1}{4} \hat{G}^a_{\mu \nu} \hat{G}^{\mu \nu a},
\end{equation}
with $\textrm{S}_{ij}$ and $\textrm{O}_{ij}$ defined as in \Eq\eqref{eq:sofields}. From these definitions it follows, in particular, that
\begin{subequations}
\begin{align}
 \textrm{S}^\dagger_{ij} i \partial_0 \textrm{S}_{ji}  & = \frac{\delta_{ij}^2}{N_c} S^\dagger i \partial_0  S =  S^\dagger i \partial_0 S, \\
 \textrm{O}^\dagger_{ij} i \partial_0  \textrm{O}_{ji}   & =  \frac{\Tr(T^a T^b)}{T_F} O^{\dagger a} i \partial_0  O^b =  O^{\dagger a} i \partial_0 O^a,
\end{align}
\end{subequations}
where $i,j$ and $a$ are fundamental and adjoint color indices respectively.

By Fourier expanding free singlet and octet fields in terms of their creation and annihilation operators we find
\begin{subequations}
\begin{align}
S(\bfr, \bfR, t) &= \int \frac{d^3 p}{(2 \pi)^3} \frac{d^3 P}{(2 \pi)^3} a_s(\bfp, \bfP) e^{ - i P \cdot R + i \bfp \cdot \bfr},  \\
 O^a (\bfr, \bfR, t) &= \int \frac{d^3 p}{(2 \pi)^3} \frac{d^3 P}{(2 \pi)^3} \sum_{c,c'} a_o(\bfp, \bfP, c, c') V^a(c,c') e^{ - i P \cdot R + i \bfp \cdot \bfr},
\end{align}
\label{eq:sofourier}
\end{subequations}
where the color structure $V^a(c,c')$ is defined as
\begin{equation}
V^a(c,c') \equiv v_{3}^i(c) \frac{ T^a_{ij}}{\sqrt{T_F}} v_{3}^j(c'),
\end{equation}
with $v_3(c)$ being the color vectors that have already been introduced in appendix \ref{sec:app-nrqcd}. As far as the kinematics is concerned, we have $P \equiv (P_0,\bfP)^T$ and $R \equiv (t,\bfR)^T$ with $P_0 = \bfp^2/m$ in \Eqs\eqref{eq:sofourier}

The nonvanishing commutation relations for the creation and annihilation operators of the bilocal fields read
\begin{subequations}
\begin{align}
[a_s(\bfp, \bfP), a^\dagger_s(\bfp', \bfP')] & = (2\pi)^6 \delta^{(3)} (\bfp - \bfp') \delta^{(3)} (\bfP - \bfP'), \\
[a_o(\bfp, \bfP, c_1, c_2), a^\dagger_o(\bfp', \bfP', c_1', c_2')] & = (2\pi)^6 \delta_{c_1 c_1'} \delta_{c_2 c_2'}  \delta^{(3)} (\bfp - \bfp') \delta^{(3)} (\bfP - \bfP').
\end{align}
\end{subequations}
As in the case of NRQCD, we can define 1-particle Fock states normalized nonrelativistically, that is
\begin{subequations}
\begin{align}
\ket{S(\bfp,\bfP)} & = a_s^\dagger (\bfp, \bfP) \ket{0}, \\
\ket{O(\bfp,\bfP,c,c')} & = a_o^\dagger (\bfp, \bfP ,c, c') \ket{0},
\end{align}
\end{subequations}
and
\begin{subequations}
\begin{align}
\braket{S(\bfp,\bfP)|S(\bfp',\bfP')} &= (2 \pi)^6 \delta^{(3)} (\bfp - \bfp') \delta^{(3)} (\bfP - \bfP'), \\
\braket{O(\bfp,\bfP, c_1, c_2)|O'(\bfp',\bfP',c_1',c_2')} &= (2 \pi)^6  \delta_{c_1 c_1'} \delta_{c_2 c_2'}  \delta^{(3)} (\bfp - \bfp') \delta^{(3)} (\bfP - \bfP').
\end{align}
\end{subequations}
The commutators of fields and operators,  our main ingredient for obtaining Feynman rules in the operator approach, turn out to be very simple
\begin{subequations}
\begin{align}
[S(\bfr,\bfR),a^\dagger_s(\bfp, \bfP)] &= e^{ - i P \cdot R + i \bfp \cdot \bfr}, \\
[O^a(\bfr,\bfR),a^\dagger_o(\bfp, \bfP, c, c')] &=  V^a(c,c') e^{ - i P \cdot R + i \bfp \cdot \bfr}.
\end{align}
\end{subequations}
Regarding the gluon fields, the corresponding formulas given in appendix \ref{sec:app-nrqcd} still apply, which is why we do not repeat them here. The only difference is that the exponential $e^{i p \cdot x}$ should be replaced with $e^{i P \cdot R}$.

The derivation of the pNRQCD Feynman rules is now straightforward. For example, we can work out the Feynman rule for the singlet-octet chromoelectric dipole interaction with one gluon emission at $\mathcal{O}(1/m^0)$ and at $\mathcal{O}(r)$. From the term $g V_A (r) \Tr (\textrm{O}^\dagger \bfr \cdot \bfE \, \textrm{S})$ in the Lagrangian we obtain
\begin{align}
& g V_A (r) \frac{ \Tr(T^a T^b)}{\sqrt{N_c T_F}}
\braket{0| a_o (\bfp',\bfP',c_1',c_2') O^{\dagger a} \bfr^i  \bfE^{bi} \, S 
a_s^\dagger (\bfp,\bfP) g^\dagger (\bfK, \lambda, d)|0} \nonumber \\
& = g V_A (r) \sqrt{\frac{T_F}{N_c}} \delta^{ab} \bfr^i
 [ a_o (\bfp',\bfP',c_1',c_2'), O^{\dagger a}]  [ \bfE^{bi},  
g^\dagger (\bfK, \lambda, d)] [S ,
a_s^\dagger (\bfp,\bfP)] \nonumber \\
& \overset{\mathcal{O}(g)}{=} i g V_A (r) \sqrt{\frac{T_F}{N_c}}  \delta^{ab} \bfr^i V^{\dagger a}_{c_1',c_2'} ( -  \bfK^i \varepsilon^{a 0}(K,\lambda,d) +  K^0 \bfeps^{a i}(K,\lambda,d)) e^{ -i (P + K - P') \cdot R + i (\bfp - \bfp') \cdot \bfr} \nonumber \\ 
& \to 
\begin{cases}
 g V_A (r) {\displaystyle \sqrt{\frac{T_F}{N_c}}} \delta^{ab} \bfr \cdot \bfK \textrm{ for temporal gluons} \\
- g V_A (r) {\displaystyle \sqrt{\frac{T_F}{N_c}}} \delta^{ab} \bfr^i K^0 \textrm{ for spatial gluons}
\end{cases},
\end{align}
which is also the Feynman rule for $g V_A (r) \Tr (\textrm{S}^\dagger \bfr \cdot \bfE \, \textrm{O})$. As far as the octet-octet sector is concerned, the treatment of $(g V_B(r) / 2) \Tr (\textrm{O}^\dagger \{ \bfr \cdot \bfE,  \textrm{O} \})$ is equally simple and boils down to
\begin{align}
& g \frac{V_B (r)}{2} \bfr^i \frac{\Tr ( T^a \{ T^b , T^c \}) }{T_F}
[a_o (\bfp',\bfP',c_1',c_2'), O^{a \dagger}]  [\bfE^{b i}, g^\dagger (\bfK, \lambda, d)] [ O^c , a^\dagger_o (\bfp,\bfP,c_1,c_2)]   \nonumber \\
& \overset{\mathcal{O}(g)}{=} i g \frac{V_B (r)}{2} \frac{\Tr ( T^a \{ T^b , T^c \}) }{T_F}  \bfr^i V^{\dagger a}_{c_1',c_2'} V^{ c}_{c_1,c_2} ( -  \bfK^i \varepsilon^{a 0}(K,\lambda,d) +  K^0 \bfeps^{a i}(K,\lambda,d)) \nonumber \\
& \hspace{3ex} \times e^{ -i (P + K - P') \cdot R + i (\bfp - \bfp') \cdot \bfr} \nonumber \\ 
& \to 
\begin{cases}
 g {\displaystyle \frac{V_B (r)}{2}  d^{abc} \bfr \cdot \bfK \textrm{ for temporal gluons}} \phantom{\displaystyle \sqrt{\frac{T_F}{N_c}}} \\
 {- g \displaystyle \frac{V_B (r)}{2} d^{abc} \bfr^i K^0 \textrm{ for spatial gluons}} \phantom{\displaystyle \sqrt{\frac{T_F}{N_c}}}
\end{cases},
\end{align}
where we used that $\Tr ( T^a \{ T^b , T^c \}) / T_F = d^{abc}$. 

Finally, we remark that in order to handle the pNRQCD singlet and octet propagators, given by $i/(P^0 - h_s)$ and $i/(P^0 - h_o)$ respectively, it is useful to write the identity operator in terms of the eigenstates $\ket{n_{s/o}}$ of the operator $h_{s/o}$ \ie
\begin{equation}
\frac{i}{P^0 - {h}_{s/o}} = \frac{i}{P^0 - {h}_{s/o}} \sum_n \ket{n_{s/o}} \bra{n_{s/o}} =\sum_n \frac{i}{P^0 - E_{s/o,n}}  \ket{n_{s/o}} \bra{n_{s/o}},
\end{equation}
where $i/(P^0 - E_{s/o,n})$ is now a c-number, with $E_{s/o,n}$ being the eigenvalues to the eigenstates $\ket{n_{s/o}}$. For example, in the calculation of the 1-loop singlet self-energy between the final state $\ket{n_{s,1}}$ and the initial state $\ket{n_{s,2}}$, the quantum mechanical part evaluates to
\begin{align}
& \bra{n_{s,1}} \bfr^i \frac{i}{P^0 - {h}_{o}} \bfr^j \ket{n_{s,2}} 
= \sum_{m_o} \frac{i}{P^0 - E_{m_o}}  \braket{n_{s,1}|\bfr^i|m_o} 
\braket{m_o|\bfr^j |n_{s,2}}.
\end{align}

\bibliographystyle{jhep}
\bibliography{inspire.bib}

\providecommand{\href}[2]{#2}\begingroup\raggedright\begin{thebibliography}{100}

\bibitem{Wilson:1973jj}
K.~Wilson and J.~B. Kogut, \emph{{The Renormalization group and the epsilon
  expansion}}, \href{https://doi.org/10.1016/0370-1573(74)90023-4}{\emph{Phys.
  Rept.} {\bfseries 12} (1974) 75}.

\bibitem{Weinberg:1978kz}
S.~Weinberg, \emph{{Phenomenological Lagrangians}},
  \href{https://doi.org/10.1016/0378-4371(79)90223-1}{\emph{Physica} {\bfseries
  A96} (1979) 327}.

\bibitem{Petrov:2016azi}
A.~A. Petrov and A.~E. Blechman, \emph{Effective Field Theories}. World
  Scientific Publishing Company, Singapore, 2016,
  \href{https://doi.org/10.1142/8619}{10.1142/8619}.

\bibitem{Pich:2018ltt}
A.~Pich, \emph{{Effective Field Theory with Nambu-Goldstone Modes}},  in
  \emph{{Les Houches summer school}: {EFT in Particle Physics and Cosmology}},
  4, 2018, \href{https://arxiv.org/abs/1804.05664}{{\ttfamily 1804.05664}}.

\bibitem{Manohar:2018aog}
A.~V. Manohar, \emph{{Introduction to Effective Field Theories}},  in
  \emph{{Les Houches summer school}: {EFT in Particle Physics and Cosmology}},
  4, 2018, \href{https://arxiv.org/abs/1804.05863}{{\ttfamily 1804.05863}}.

\bibitem{Caswell:1985ui}
W.~E. Caswell and G.~P. Lepage, \emph{{Effective Lagrangians for Bound State
  Problems in QED, QCD, and Other Field Theories}},
  \href{https://doi.org/10.1016/0370-2693(86)91297-9}{\emph{Phys. Lett.}
  {\bfseries 167B} (1986) 437}.

\bibitem{Bodwin:1994jh}
G.~T. Bodwin, E.~Braaten and G.~P. Lepage, \emph{{Rigorous QCD analysis of
  inclusive annihilation and production of heavy quarkonium}},
  \href{https://doi.org/10.1103/PhysRevD.55.5853,
  10.1103/PhysRevD.51.1125}{\emph{Phys. Rev.} {\bfseries D51} (1995) 1125}
  [\href{https://arxiv.org/abs/hep-ph/9407339}{{\ttfamily hep-ph/9407339}}].

\bibitem{Brambilla:2004jw}
N.~Brambilla, A.~Pineda, J.~Soto and A.~Vairo, \emph{{Effective Field Theories
  for Heavy Quarkonium}},
  \href{https://doi.org/10.1103/RevModPhys.77.1423}{\emph{Rev. Mod. Phys.}
  {\bfseries 77} (2005) 1423}
  [\href{https://arxiv.org/abs/hep-ph/0410047}{{\ttfamily hep-ph/0410047}}].

\bibitem{Brambilla:2017ffe}
N.~Brambilla, V.~Shtabovenko, J.~Tarrús~Castellà and A.~Vairo,
  \emph{{Effective field theories for van der Waals interactions}},
  \href{https://doi.org/10.1103/PhysRevD.95.116004}{\emph{Phys. Rev.}
  {\bfseries D95} (2017) 116004}
  [\href{https://arxiv.org/abs/1704.03476}{{\ttfamily 1704.03476}}].

\bibitem{Brambilla:2017uyf}
N.~Brambilla, G.~Krein, J.~Tarrús~Castellà and A.~Vairo,
  \emph{{Born-Oppenheimer approximation in an effective field theory
  language}}, \href{https://doi.org/10.1103/PhysRevD.97.016016}{\emph{Phys.
  Rev.} {\bfseries D97} (2018) 016016}
  [\href{https://arxiv.org/abs/1707.09647}{{\ttfamily 1707.09647}}].

\bibitem{Hisano:2002fk}
J.~Hisano, S.~Matsumoto and M.~M. Nojiri, \emph{{Unitarity and higher order
  corrections in neutralino dark matter annihilation into two photons}},
  \href{https://doi.org/10.1103/PhysRevD.67.075014}{\emph{Phys. Rev. D}
  {\bfseries 67} (2003) 075014}
  [\href{https://arxiv.org/abs/hep-ph/0212022}{{\ttfamily hep-ph/0212022}}].

\bibitem{Hisano:2003ec}
J.~Hisano, S.~Matsumoto and M.~M. Nojiri, \emph{{Explosive dark matter
  annihilation}},
  \href{https://doi.org/10.1103/PhysRevLett.92.031303}{\emph{Phys. Rev. Lett.}
  {\bfseries 92} (2004) 031303}
  [\href{https://arxiv.org/abs/hep-ph/0307216}{{\ttfamily hep-ph/0307216}}].

\bibitem{Hisano:2004ds}
J.~Hisano, S.~Matsumoto, M.~M. Nojiri and O.~Saito, \emph{{Non-perturbative
  effect on dark matter annihilation and gamma ray signature from galactic
  center}}, \href{https://doi.org/10.1103/PhysRevD.71.063528}{\emph{Phys. Rev.
  D} {\bfseries 71} (2005) 063528}
  [\href{https://arxiv.org/abs/hep-ph/0412403}{{\ttfamily hep-ph/0412403}}].

\bibitem{Shepherd:2009sa}
W.~Shepherd, T.~M. Tait and G.~Zaharijas, \emph{{Bound states of weakly
  interacting dark matter}},
  \href{https://doi.org/10.1103/PhysRevD.79.055022}{\emph{Phys. Rev. D}
  {\bfseries 79} (2009) 055022}
  [\href{https://arxiv.org/abs/0901.2125}{{\ttfamily 0901.2125}}].

\bibitem{An:2015pva}
H.~An, B.~Echenard, M.~Pospelov and Y.~Zhang, \emph{{Probing the Dark Sector
  with Dark Matter Bound States}},
  \href{https://doi.org/10.1103/PhysRevLett.116.151801}{\emph{Phys. Rev. Lett.}
  {\bfseries 116} (2016) 151801}
  [\href{https://arxiv.org/abs/1510.05020}{{\ttfamily 1510.05020}}].

\bibitem{Biondini:2018pwp}
S.~Biondini and M.~Laine, \emph{{Thermal dark matter co-annihilating with a
  strongly interacting scalar}},
  \href{https://doi.org/10.1007/JHEP04(2018)072}{\emph{JHEP} {\bfseries 04}
  (2018) 072} [\href{https://arxiv.org/abs/1801.05821}{{\ttfamily
  1801.05821}}].

\bibitem{Biondini:2018xor}
S.~Biondini, \emph{{Bound-state effects for dark matter with Higgs-like
  mediators}}, \href{https://doi.org/10.1007/JHEP06(2018)104}{\emph{JHEP}
  {\bfseries 06} (2018) 104}
  [\href{https://arxiv.org/abs/1805.00353}{{\ttfamily 1805.00353}}].

\bibitem{Beneke:2019vhz}
M.~Beneke, A.~Broggio, C.~Hasner, K.~Urban and M.~Vollmann, \emph{{Resummed
  photon spectrum from dark matter annihilation for intermediate and narrow
  energy resolution}},
  \href{https://doi.org/10.1007/JHEP08(2019)103}{\emph{JHEP} {\bfseries 08}
  (2019) 103} [\href{https://arxiv.org/abs/1903.08702}{{\ttfamily
  1903.08702}}].

\bibitem{Biondini:2013xua}
S.~Biondini, N.~Brambilla, M.~A. Escobedo and A.~Vairo, \emph{{An effective
  field theory for non-relativistic Majorana neutrinos}},
  \href{https://doi.org/10.1007/JHEP12(2013)028}{\emph{JHEP} {\bfseries 12}
  (2013) 028} [\href{https://arxiv.org/abs/1307.7680}{{\ttfamily 1307.7680}}].

\bibitem{Mertig:1990an}
R.~Mertig, M.~Bohm and A.~Denner, \emph{{FEYN CALC: Computer algebraic
  calculation of Feynman amplitudes}},
  \href{https://doi.org/10.1016/0010-4655(91)90130-D}{\emph{Comput. Phys.
  Commun.} {\bfseries 64} (1991) 345}.

\bibitem{Shtabovenko:2016sxi}
V.~Shtabovenko, R.~Mertig and F.~Orellana, \emph{{New Developments in FeynCalc
  9.0}}, \href{https://doi.org/10.1016/j.cpc.2016.06.008}{\emph{Comput. Phys.
  Commun.} {\bfseries 207} (2016) 432}
  [\href{https://arxiv.org/abs/1601.01167}{{\ttfamily 1601.01167}}].

\bibitem{Hahn:1998yk}
T.~Hahn and M.~Perez-Victoria, \emph{{Automatized one loop calculations in
  four-dimensions and D-dimensions}},
  \href{https://doi.org/10.1016/S0010-4655(98)00173-8}{\emph{Comput. Phys.
  Commun.} {\bfseries 118} (1999) 153}
  [\href{https://arxiv.org/abs/hep-ph/9807565}{{\ttfamily hep-ph/9807565}}].

\bibitem{Patel:2015tea}
H.~H. Patel, \emph{{Package-X: A Mathematica package for the analytic
  calculation of one-loop integrals}},
  \href{https://doi.org/10.1016/j.cpc.2015.08.017}{\emph{Comput. Phys. Commun.}
  {\bfseries 197} (2015) 276}
  [\href{https://arxiv.org/abs/1503.01469}{{\ttfamily 1503.01469}}].

\bibitem{Patel:2016fam}
H.~H. Patel, \emph{{Package-X 2.0: A Mathematica package for the analytic
  calculation of one-loop integrals}},
  \href{https://doi.org/10.1016/j.cpc.2017.04.015}{\emph{Comput. Phys. Commun.}
  {\bfseries 218} (2017) 66}
  [\href{https://arxiv.org/abs/1612.00009}{{\ttfamily 1612.00009}}].

\bibitem{Wiebusch:2014qba}
M.~Wiebusch, \emph{{HEPMath 1.4: A mathematica package for semi-automatic
  computations in high energy physics}},
  \href{https://doi.org/10.1016/j.cpc.2015.04.022}{\emph{Comput. Phys. Commun.}
  {\bfseries 195} (2015) 172}
  [\href{https://arxiv.org/abs/1412.6102}{{\ttfamily 1412.6102}}].

\bibitem{Cyrol:2016zqb}
A.~K. Cyrol, M.~Mitter and N.~Strodthoff, \emph{{FormTracer - A Mathematica
  Tracing Package Using FORM}},
  \href{https://doi.org/10.1016/j.cpc.2017.05.024}{\emph{Comput. Phys. Commun.}
  {\bfseries 219} (2017) 346}
  [\href{https://arxiv.org/abs/1610.09331}{{\ttfamily 1610.09331}}].

\bibitem{Vermaseren:2000nd}
J.~A.~M. Vermaseren, \emph{{New features of FORM}},
  \href{https://arxiv.org/abs/math-ph/0010025}{{\ttfamily math-ph/0010025}}.

\bibitem{Kuipers:2012rf}
J.~Kuipers, T.~Ueda, J.~A.~M. Vermaseren and J.~Vollinga, \emph{{FORM version
  4.0}}, \href{https://doi.org/10.1016/j.cpc.2012.12.028}{\emph{Comput. Phys.
  Commun.} {\bfseries 184} (2013) 1453}
  [\href{https://arxiv.org/abs/1203.6543}{{\ttfamily 1203.6543}}].

\bibitem{Shtabovenko:2017iqw}
V.~Shtabovenko, \emph{{Nonrelativistic Effective Field Theories of QED and QCD:
  Applications and Automatic Calculations}}, Ph.D. thesis, Munich, Tech. U.,
  2017.

\bibitem{Pineda:1997bj}
A.~Pineda and J.~Soto, \emph{{Effective field theory for ultrasoft momenta in
  NRQCD and NRQED}},
  \href{https://doi.org/10.1016/S0920-5632(97)01102-X}{\emph{Nucl. Phys. Proc.
  Suppl.} {\bfseries 64} (1998) 428}
  [\href{https://arxiv.org/abs/hep-ph/9707481}{{\ttfamily hep-ph/9707481}}].

\bibitem{Brambilla:1999xf}
N.~Brambilla, A.~Pineda, J.~Soto and A.~Vairo, \emph{{Potential NRQCD: An
  Effective theory for heavy quarkonium}},
  \href{https://doi.org/10.1016/S0550-3213(99)00693-8}{\emph{Nucl. Phys.}
  {\bfseries B566} (2000) 275}
  [\href{https://arxiv.org/abs/hep-ph/9907240}{{\ttfamily hep-ph/9907240}}].

\bibitem{Shtabovenko:2020gxv}
V.~Shtabovenko, R.~Mertig and F.~Orellana, \emph{Feyncalc 9.3: New features and
  improvements},  \href{https://arxiv.org/abs/2001.04407}{{\ttfamily
  2001.04407}}.

\bibitem{Isgur:1989vq}
N.~Isgur and M.~B. Wise, \emph{{Weak Decays of Heavy Mesons in the Static Quark
  Approximation}},
  \href{https://doi.org/10.1016/0370-2693(89)90566-2}{\emph{Phys. Lett. B}
  {\bfseries 232} (1989) 113}.

\bibitem{Isgur:1989ed}
N.~Isgur and M.~B. Wise, \emph{{Weak transition form factors between heavy
  mesons}}, \href{https://doi.org/10.1016/0370-2693(90)91219-2}{\emph{Phys.
  Lett. B} {\bfseries 237} (1990) 527}.

\bibitem{Eichten:1990vp}
E.~Eichten and B.~R. Hill, \emph{{Static effective field theory: $1/m$
  corrections}},
  \href{https://doi.org/10.1016/0370-2693(90)91408-4}{\emph{Phys. Lett. B}
  {\bfseries 243} (1990) 427}.

\bibitem{Georgi:1990um}
H.~Georgi, \emph{{An Effective Field Theory for Heavy Quarks at Low-energies}},
  \href{https://doi.org/10.1016/0370-2693(90)91128-X}{\emph{Phys. Lett. B}
  {\bfseries 240} (1990) 447}.

\bibitem{Grinstein:1990mj}
B.~Grinstein, \emph{{The Static Quark Effective Theory}},
  \href{https://doi.org/10.1016/0550-3213(90)90349-I}{\emph{Nucl. Phys. B}
  {\bfseries 339} (1990) 253}.

\bibitem{Buchmuller:1985jz}
W.~Buchmüller and D.~Wyler, \emph{{Effective Lagrangian Analysis of New
  Interactions and Flavor Conservation}},
  \href{https://doi.org/10.1016/0550-3213(86)90262-2}{\emph{Nucl.\ Phys.\ B}
  {\bfseries 268} (1986) 621}.

\bibitem{Grzadkowski:2010es}
B.~Grzadkowski, M.~Iskrzynski, M.~Misiak and J.~Rosiek, \emph{{Dimension-Six
  Terms in the Standard Model Lagrangian}},
  \href{https://doi.org/10.1007/JHEP10(2010)085}{\emph{JHEP} {\bfseries 10}
  (2010) 085} [\href{https://arxiv.org/abs/1008.4884}{{\ttfamily 1008.4884}}].

\bibitem{Falkowski:2015wza}
A.~Falkowski, B.~Fuks, K.~Mawatari, K.~Mimasu, F.~Riva and V.~Sanz,
  \emph{{Rosetta: an operator basis translator for Standard Model effective
  field theory}},
  \href{https://doi.org/10.1140/epjc/s10052-015-3806-x}{\emph{Eur. Phys. J.}
  {\bfseries C75} (2015) 583}
  [\href{https://arxiv.org/abs/1508.05895}{{\ttfamily 1508.05895}}].

\bibitem{Brivio:2017btx}
I.~Brivio, Y.~Jiang and M.~Trott, \emph{{The SMEFTsim package, theory and
  tools}}, \href{https://doi.org/10.1007/JHEP12(2017)070}{\emph{JHEP}
  {\bfseries 12} (2017) 070}
  [\href{https://arxiv.org/abs/1709.06492}{{\ttfamily 1709.06492}}].

\bibitem{Criado:2017khh}
J.~C. Criado, \emph{{MatchingTools: a Python library for symbolic effective
  field theory calculations}},
  \href{https://doi.org/10.1016/j.cpc.2018.02.016}{\emph{Comput. Phys. Commun.}
  {\bfseries 227} (2018) 42}
  [\href{https://arxiv.org/abs/1710.06445}{{\ttfamily 1710.06445}}].

\bibitem{Bakshi:2018ics}
S.~Das~Bakshi, J.~Chakrabortty and S.~K. Patra, \emph{{CoDEx: Wilson
  coefficient calculator connecting SMEFT to UV theory}},
  \href{https://doi.org/10.1140/epjc/s10052-018-6444-2}{\emph{Eur. Phys. J.}
  {\bfseries C79} (2019) 21}
  [\href{https://arxiv.org/abs/1808.04403}{{\ttfamily 1808.04403}}].

\bibitem{Aebischer:2018bkb}
J.~Aebischer, J.~Kumar and D.~M. Straub, \emph{{Wilson: a Python package for
  the running and matching of Wilson coefficients above and below the
  electroweak scale}},
  \href{https://doi.org/10.1140/epjc/s10052-018-6492-7}{\emph{Eur. Phys. J.}
  {\bfseries C78} (2018) 1026}
  [\href{https://arxiv.org/abs/1804.05033}{{\ttfamily 1804.05033}}].

\bibitem{Gripaios:2018zrz}
B.~Gripaios and D.~Sutherland, \emph{{DEFT: A program for operators in EFT}},
  \href{https://doi.org/10.1007/JHEP01(2019)128}{\emph{JHEP} {\bfseries 01}
  (2019) 128} [\href{https://arxiv.org/abs/1807.07546}{{\ttfamily
  1807.07546}}].

\bibitem{Dedes:2019uzs}
A.~Dedes, M.~Paraskevas, J.~Rosiek, K.~Suxho and L.~Trifyllis, \emph{{SmeftFR
  -- Feynman rules generator for the Standard Model Effective Field Theory}},
  \href{https://doi.org/10.1016/j.cpc.2019.106931}{\emph{Comput. Phys. Commun.}
  {\bfseries 247} (2020) 106931}
  [\href{https://arxiv.org/abs/1904.03204}{{\ttfamily 1904.03204}}].

\bibitem{Criado:2019ugp}
J.~C. Criado, \emph{{BasisGen: automatic generation of operator bases}},
  \href{https://doi.org/10.1140/epjc/s10052-019-6769-5}{\emph{Eur. Phys. J.}
  {\bfseries C79} (2019) 256}
  [\href{https://arxiv.org/abs/1901.03501}{{\ttfamily 1901.03501}}].

\bibitem{Fonseca:2017lem}
R.~M. Fonseca, \emph{{The Sym2Int program: going from symmetries to
  interactions}},
  \href{https://doi.org/10.1088/1742-6596/873/1/012045}{\emph{J. Phys. Conf.
  Ser.} {\bfseries 873} (2017) 012045}
  [\href{https://arxiv.org/abs/1703.05221}{{\ttfamily 1703.05221}}].

\bibitem{Marinissen:2020jmb}
C.~B. Marinissen, R.~Rahn and W.~J. Waalewijn, \emph{{..., 83106786, 114382724,
  1509048322, 2343463290, 27410087742, ... Efficient Hilbert Series for
  Effective Theories}},  \href{https://arxiv.org/abs/2004.09521}{{\ttfamily
  2004.09521}}.

\bibitem{Banerjee:2020bym}
U.~Banerjee, J.~Chakrabortty, S.~Prakash and S.~U. Rahaman, \emph{{Characters
  and Group Invariant Polynomials of (Super)fields: Road to "Lagrangian"}},
  \href{https://arxiv.org/abs/2004.12830}{{\ttfamily 2004.12830}}.

\bibitem{Aebischer:2017ugx}
J.~Aebischer et~al., \emph{{WCxf: an exchange format for Wilson coefficients
  beyond the Standard Model}},
  \href{https://doi.org/10.1016/j.cpc.2018.05.022}{\emph{Comput. Phys. Commun.}
  {\bfseries 232} (2018) 71}
  [\href{https://arxiv.org/abs/1712.05298}{{\ttfamily 1712.05298}}].

\bibitem{Christensen:2008py}
N.~D. Christensen and C.~Duhr, \emph{{FeynRules - Feynman rules made easy}},
  \href{https://doi.org/10.1016/j.cpc.2009.02.018}{\emph{Comput. Phys. Commun.}
  {\bfseries 180} (2009) 1614}
  [\href{https://arxiv.org/abs/0806.4194}{{\ttfamily 0806.4194}}].

\bibitem{Alloul:2013bka}
A.~Alloul, N.~D. Christensen, C.~Degrande, C.~Duhr and B.~Fuks,
  \emph{{FeynRules 2.0 - A complete toolbox for tree-level phenomenology}},
  \href{https://doi.org/10.1016/j.cpc.2014.04.012}{\emph{Comput. Phys. Commun.}
  {\bfseries 185} (2014) 2250}
  [\href{https://arxiv.org/abs/1310.1921}{{\ttfamily 1310.1921}}].

\bibitem{Degrande:2014vpa}
C.~Degrande, \emph{{Automatic evaluation of UV and R2 terms for beyond the
  Standard Model Lagrangians: a proof-of-principle}},
  \href{https://doi.org/10.1016/j.cpc.2015.08.015}{\emph{Comput. Phys. Commun.}
  {\bfseries 197} (2015) 239}
  [\href{https://arxiv.org/abs/1406.3030}{{\ttfamily 1406.3030}}].

\bibitem{Degrande:2011ua}
C.~Degrande, C.~Duhr, B.~Fuks, D.~Grellscheid, O.~Mattelaer and T.~Reiter,
  \emph{{UFO - The Universal FeynRules Output}},
  \href{https://doi.org/10.1016/j.cpc.2012.01.022}{\emph{Comput. Phys. Commun.}
  {\bfseries 183} (2012) 1201}
  [\href{https://arxiv.org/abs/1108.2040}{{\ttfamily 1108.2040}}].

\bibitem{Alwall:2014hca}
J.~Alwall, R.~Frederix, S.~Frixione, V.~Hirschi, F.~Maltoni, O.~Mattelaer
  et~al., \emph{{The automated computation of tree-level and next-to-leading
  order differential cross sections, and their matching to parton shower
  simulations}}, \href{https://doi.org/10.1007/JHEP07(2014)079}{\emph{JHEP}
  {\bfseries 07} (2014) 079} [\href{https://arxiv.org/abs/1405.0301}{{\ttfamily
  1405.0301}}].

\bibitem{Cullen:2011ac}
G.~Cullen, N.~Greiner, G.~Heinrich, G.~Luisoni, P.~Mastrolia, G.~Ossola et~al.,
  \emph{{Automated One-Loop Calculations with GoSam}},
  \href{https://doi.org/10.1140/epjc/s10052-012-1889-1}{\emph{Eur. Phys. J.}
  {\bfseries C72} (2012) 1889}
  [\href{https://arxiv.org/abs/1111.2034}{{\ttfamily 1111.2034}}].

\bibitem{Cullen:2014yla}
G.~Cullen et~al., \emph{{GOSAM-2.0: a tool for automated one-loop calculations
  within the Standard Model and beyond}},
  \href{https://doi.org/10.1140/epjc/s10052-014-3001-5}{\emph{Eur. Phys. J.}
  {\bfseries C74} (2014) 3001}
  [\href{https://arxiv.org/abs/1404.7096}{{\ttfamily 1404.7096}}].

\bibitem{Bahr:2008pv}
M.~Bahr et~al., \emph{{Herwig++ Physics and Manual}},
  \href{https://doi.org/10.1140/epjc/s10052-008-0798-9}{\emph{Eur. Phys. J.}
  {\bfseries C58} (2008) 639}
  [\href{https://arxiv.org/abs/0803.0883}{{\ttfamily 0803.0883}}].

\bibitem{Gleisberg:2008ta}
T.~Gleisberg, S.~Hoeche, F.~Krauss, M.~Schonherr, S.~Schumann, F.~Siegert
  et~al., \emph{{Event generation with SHERPA 1.1}},
  \href{https://doi.org/10.1088/1126-6708/2009/02/007}{\emph{JHEP} {\bfseries
  02} (2009) 007} [\href{https://arxiv.org/abs/0811.4622}{{\ttfamily
  0811.4622}}].

\bibitem{Moretti:2001zz}
M.~Moretti, T.~Ohl and J.~Reuter, \emph{{O'Mega: An Optimizing matrix element
  generator}},  \href{https://arxiv.org/abs/hep-ph/0102195}{{\ttfamily
  hep-ph/0102195}}.

\bibitem{Kilian:2007gr}
W.~Kilian, T.~Ohl and J.~Reuter, \emph{{WHIZARD: Simulating Multi-Particle
  Processes at LHC and ILC}},
  \href{https://doi.org/10.1140/epjc/s10052-011-1742-y}{\emph{Eur. Phys. J.}
  {\bfseries C71} (2011) 1742}
  [\href{https://arxiv.org/abs/0708.4233}{{\ttfamily 0708.4233}}].

\bibitem{Belyaev:2012qa}
A.~Belyaev, N.~D. Christensen and A.~Pukhov, \emph{{CalcHEP 3.4 for collider
  physics within and beyond the Standard Model}},
  \href{https://doi.org/10.1016/j.cpc.2013.01.014}{\emph{Comput. Phys. Commun.}
  {\bfseries 184} (2013) 1729}
  [\href{https://arxiv.org/abs/1207.6082}{{\ttfamily 1207.6082}}].

\bibitem{Boos:2004kh}
{\scshape CompHEP} collaboration, \emph{{CompHEP 4.4: Automatic computations
  from Lagrangians to events}},
  \href{https://doi.org/10.1016/j.nima.2004.07.096}{\emph{Nucl. Instrum. Meth.}
  {\bfseries A534} (2004) 250}
  [\href{https://arxiv.org/abs/hep-ph/0403113}{{\ttfamily hep-ph/0403113}}].

\bibitem{Hahn:2000kx}
T.~Hahn, \emph{{Generating Feynman diagrams and amplitudes with FeynArts 3}},
  \href{https://doi.org/10.1016/S0010-4655(01)00290-9}{\emph{Comput. Phys.
  Commun.} {\bfseries 140} (2001) 418}
  [\href{https://arxiv.org/abs/hep-ph/0012260}{{\ttfamily hep-ph/0012260}}].

\bibitem{Bauer:2000ew}
C.~W. Bauer, S.~Fleming and M.~E. Luke, \emph{{Summing Sudakov logarithms in B
  ---> X(s gamma) in effective field theory}},
  \href{https://doi.org/10.1103/PhysRevD.63.014006}{\emph{Phys. Rev. D}
  {\bfseries 63} (2000) 014006}
  [\href{https://arxiv.org/abs/hep-ph/0005275}{{\ttfamily hep-ph/0005275}}].

\bibitem{Bauer:2000yr}
C.~W. Bauer, S.~Fleming, D.~Pirjol and I.~W. Stewart, \emph{{An Effective field
  theory for collinear and soft gluons: Heavy to light decays}},
  \href{https://doi.org/10.1103/PhysRevD.63.114020}{\emph{Phys. Rev. D}
  {\bfseries 63} (2001) 114020}
  [\href{https://arxiv.org/abs/hep-ph/0011336}{{\ttfamily hep-ph/0011336}}].

\bibitem{Bauer:2001ct}
C.~W. Bauer and I.~W. Stewart, \emph{{Invariant operators in collinear
  effective theory}},
  \href{https://doi.org/10.1016/S0370-2693(01)00902-9}{\emph{Phys. Lett. B}
  {\bfseries 516} (2001) 134}
  [\href{https://arxiv.org/abs/hep-ph/0107001}{{\ttfamily hep-ph/0107001}}].

\bibitem{Bauer:2001yt}
C.~W. Bauer, D.~Pirjol and I.~W. Stewart, \emph{{Soft collinear factorization
  in effective field theory}},
  \href{https://doi.org/10.1103/PhysRevD.65.054022}{\emph{Phys. Rev. D}
  {\bfseries 65} (2002) 054022}
  [\href{https://arxiv.org/abs/hep-ph/0109045}{{\ttfamily hep-ph/0109045}}].

\bibitem{Bauer:2002nz}
C.~W. Bauer, S.~Fleming, D.~Pirjol, I.~Z. Rothstein and I.~W. Stewart,
  \emph{{Hard scattering factorization from effective field theory}},
  \href{https://doi.org/10.1103/PhysRevD.66.014017}{\emph{Phys. Rev. D}
  {\bfseries 66} (2002) 014017}
  [\href{https://arxiv.org/abs/hep-ph/0202088}{{\ttfamily hep-ph/0202088}}].

\bibitem{Beneke:2002ph}
M.~Beneke, A.~Chapovsky, M.~Diehl and T.~Feldmann, \emph{{Soft collinear
  effective theory and heavy to light currents beyond leading power}},
  \href{https://doi.org/10.1016/S0550-3213(02)00687-9}{\emph{Nucl. Phys. B}
  {\bfseries 643} (2002) 431}
  [\href{https://arxiv.org/abs/hep-ph/0206152}{{\ttfamily hep-ph/0206152}}].

\bibitem{Gasser:1983yg}
J.~Gasser and H.~Leutwyler, \emph{{Chiral Perturbation Theory to One Loop}},
  \href{https://doi.org/10.1016/0003-4916(84)90242-2}{\emph{Annals Phys.}
  {\bfseries 158} (1984) 142}.

\bibitem{Gasser:1984gg}
J.~Gasser and H.~Leutwyler, \emph{{Chiral Perturbation Theory: Expansions in
  the Mass of the Strange Quark}},
  \href{https://doi.org/10.1016/0550-3213(85)90492-4}{\emph{Nucl. Phys. B}
  {\bfseries 250} (1985) 465}.

\bibitem{Bishara:2017nnn}
F.~Bishara, J.~Brod, B.~Grinstein and J.~Zupan, \emph{{DirectDM: a tool for
  dark matter direct detection}},
  \href{https://arxiv.org/abs/1708.02678}{{\ttfamily 1708.02678}}.

\bibitem{Bishara:2016hek}
F.~Bishara, J.~Brod, B.~Grinstein and J.~Zupan, \emph{{Chiral Effective Theory
  of Dark Matter Direct Detection}},
  \href{https://doi.org/10.1088/1475-7516/2017/02/009}{\emph{JCAP} {\bfseries
  1702} (2017) 009} [\href{https://arxiv.org/abs/1611.00368}{{\ttfamily
  1611.00368}}].

\bibitem{Bishara:2017pfq}
F.~Bishara, J.~Brod, B.~Grinstein and J.~Zupan, \emph{{From quarks to nucleons
  in dark matter direct detection}},
  \href{https://doi.org/10.1007/JHEP11(2017)059}{\emph{JHEP} {\bfseries 11}
  (2017) 059} [\href{https://arxiv.org/abs/1707.06998}{{\ttfamily
  1707.06998}}].

\bibitem{Unterdorfer:2005au}
R.~Unterdorfer and G.~Ecker, \emph{{Generating functional for strong and
  nonleptonic weak interactions}},
  \href{https://doi.org/10.1088/1126-6708/2005/10/017}{\emph{JHEP} {\bfseries
  10} (2005) 017} [\href{https://arxiv.org/abs/hep-ph/0507173}{{\ttfamily
  hep-ph/0507173}}].

\bibitem{Grozin:2000jv}
A.~G. Grozin, \emph{{Calculating three loop diagrams in heavy quark effective
  theory with integration by parts recurrence relations}},
  \href{https://doi.org/10.1088/1126-6708/2000/03/013}{\emph{JHEP} {\bfseries
  03} (2000) 013} [\href{https://arxiv.org/abs/hep-ph/0002266}{{\ttfamily
  hep-ph/0002266}}].

\bibitem{Bell:2018oqa}
G.~Bell, R.~Rahn and J.~Talbert, \emph{{Generic dijet soft functions at
  two-loop order: correlated emissions}},
  \href{https://doi.org/10.1007/JHEP07(2019)101}{\emph{JHEP} {\bfseries 07}
  (2019) 101} [\href{https://arxiv.org/abs/1812.08690}{{\ttfamily
  1812.08690}}].

\bibitem{Artoisenet:2007qm}
P.~Artoisenet, F.~Maltoni and T.~Stelzer, \emph{{Automatic generation of
  quarkonium amplitudes in NRQCD}},
  \href{https://doi.org/10.1088/1126-6708/2008/02/102}{\emph{JHEP} {\bfseries
  02} (2008) 102} [\href{https://arxiv.org/abs/0712.2770}{{\ttfamily
  0712.2770}}].

\bibitem{Shao:2012iz}
H.-S. Shao, \emph{{HELAC-Onia: An automatic matrix element generator for heavy
  quarkonium physics}},
  \href{https://doi.org/10.1016/j.cpc.2013.05.023}{\emph{Comput. Phys. Commun.}
  {\bfseries 184} (2013) 2562}
  [\href{https://arxiv.org/abs/1212.5293}{{\ttfamily 1212.5293}}].

\bibitem{Shao:2015vga}
H.-S. Shao, \emph{{HELAC-Onia 2.0: an upgraded matrix-element and event
  generator for heavy quarkonium physics}},
  \href{https://doi.org/10.1016/j.cpc.2015.09.011}{\emph{Comput. Phys. Commun.}
  {\bfseries 198} (2016) 238}
  [\href{https://arxiv.org/abs/1507.03435}{{\ttfamily 1507.03435}}].

\bibitem{Wang:2004du}
J.-X. Wang, \emph{{Progress in FDC project}},
  \href{https://doi.org/10.1016/j.nima.2004.07.094}{\emph{Nucl. Instrum. Meth.}
  {\bfseries A534} (2004) 241}
  [\href{https://arxiv.org/abs/hep-ph/0407058}{{\ttfamily hep-ph/0407058}}].

\bibitem{Wan:2014vka}
L.-P. Wan and J.-X. Wang, \emph{{FDCHQHP: A Fortran package for heavy
  quarkonium hadroproduction}},
  \href{https://doi.org/10.1016/j.cpc.2014.06.022}{\emph{Comput. Phys. Commun.}
  {\bfseries 185} (2014) 2939}
  [\href{https://arxiv.org/abs/1405.2143}{{\ttfamily 1405.2143}}].

\bibitem{Nogueira:1991ex}
P.~Nogueira, \emph{{Automatic Feynman graph generation}},
  \href{https://doi.org/10.1006/jcph.1993.1074}{\emph{J. Comput. Phys.}
  {\bfseries 105} (1993) 279}.

\bibitem{Bolotin:2013qgr}
D.~A. Bolotin and S.~V. Poslavsky, \emph{{Introduction to Redberry: a computer
  algebra system designed for tensor manipulation}},
  \href{https://arxiv.org/abs/1302.1219}{{\ttfamily 1302.1219}}.

\bibitem{Feng:2012tk}
F.~Feng and R.~Mertig, \emph{{FormLink/FeynCalcFormLink : Embedding FORM in
  Mathematica and FeynCalc}},
  \href{https://arxiv.org/abs/1212.3522}{{\ttfamily 1212.3522}}.

\bibitem{Smirnov:2006ry}
V.~A. Smirnov, \emph{{Feynman integral calculus}}. Springer-Verlag Berlin
  Heidelberg, 2006,
  \href{https://doi.org/10.1007/3-540-30611-0}{10.1007/3-540-30611-0}.

\bibitem{Chetyrkin:1981qh}
K.~G. Chetyrkin and F.~V. Tkachov, \emph{{Integration by Parts: The Algorithm
  to Calculate beta Functions in 4 Loops}},
  \href{https://doi.org/10.1016/0550-3213(81)90199-1}{\emph{Nucl. Phys.}
  {\bfseries B192} (1981) 159}.

\bibitem{Tkachov:1981wb}
F.~V. Tkachov, \emph{{A Theorem on Analytical Calculability of Four Loop
  Renormalization Group Functions}},
  \href{https://doi.org/10.1016/0370-2693(81)90288-4}{\emph{Phys. Lett.}
  {\bfseries 100B} (1981) 65}.

\bibitem{Kotikov:1991pm}
A.~V. Kotikov, \emph{{Differential equation method: The Calculation of N point
  Feynman diagrams}}, \href{https://doi.org/10.1016/0370-2693(91)90536-Y,
  10.1016/0370-2693(92)91582-T}{\emph{Phys. Lett.} {\bfseries B267} (1991)
  123}.

\bibitem{Kotikov:1990kg}
A.~V. Kotikov, \emph{{Differential equations method: New technique for massive
  Feynman diagrams calculation}},
  \href{https://doi.org/10.1016/0370-2693(91)90413-K}{\emph{Phys. Lett.}
  {\bfseries B254} (1991) 158}.

\bibitem{Kotikov:1991hm}
A.~V. Kotikov, \emph{{Differential equations method: The Calculation of vertex
  type Feynman diagrams}},
  \href{https://doi.org/10.1016/0370-2693(91)90834-D}{\emph{Phys. Lett.}
  {\bfseries B259} (1991) 314}.

\bibitem{Bern:1993kr}
Z.~Bern, L.~J. Dixon and D.~A. Kosower, \emph{{Dimensionally regulated pentagon
  integrals}}, \href{https://doi.org/10.1016/0550-3213(94)90398-0}{\emph{Nucl.
  Phys.} {\bfseries B412} (1994) 751}
  [\href{https://arxiv.org/abs/hep-ph/9306240}{{\ttfamily hep-ph/9306240}}].

\bibitem{Remiddi:1997ny}
E.~Remiddi, \emph{{Differential equations for Feynman graph amplitudes}},
  {\emph{Nuovo Cim.} {\bfseries A110} (1997) 1435}
  [\href{https://arxiv.org/abs/hep-th/9711188}{{\ttfamily hep-th/9711188}}].

\bibitem{Gehrmann:1999as}
T.~Gehrmann and E.~Remiddi, \emph{{Differential equations for two loop four
  point functions}},
  \href{https://doi.org/10.1016/S0550-3213(00)00223-6}{\emph{Nucl. Phys.}
  {\bfseries B580} (2000) 485}
  [\href{https://arxiv.org/abs/hep-ph/9912329}{{\ttfamily hep-ph/9912329}}].

\bibitem{Binoth:2000ps}
T.~Binoth and G.~Heinrich, \emph{{An automatized algorithm to compute infrared
  divergent multiloop integrals}},
  \href{https://doi.org/10.1016/S0550-3213(00)00429-6}{\emph{Nucl. Phys.}
  {\bfseries B585} (2000) 741}
  [\href{https://arxiv.org/abs/hep-ph/0004013}{{\ttfamily hep-ph/0004013}}].

\bibitem{Binoth:2003ak}
T.~Binoth and G.~Heinrich, \emph{{Numerical evaluation of multiloop integrals
  by sector decomposition}},
  \href{https://doi.org/10.1016/j.nuclphysb.2003.12.023}{\emph{Nucl. Phys.}
  {\bfseries B680} (2004) 375}
  [\href{https://arxiv.org/abs/hep-ph/0305234}{{\ttfamily hep-ph/0305234}}].

\bibitem{Binoth:2004jv}
T.~Binoth and G.~Heinrich, \emph{{Numerical evaluation of phase space integrals
  by sector decomposition}},
  \href{https://doi.org/10.1016/j.nuclphysb.2004.06.005}{\emph{Nucl. Phys.}
  {\bfseries B693} (2004) 134}
  [\href{https://arxiv.org/abs/hep-ph/0402265}{{\ttfamily hep-ph/0402265}}].

\bibitem{Smirnov:1999gc}
V.~A. Smirnov, \emph{{Analytical result for dimensionally regularized massless
  on shell double box}},
  \href{https://doi.org/10.1016/S0370-2693(99)00777-7}{\emph{Phys. Lett.}
  {\bfseries B460} (1999) 397}
  [\href{https://arxiv.org/abs/hep-ph/9905323}{{\ttfamily hep-ph/9905323}}].

\bibitem{Tausk:1999vh}
J.~B. Tausk, \emph{{Nonplanar massless two loop Feynman diagrams with four
  on-shell legs}},
  \href{https://doi.org/10.1016/S0370-2693(99)01277-0}{\emph{Phys. Lett.}
  {\bfseries B469} (1999) 225}
  [\href{https://arxiv.org/abs/hep-ph/9909506}{{\ttfamily hep-ph/9909506}}].

\bibitem{Anastasiou:2005cb}
C.~Anastasiou and A.~Daleo, \emph{{Numerical evaluation of loop integrals}},
  \href{https://doi.org/10.1088/1126-6708/2006/10/031}{\emph{JHEP} {\bfseries
  10} (2006) 031} [\href{https://arxiv.org/abs/hep-ph/0511176}{{\ttfamily
  hep-ph/0511176}}].

\bibitem{Czakon:2005rk}
M.~Czakon, \emph{{Automatized analytic continuation of Mellin-Barnes
  integrals}}, \href{https://doi.org/10.1016/j.cpc.2006.07.002}{\emph{Comput.
  Phys. Commun.} {\bfseries 175} (2006) 559}
  [\href{https://arxiv.org/abs/hep-ph/0511200}{{\ttfamily hep-ph/0511200}}].

\bibitem{Anastasiou:2002yz}
C.~Anastasiou and K.~Melnikov, \emph{{Higgs boson production at hadron
  colliders in NNLO QCD}},
  \href{https://doi.org/10.1016/S0550-3213(02)00837-4}{\emph{Nucl. Phys.}
  {\bfseries B646} (2002) 220}
  [\href{https://arxiv.org/abs/hep-ph/0207004}{{\ttfamily hep-ph/0207004}}].

\bibitem{Anastasiou:2003yy}
C.~Anastasiou, L.~J. Dixon, K.~Melnikov and F.~Petriello, \emph{{Dilepton
  rapidity distribution in the Drell-Yan process at NNLO in QCD}},
  \href{https://doi.org/10.1103/PhysRevLett.91.182002}{\emph{Phys. Rev. Lett.}
  {\bfseries 91} (2003) 182002}
  [\href{https://arxiv.org/abs/hep-ph/0306192}{{\ttfamily hep-ph/0306192}}].

\bibitem{Passarino:1978jh}
G.~Passarino and M.~J.~G. Veltman, \emph{{One Loop Corrections for e+ e-
  Annihilation Into mu+ mu- in the Weinberg Model}},
  \href{https://doi.org/10.1016/0550-3213(79)90234-7}{\emph{Nucl. Phys.}
  {\bfseries B160} (1979) 151}.

\bibitem{Chang:2020hii}
H.-R. Chang, \emph{{Generalized Passarino-Veltman reduction scheme in the
  absence of Lorentz invariance}},
  \href{https://arxiv.org/abs/2008.11314}{{\ttfamily 2008.11314}}.

\bibitem{Devaraj:1997es}
G.~Devaraj and R.~G. Stuart, \emph{{Reduction of one loop tensor form-factors
  to scalar integrals: A General scheme}},
  \href{https://doi.org/10.1016/S0550-3213(98)00035-2}{\emph{Nucl. Phys. B}
  {\bfseries 519} (1998) 483}
  [\href{https://arxiv.org/abs/hep-ph/9704308}{{\ttfamily hep-ph/9704308}}].

\bibitem{Denner:2005nn}
A.~Denner and S.~Dittmaier, \emph{{Reduction schemes for one-loop tensor
  integrals}},
  \href{https://doi.org/10.1016/j.nuclphysb.2005.11.007}{\emph{Nucl. Phys. B}
  {\bfseries 734} (2006) 62}
  [\href{https://arxiv.org/abs/hep-ph/0509141}{{\ttfamily hep-ph/0509141}}].

\bibitem{Shtabovenko:2016whf}
V.~Shtabovenko, \emph{{FeynHelpers: Connecting FeynCalc to FIRE and
  Package-X}}, \href{https://doi.org/10.1016/j.cpc.2017.04.014}{\emph{Comput.
  Phys. Commun.} {\bfseries 218} (2017) 48}
  [\href{https://arxiv.org/abs/1611.06793}{{\ttfamily 1611.06793}}].

\bibitem{Buras:1989xd}
A.~J. Buras and P.~H. Weisz, \emph{{QCD Nonleading Corrections to Weak Decays
  in Dimensional Regularization and 't Hooft-Veltman Schemes}},
  \href{https://doi.org/10.1016/0550-3213(90)90223-Z}{\emph{Nucl. Phys.}
  {\bfseries B333} (1990) 66}.

\bibitem{Li:2013nna}
Y.-J. Li, G.-Z. Xu, K.-Y. Liu and Y.-J. Zhang, \emph{{Search for $C=+$
  charmonium and XYZ states in $e^+e^-\to \gamma+ H$ at BESIII}},
  \href{https://doi.org/10.1007/JHEP01(2014)022}{\emph{JHEP} {\bfseries 01}
  (2014) 022} [\href{https://arxiv.org/abs/1310.0374}{{\ttfamily 1310.0374}}].

\bibitem{Chao:2013cca}
K.-T. Chao, Z.-G. He, D.~Li and C.~Meng, \emph{{Search for $C=+$ charmonium
  states in $e^+e^-\to \gamma+~X$ at BEPCII/BESIII}},
  \href{https://arxiv.org/abs/1310.8597}{{\ttfamily 1310.8597}}.

\bibitem{Brambilla:2017kgw}
N.~Brambilla, W.~Chen, Y.~Jia, V.~Shtabovenko and A.~Vairo, \emph{{Relativistic
  corrections to exclusive $\chi_{cJ} + \gamma$ production from $e^+ e^-$
  annihilation}}, \href{https://doi.org/10.1103/PhysRevD.97.096001}{\emph{Phys.
  Rev.} {\bfseries D97} (2018) 096001}
  [\href{https://arxiv.org/abs/1712.06165}{{\ttfamily 1712.06165}}].

\bibitem{Hoang:2006ty}
A.~H. Hoang and P.~Ruiz-Femenia, \emph{{Heavy pair production currents with
  general quantum numbers in dimensionally regularized NRQCD}},
  \href{https://doi.org/10.1103/PhysRevD.74.114016}{\emph{Phys. Rev.}
  {\bfseries D74} (2006) 114016}
  [\href{https://arxiv.org/abs/hep-ph/0609151}{{\ttfamily hep-ph/0609151}}].

\bibitem{Dugan:1990df}
M.~J. Dugan and B.~Grinstein, \emph{{On the vanishing of evanescent
  operators}}, \href{https://doi.org/10.1016/0370-2693(91)90680-O}{\emph{Phys.
  Lett.} {\bfseries B256} (1991) 239}.

\bibitem{Herrlich:1994kh}
S.~Herrlich and U.~Nierste, \emph{{Evanescent operators, scheme dependences and
  double insertions}},
  \href{https://doi.org/10.1016/0550-3213(95)00474-7}{\emph{Nucl. Phys.}
  {\bfseries B455} (1995) 39}
  [\href{https://arxiv.org/abs/hep-ph/9412375}{{\ttfamily hep-ph/9412375}}].

\bibitem{Braaten:1996rp}
E.~Braaten and Y.-Q. Chen, \emph{{Dimensional regularization in quarkonium
  calculations}}, \href{https://doi.org/10.1103/PhysRevD.55.2693}{\emph{Phys.
  Rev.} {\bfseries D55} (1997) 2693}
  [\href{https://arxiv.org/abs/hep-ph/9610401}{{\ttfamily hep-ph/9610401}}].

\bibitem{Pineda:1998kj}
A.~Pineda and J.~Soto, \emph{{Matching at one loop for the four quark operators
  in NRQCD}}, \href{https://doi.org/10.1103/PhysRevD.58.114011}{\emph{Phys.
  Rev.} {\bfseries D58} (1998) 114011}
  [\href{https://arxiv.org/abs/hep-ph/9802365}{{\ttfamily hep-ph/9802365}}].

\bibitem{Gerlach:2019bso}
M.~Gerlach, G.~Mishima and M.~Steinhauser, \emph{{Matching coefficients in
  nonrelativistic QCD to two-loop accuracy}},
  \href{https://doi.org/10.1103/PhysRevD.100.054016}{\emph{Phys. Rev.}
  {\bfseries D100} (2019) 054016}
  [\href{https://arxiv.org/abs/1907.08227}{{\ttfamily 1907.08227}}].

\bibitem{Braaten:1996jt}
E.~Braaten and Y.-Q. Chen, \emph{{Helicity decomposition for inclusive J / psi
  production}}, \href{https://doi.org/10.1103/PhysRevD.54.3216}{\emph{Phys.
  Rev.} {\bfseries D54} (1996) 3216}
  [\href{https://arxiv.org/abs/hep-ph/9604237}{{\ttfamily hep-ph/9604237}}].

\bibitem{Bodwin:2002hg}
G.~T. Bodwin and A.~Petrelli, \emph{{Order-$v^4$ corrections to $S$-wave
  quarkonium decay}}, \href{https://doi.org/10.1103/PhysRevD.87.039902,
  10.1103/PhysRevD.66.094011}{\emph{Phys. Rev.} {\bfseries D66} (2002) 094011}
  [\href{https://arxiv.org/abs/hep-ph/0205210}{{\ttfamily hep-ph/0205210}}].

\bibitem{Heisenberg:1935qt}
W.~Heisenberg and H.~Euler, \emph{{Consequences of Dirac's theory of
  positrons}}, \href{https://doi.org/10.1007/BF01343663,
  10.1007/978-3-642-70078-1_9}{\emph{Z. Phys.} {\bfseries 98} (1936) 714}
  [\href{https://arxiv.org/abs/physics/0605038}{{\ttfamily physics/0605038}}].

\bibitem{Furry:1937zz}
W.~H. Furry, \emph{{A Symmetry Theorem in the Positron Theory}},
  \href{https://doi.org/10.1103/PhysRev.51.125}{\emph{Phys. Rev.} {\bfseries
  51} (1937) 125}.

\bibitem{Kaplan:2005es}
D.~B. Kaplan, \emph{{Five lectures on effective field theory}},  in
  \emph{Lectures delivered at the 17th National Nuclear Physics Summer School
  2005, Berkeley, CA}, 2005,
  \href{https://arxiv.org/abs/nucl-th/0510023}{{\ttfamily nucl-th/0510023}}.

\bibitem{Grozin:2009an}
A.~G. Grozin, \emph{{Introduction to effective field theories. 1.
  Heisenberg-Euler effective theory, decoupling of heavy flavours}},  in
  \emph{{Helmholtz International School - Workshop on Calculations for Modern
  and Future Colliders (CALC 2009) Dubna, Russia, July 10-20, 2009}}, 2009,
  \href{https://arxiv.org/abs/0908.4392}{{\ttfamily 0908.4392}}.

\bibitem{Smirnov:2014hma}
A.~V. Smirnov, \emph{{FIRE5: a C++ implementation of Feynman Integral
  REduction}}, \href{https://doi.org/10.1016/j.cpc.2014.11.024}{\emph{Comput.
  Phys. Commun.} {\bfseries 189} (2015) 182}
  [\href{https://arxiv.org/abs/1408.2372}{{\ttfamily 1408.2372}}].

\bibitem{Quevillon:2018mfl}
J.~Quevillon, C.~Smith and S.~Touati, \emph{{Effective action for gauge
  bosons}}, \href{https://doi.org/10.1103/PhysRevD.99.013003}{\emph{Phys. Rev.}
  {\bfseries D99} (2019) 013003}
  [\href{https://arxiv.org/abs/1810.06994}{{\ttfamily 1810.06994}}].

\bibitem{Ecker:1995rk}
G.~Ecker and M.~Mojzis, \emph{{Low-energy expansion of the pion - nucleon
  Lagrangian}}, \href{https://doi.org/10.1016/0370-2693(95)01275-3}{\emph{Phys.
  Lett.} {\bfseries B365} (1996) 312}
  [\href{https://arxiv.org/abs/hep-ph/9508204}{{\ttfamily hep-ph/9508204}}].

\bibitem{Scherer:2002tk}
S.~Scherer, \emph{{Introduction to chiral perturbation theory}}, {\emph{Adv.
  Nucl. Phys.} {\bfseries 27} (2003) 277}
  [\href{https://arxiv.org/abs/hep-ph/0210398}{{\ttfamily hep-ph/0210398}}].

\bibitem{Brambilla:2005yk}
N.~Brambilla, A.~Vairo and T.~Rosch, \emph{{Effective field theory Lagrangians
  for baryons with two and three heavy quarks}},
  \href{https://doi.org/10.1103/PhysRevD.72.034021}{\emph{Phys. Rev. D}
  {\bfseries 72} (2005) 034021}
  [\href{https://arxiv.org/abs/hep-ph/0506065}{{\ttfamily hep-ph/0506065}}].

\bibitem{Ore:1949te}
A.~Ore and J.~Powell, \emph{{Three photon annihilation of an electron -
  positron pair}}, \href{https://doi.org/10.1103/PhysRev.75.1696}{\emph{Phys.
  Rev.} {\bfseries 75} (1949) 1696}.

\bibitem{Landau:1948kw}
L.~Landau, \emph{{On the angular momentum of a system of two photons}},
  \href{https://doi.org/10.1016/B978-0-08-010586-4.50070-5}{\emph{Dokl. Akad.
  Nauk SSSR} {\bfseries 60} (1948) 207}.

\bibitem{Yang:1950rg}
C.-N. Yang, \emph{{Selection Rules for the Dematerialization of a Particle Into
  Two Photons}}, \href{https://doi.org/10.1103/PhysRev.77.242}{\emph{Phys.
  Rev.} {\bfseries 77} (1950) 242}.

\bibitem{Brambilla:2006ph}
N.~Brambilla, E.~Mereghetti and A.~Vairo, \emph{{Electromagnetic quarkonium
  decays at order $v^7$}},
  \href{https://doi.org/10.1088/1126-6708/2006/08/039}{\emph{JHEP} {\bfseries
  08} (2006) 039} [\href{https://arxiv.org/abs/hep-ph/0604190}{{\ttfamily
  hep-ph/0604190}}].

\bibitem{Barbieri:1980yp}
R.~Barbieri, M.~Caffo, R.~Gatto and E.~Remiddi, \emph{{Strong QCD Corrections
  to p Wave Quarkonium Decays}},
  \href{https://doi.org/10.1016/0370-2693(80)90407-4}{\emph{Phys. Lett.}
  {\bfseries 95B} (1980) 93}.

\bibitem{Petrelli:1997ge}
A.~Petrelli, M.~Cacciari, M.~Greco, F.~Maltoni and M.~L. Mangano, \emph{{NLO
  production and decay of quarkonium}},
  \href{https://doi.org/10.1016/S0550-3213(97)00801-8}{\emph{Nucl. Phys.}
  {\bfseries B514} (1998) 245}
  [\href{https://arxiv.org/abs/hep-ph/9707223}{{\ttfamily hep-ph/9707223}}].

\bibitem{Butenschon:2009zza}
M.~Butenschoen, \emph{{Photoproduction of the J/psi meson at HERA at
  next-to-leading order within the frmework of nonrelativistic QCD}}, Ph.D.
  thesis, Hamburg U., 2009, DESY-THESIS-2009-021, 10.3204/DESY-THESIS-2009-021.

\bibitem{Brambilla:2019fmu}
N.~Brambilla, H.~S. Chung, W.~K. Lai, V.~Shtabovenko and A.~Vairo, \emph{{Order
  $v^4$ corrections to Higgs boson decay into $J/\psi + \gamma$}},
  \href{https://doi.org/10.1103/PhysRevD.100.054038}{\emph{Phys.\ Rev.\ D}
  {\bfseries 100} (2019) 054038}
  [\href{https://arxiv.org/abs/1907.06473}{{\ttfamily 1907.06473}}].

\bibitem{Coope1970}
J.~A.~R. Coope and R.~F. Snider, \emph{Irreducible cartesian tensors. ii.
  general formulation}, \href{https://doi.org/10.1063/1.1665190}{\emph{J. Math.
  Phys.} {\bfseries 11} (1970) 1003}.

\bibitem{Pineda:2000gza}
A.~Pineda and J.~Soto, \emph{{The Renormalization group improvement of the QCD
  static potentials}},
  \href{https://doi.org/10.1016/S0370-2693(00)01261-2}{\emph{Phys. Lett. B}
  {\bfseries 495} (2000) 323}
  [\href{https://arxiv.org/abs/hep-ph/0007197}{{\ttfamily hep-ph/0007197}}].

\bibitem{Semenov:1996es}
A.~V. Semenov, \emph{{LanHEP: A Package for automatic generation of Feynman
  rules in gauge models}},
  \href{https://arxiv.org/abs/hep-ph/9608488}{{\ttfamily hep-ph/9608488}}.

\bibitem{Semenov:1997qm}
A.~V. Semenov, \emph{{Automatic generation of Feynman rules from the Lagrangian
  by means of LanHEP package}},
  \href{https://doi.org/10.1016/S0168-9002(97)00096-X}{\emph{Nucl. Instrum.
  Meth.} {\bfseries A389} (1997) 293}.

\bibitem{Staub:2009bi}
F.~Staub, \emph{{From Superpotential to Model Files for FeynArts and
  CalcHep/CompHep}},
  \href{https://doi.org/10.1016/j.cpc.2010.01.011}{\emph{Comput. Phys. Commun.}
  {\bfseries 181} (2010) 1077}
  [\href{https://arxiv.org/abs/0909.2863}{{\ttfamily 0909.2863}}].

\bibitem{Staub:2010jh}
F.~Staub, \emph{{Automatic Calculation of supersymmetric Renormalization Group
  Equations and Self Energies}},
  \href{https://doi.org/10.1016/j.cpc.2010.11.030}{\emph{Comput. Phys. Commun.}
  {\bfseries 182} (2011) 808}
  [\href{https://arxiv.org/abs/1002.0840}{{\ttfamily 1002.0840}}].

\bibitem{Staub:2012pb}
F.~Staub, \emph{{SARAH 3.2: Dirac Gauginos, UFO output, and more}},
  \href{https://doi.org/10.1016/j.cpc.2013.02.019}{\emph{Comput. Phys. Commun.}
  {\bfseries 184} (2013) 1792}
  [\href{https://arxiv.org/abs/1207.0906}{{\ttfamily 1207.0906}}].

\bibitem{Staub:2013tta}
F.~Staub, \emph{{SARAH 4 : A tool for (not only SUSY) model builders}},
  \href{https://doi.org/10.1016/j.cpc.2014.02.018}{\emph{Comput. Phys. Commun.}
  {\bfseries 185} (2014) 1773}
  [\href{https://arxiv.org/abs/1309.7223}{{\ttfamily 1309.7223}}].

\bibitem{Becchi:1974md}
C.~Becchi, A.~Rouet and R.~Stora, \emph{{Renormalization of the Abelian
  Higgs-Kibble Model}}, \href{https://doi.org/10.1007/BF01614158}{\emph{Commun.
  Math. Phys.} {\bfseries 42} (1975) 127}.

\bibitem{Tyutin:1975qk}
I.~Tyutin, \emph{{Gauge Invariance in Field Theory and Statistical Physics in
  Operator Formalism}}, {\emph{preprint of P.N. Lebedev Physical Institute}
  {\bfseries No. 39} (1975) }
  [\href{https://arxiv.org/abs/0812.0580}{{\ttfamily 0812.0580}}].

\bibitem{Pineda:2011dg}
A.~Pineda, \emph{{Review of Heavy Quarkonium at weak coupling}},
  \href{https://doi.org/10.1016/j.ppnp.2012.01.038}{\emph{Prog. Part. Nucl.
  Phys.} {\bfseries 67} (2012) 735}
  [\href{https://arxiv.org/abs/1111.0165}{{\ttfamily 1111.0165}}].

\end{thebibliography}\endgroup

\end{document}